\documentclass{jpp}
\usepackage{graphicx}
\usepackage{epstopdf, epsfig}
\usepackage{hyperref}
\usepackage{amsmath,amsfonts,amssymb}	
\usepackage{subfigure}
\usepackage{mathrsfs} 
\usepackage{upgreek}	
\usepackage{xcolor}		
\usepackage{cancel}

\DeclareMathOperator{\Tr}{Tr}

\shorttitle{Beam focusing in Doppler Backscattering}
\shortauthor{Ruiz Ruiz, Parra, Hall-Chen, Belrhali, Giroud, Hillesheim, Lopez}

\title{Beam focusing and consequences for Doppler Backscattering measurements}

\author{J. Ruiz Ruiz\aff{1}\corresp{\email{juan.ruiz@physics.ox.ac.uk}}, F. I. Parra\aff{2}, V. H. Hall-Chen\aff{3}, N. Belrhali\aff{4}, C. Giroud\aff{5}, J. C. Hillesheim\aff{6}, N. A. Lopez\aff{1} and JET contributors\aff{7}}

\affiliation{
\aff{1}Rudolf Peierls Centre for Theoretical Physics, University of Oxford, OX1 3NP, UK
\aff{2}Princeton Plasma Physics Laboratory, Princeton, New Jersey 08543, USA
\aff{3}Institute of High Performance Computing, A*STAR, Singapore 138632, Singapore
\aff{4}Ecole Normale Supérieure (ENS), Paris, France
\aff{5}CCFE, Culham Science Centre, Abingdon, Oxon OX14 3DB, UK
\aff{6}Commonwealth Fusion Systems, Devens, MA, USA
\aff{7}*See the author list of “Overview of T and D-T results in JET with ITER-like wall” by CF Maggi et al. to be published in Nuclear Fusion Special Issue: Overview and Summary Papers from the 29th Fusion Energy Conference (London, UK, 16-21 October 2023).
}

\begin{document}

\maketitle

\begin{abstract}
The phenomenon of beam focusing of microwaves in a plasma near a turning-point caustic is discussed in the context of the analytical solution to the Gaussian beam-tracing equations in the 2D linear-layer problem. The location of maximum beam focusing and the beam width at that location are studied in terms of the beam initial conditions. The analytic solution is used to study the effect of this focusing on Doppler backscattering (DBS). We find that the filter function that characterises the scattering intensity contributions along the beam path through the plasma is inversely proportional to the beam width, predicting enhanced scattering contributions from the beam focusing region. We show that the DBS signal enhancement for small incident angles between the beam path and the density gradient is due to beam focusing and not due to \emph{forward scattering}. The analytic beam model is used to predict the measurement of the $k_y$ density-fluctuation wavenumber power spectrum via DBS, showing that the spectral exponent of the turbulent, intermediate-to-high $k_y$ density-fluctuation spectrum might be quantitatively measurable via DBS, but not the spectral peak corresponding to the driving scale of the turbulent cascade.
\end{abstract}


\section{Introduction}

The confinement of plasmas in magnetically confined fusion experiments, such as tokamaks and stellarators, is dictated by small-scale microturbulence fluctuations. The turbulence produces anomalous transport of particles and heat, determining the background equilibrium profiles of density and temperature. In the past decades, we have improved our understanding of microturbulence fluctuations and their related anomalous transport using experimental measurements \citep{liewer_nf_1985, tynan_ppcf_2009}, analytical calculations of the micro-instabilities driving the turbulence \citep{horton_revmodphys_1999, garbet_ppcf_2001}, and using direct numerical turbulence simulations \citep{garbet_nf_2010} and reduced fluid models \citep{staebler_pop_2005, ivanov_jpp_2020}. To study turbulence in magnetic confinement fusion devices, the theory of gyrokinetics \citep{catto_pp_1978, frieman_chen_physflu_1982} has been developed. Gyrokinetics has been highly successful at predicting the linear micro-instabilities driving turbulence and the associated transport. Due to the complexity of the equations, only linear calculations are analytically tractable in certain limits, but these cannot predict the saturated turbulence. To systematically study the turbulence saturated state and transport, the nonlinear gyrokinetic system of equations \citep{frieman_chen_physflu_1982} has been implemented over the years in performance codes such as GENE \citep{gene_jenko_pop_2000}, GS2 \citep{kotschenreuther_cpc_1995}, GYRO \citep{gyro, gyro_guide}, CGYRO \citep{candy_jcp_2016}, STELLA \citep{barnes_jcp_2019} etc. These codes have been well benchmarked \citep{dimits_pop_2000, nevins_pop_2006, bravenec_pop_2013} and have proven successful at predicting the experimentally inferred transport levels \citep{candy_prl_2003, howard_nf_2013}. 

Gyrokinetic codes can also calculate intrinsic turbulence characteristics, some of which can be measured by fluctuation diagnostics. Detailed comparisons of the intrinsic turbulence characteristics (fluctuation spectrum, correlation length, etc.) between the measurements and simulations remain difficult but are becoming more common practice \citep{white_pop_2008, holland_pop_2009, hillesheim_rsi_2012, holland_nf_2012, leerink_prl_2012, gusakov_ppcf_2013, stroth_nf_2015, lechte_ppcf_2017, happel_ppcf_2017, ruizruiz_ppcf_2019, krutkin_nf_2019}. As we approach burning-plasma scenarios in the coming decade (ITER \citep{ikeda_nf_2007}, SPARC \citep{creely_jpp_2020}, STEP \citep{step_2020}), it is important to measure and characterise detailed physical turbulence processes in today's tokamaks to validate our models and to build confidence in the model's predictions for next generation fusion devices \citep{terry_pop_2008, greenwald_pop_2010, holland_pop_2016, white_jpp_2019}. This motivates a detailed understanding of the complex turbulent phenomena being measured, as well as the detailed physical mechanisms in the measurement process itself. Both should be understood from a fundamental level. 

In order to make quantitative comparisons between fluctuation diagnostics and numerical turbulence simulations, synthetic diagnostics are needed. Synthetic diagnostics enable the understanding of the diagnostic effects on the measurement \citep{bravenec_rsi_2006, shafer_rsi_2006}, and are a natural tool to help design new fluctuation diagnostics. These require a detailed understanding of the physical process behind the experimental measurement: collisional excitation and charge exchange rates in the case of beam emission spectroscopy (BES) \citep{fonck_rsi_1990, hutchinson_2002_book}, electron-cyclotron emission (ECE) physics in the case of ECE \citep{sattler_prl_1994, cima_pop_1995}, plasma sheath physics for magnetic probe measurements \citep{hutchinson_2002_book}, wave diffraction physics in phase-contrast imaging \citep{weisen_rsi_1988, coda_rsi_1992}, and microwave scattering physics in the case of microwave diagnostics such as reflectometry \citep{cripwell_eps_1989, costley_rsi_1990}, Doppler backscattering (DBS) \citep{holzhauer_ppcf_1998, hirsch_rsi_2001} and high-k scattering \citep{mazzucato_prl_1976, surko_prl_1976, slusher_physflu_1980, peebles_rsi_1981}, to name a few. In this manuscript, we focus on the measurement of the turbulence wavenumber spectrum via DBS. We use a linear response beam-tracing model for the propagation and scattering of the microwaves inside the plasma to assess the impact of the beam properties on the backscattered power spectrum measured by DBS.

The Doppler backscattering technique \citep{holzhauer_ppcf_1998, hirsch_rsi_2001} can measure the turbulent wavenumber spectrum \citep{hennequin_nf_2006, hillesheim_nf_2015b}, zonal and equilibrium flows \citep{hirsch_ppcf_2004, hennequin_rsi_2004, hillesheim_prl_2016} as well as the turbulent correlation length \citep{schirmer_ppcf_2007}. To perform the measurement, a beam of microwaves is launched into the core plasma with a finite incidence angle $\alpha_0$ with respect to the density gradient $\nabla n$. The beam propagates in the plasma until it encounters a cutoff surface, following which the forward beam is deviated away from the detector. The detector receives backscattered radiation from all along the beam path from turbulent fluctuations with characteristic wavevector $\mathbf{k}$, which are related to the incident beam wavevector $\mathbf{K}$ via the Bragg condition for backscattering $\mathbf{k} = - 2 \mathbf{K}$. Despite the simple qualitative idea behind the measurement, there is very rich physics impacting the scattering measurement, such as scattering along the path \citep{gusakov_ppcf_2004}, the mismatch angle between the incident beam $\mathbf{K}$ and the turbulence wavenumber $\mathbf{k}$ \citep{rhodes_rsi_2006, hillesheim_nf_2015b, valerian_ppcf_2022, valerian_rsi_2022, valerian_arxiv_2022}, the Doppler shift, and a nonlinear response of the diagnostic for sufficiently large fluctuations \citep{gusakov_ppcf_2005, blanco_estrada_ppcf_2013, fernandezmarina_nf_2014, krutkin_ppcf_2019}. One option to understand some of these effects is to couple high-fidelity full-wave models with direct nonlinear gyrokinetic simulations. This is a daunting task, but has been successfully carried out by some authors in the context of DBS \citep{stroth_nf_2015, happel_ppcf_2017, lechte_ppcf_2017, lechte_pst_2020}. This critical and necessary exercise validates full-physics turbulence simulations as well as full-wave codes, where agreement should be reached between the full-physics modelling and experimental measurements. When agreement is not reached, the difference might be due to insufficient physics or resolution in the simulated turbulence, full-wave simulations, or both. These exercises make it difficult to isolate particular physics phenomena affecting the measurement, in many cases failing to provide fundamental understanding. A second option to understand DBS is to split the problem into smaller pieces and to use reduced models for the turbulence, the wave propagation, or both. With reduced models, one can make analytical progress, have a strong handle on the hypotheses and limitations, and establish a clear origin of the results and predictions. The latter is the approach we have opted for in this manuscript.

First principles theory and analytical calculations of the DBS scattered power spectra have been carried out for almost two decades by Gusakov \emph{et al}. \citep{gusakov_ppcf_2004, gusakov_eps_2011, gusakov_ppcf_2014, gusakov_pop_2017}. These calculations start from the full-wave analytical solution to the Helmholtz wave equation, and they are restricted to 2D geometry in a Cartesian slab, although some preliminary calculations have been performed in a cylinder \citep{gusakov_ppr_2017}. The 2D calculations have highlighted the importance of a so-called "forward scattering" contribution, in addition to the backscattering contribution, for the measured DBS power spectrum. Gusakov \emph{et al}. also analysed the measurement locality, providing insight beyond the traditional understanding that DBS is sensitive to fluctuations from the cutoff (in this manuscript, we will interchangeably refer to this location as the cutoff, or turning point). These calculations use a realistic representation of the wave scattering process and a simple turbulence spectrum (Gaussian) in their analytical calculations. 

In this work, we adopt Gaussian beam tracing to model the propagation of the electric field in a simple 2D slab geometry as in \citep{gusakov_ppcf_2004, gusakov_eps_2011, gusakov_ppcf_2014, gusakov_pop_2017}. The theory of Gaussian beam-tracing has been developed in different fields of physics \citep{casperson_73, cernevy_geoji_1982, kravtsov_berczynski_2007}. In magnetic confinement fusion, the theory of Gaussian beam propagation in anisotropic media has been used for approximately three decades \citep{pereverzev_nf_1992, pereverzev_1993, pereverzev_1996, pereverzev_pop_1998, poli_pop_1999, poli_pop_2001, poli_fed_2001} and has been successfully implemented in numerical codes such as TORBEAM \citep{poli_cpc_2001, poli_cpc_2018} and more recently in Scotty \citep{valerian_ppcf_2022} to model reflectometry/DBS and EC beam absorption. The Gaussian beam-tracing equations are a set of ODEs, which present great computational advantage with respect to full-wave codes, even in 2D. Although scarce, analytic solutions for beam-tracing exist \citep{maj_pop_2009, maj_ppcf_2010, weber_jpcs_2018} and are exploited in this work. In particular, \cite{maj_pop_2009, maj_ppcf_2010} found that for critical cases near the turning point, beam tracing remains a good approximation, contrary previous concerns raised by \cite{balakin_jphys_2007, balakin_nf_2008}. Some of these concerns are revisited in this manuscript.

The rest of the manuscript proceeds as follows. In section \ref{section_beammodel}, we introduce Gaussian beam tracing. We use the analytical solution by \cite{maj_pop_2009, maj_ppcf_2010} that exhibits beam focusing (also referred to as pinching, or lensing) near the turning point. This phenomenon of beam focusing was already observed in past work using numerical simulations \citep{poli_pop_1999, poli_fed_2001, kravtsov_berczynski_2007, conway_irw_2007}. We characterise how the beam focuses in a 2D Cartesian slab for the O-mode and a linear density gradient profile as a function of the beam initial conditions. In section \ref{gusakov_section}, we use the analytic solution for beam tracing in conjunction with a recently developed linear-response beam model of DBS \citep{valerian_ppcf_2022} to find an analytic expression for the DBS filter function $|F_{x\mu}|^2$. With this filter function, we develop and implement an analytic synthetic diagnostic for DBS, which can be used to analyse fluctuation data from gyrokinetic codes. In section \ref{conseq_dbs}, we show that by parametrising the filter function $|F_{x\mu}|^2$ by the scattered $k_x$ component of the turbulent wavevector along the path, we are able to recover previous known results for the scattered power that were believed to include full-wave effects (through the Airy function) \citep{gusakov_ppcf_2004, gusakov_ppcf_2014, gusakov_pop_2017}. Gusakov \emph{et al}. argued that the enhancement of the DBS signal power was due to a forward-scattering component, which is absent in our model by design. We find the exact same analytical formulas as Gusakov \emph{et al}. for the DBS filter function $|F_{x\mu}|^2$ by using a model only based on backscattering and Gaussian beam tracing, instead of the full-wave solution. This suggests that forward scattering is not the physical phenomenon leading to the enhancement of the DBS power for finite $k_x$ values near the turning point, but rather that beam focusing is. This section, in conjunction with appendices \ref{mapping_density} and \ref{app_amplitude_2d}, establish the equivalence between the beam model of DBS by \cite{valerian_ppcf_2022} and the 2D DBS model by Gusakov \emph{et al}. In the last section of this manuscript, we apply the analytical filter function $|F_{x\mu}|^2$ from the beam model to understand its effect on scattered power measurements from DBS. We use a realistic turbulence spectrum obtained from gyrokinetic simulations, following recent work by \cite{ruizruiz_ppcf_2022}. Our model predicts that the Doppler-backscattered power spectrum cannot reproduce the peak in $k_y$ from the true density fluctuation spectrum, but is accurate at predicting the characteristic spectral exponent of the turbulent cascade.


\section{Beam focusing and beam tracing}
\label{section_beammodel}

In DBS experiments, one can model the propagation of the microwaves launched externally from the plasma as a Gaussian beam. The behaviour of a Gaussian beam can be complex due to the inhomogeneities in density and magnetic field. In order to gain physical understanding of the phenomena affecting beam propagation, we first discuss vacuum propagation. 

In vacuum, a Gaussian beam will propagate in a straight line. If the beam is not focusing, then it will constantly expand. Asymptotically, the width of the beam will grow linearly with the propagation path length. If the beam is initially focusing, it will focus to a minimum width, called the waist, following which the beam will expand, as previously described by \cite{goldsmith_book}. This behaviour is due to diffraction: a wave packet of finite extent does not want to be confined: if initially expanding, it will not cease to expand; if initially focusing, it will focus before expanding. 

For a beam propagating in an anisotropic, inhomogeneous medium, its trajectory can substantially differ from the one in vacuum. The inhomogeneity in space causes refraction of the central ray, which is well accounted for by ray tracing. When this happens, the inhomogeneity experienced by the central ray changes as the ray propagates. This causes changes in the group velocity $\mathbf{g}$, which affects the intensity of the electric field, which scales as $E \propto 1/|\mathbf{g}|^\frac{1}{2}$ \citep{lopez_jopt_2021} \footnote{ In contrast to beam tracing, the electric field intensity in ray tracing reads $E \sim 1 / |\textbf{g} \cos \alpha |^\frac{1}{2}$ for the 2D linear layer \citep{lopez_jopt_2021}, where $\alpha$ is the angle between $\mathbf{g}$ and the density gradient, see figure \ref{frames} and equation \ref{sincos}. This divergence is integrable in ray tracing and resolved in beam tracing. }. Of particular interest is the problem of a wave encountering a cutoff surface at normal incidence. This is the situation encountered by diagnostics such as reflectometry, where the group velocity along the inhomogeneity approaches zero at the cutoff and the electric field diverges. This is a well known limitation of ray tracing when close to normal incidence to the inhomogeneity. Close to a turning point, one cannot assume slow variation of the background experienced by the wave, and the WKB approximation fails. To solve this problem, one needs to resort to solving the full Helmholtz wave equation. This limitation disappears when the angle of incidence with respect to the inhomogeneity is finite (or more precisely, large enough, as we will discuss), such as encountered in DBS. The group velocity still decreases approaching the cutoff (the component of $\mathbf{g}$ along the inhomogeneity vanishes, while the $\mathbf{g}$ component perpendicular to it remains approximately constant). This produces an enhancement of the electric field $E$ near the turning point, which should remain finite due to a finite $|\mathbf{g}|$. The decrease of the group velocity near the turning point is one factor that enhances the electric-field amplitude in DBS, and will be discussed in this manuscript. This will show up as a \emph{ray} term affecting the filter function $|F_{x\mu}|^2$ for backscattering. As we will see, $|F_{x\mu}|^2 \propto 1/K \propto 1/|\mathbf{g}|$. Importantly, there is another factor that can enhance the intensity of the electric field, and that is the beam width.

For a beam propagating in an anisotropic, inhomogeneous medium, the behaviour of the beam width can also be substantially different from the behaviour in vacuum. Beam tracing can describe diffraction effects due to the finite extent $\sim W$ of a wave packet. In addition to the beam waist present in vacuum, other authors have shown that the beam can experience additional focusing, or lensing, in the presence of inhomogeneity \citep{poli_pop_1999, poli_fed_2001, bornatici_maj_ppcf_2003, maj_jmathphys_2005, kravtsov_berczynski_2007, conway_irw_2007, berczynski_CEJP_2008aa, maj_pop_2009, maj_ppcf_2010}. This phenomenon tends to happen close to the turning point. In the beam-tracing formulation, one can show that the electric-field amplitude $E$ scales inversely proportional to the beam width $W$, as $E \sim 1/W^\frac{1}{2}$ \citep{valerian_ppcf_2022}. This shows that when the beam focuses, the electric field is enhanced. This is a different effect from the enhancement due to the decreasing group velocity of the central ray. In fact, $E \sim 1/(|\mathbf{g}|W)^\frac{1}{2}$. It is this phenomenon, the beam focusing in an inhomogeneous medium, that we explore in this manuscript. In full toroidal geometry, this results in the filter function $\sim 1/KW$ \citep{valerian_ppcf_2022}. In the 2D linear-layer model, we will find $|F_{x\mu}|^2/|F_0|^2 \equiv K_0W_0/KW$, where $|F_0|^2$, $K_0$ and $W_0$ are the filter function $|F_{x\mu}|^2$, the magnitude of the beam central wavenumber $\mathbf{K}$ and the beam width $W$ evaluated at the initial condition. We call the $1/W$ term the \emph{beam} term in the filter function \citep{valerian_ppcf_2022}. Importantly, the beam term and the ray term can both \emph{separately} contribute to the enhancement of the electric field in the vicinity of a turning point. This motivates studying the propagation of a ray and a beam in inhomogeneous, anisotropic media. For this, we adopt the ray-tracing and beam-tracing formulation, that we describe next.

\subsection{Ray tracing and Gaussian beam tracing}

This section closely follows notation introduced by \cite{valerian_ppcf_2022}, where Gaussian beam tracing is rederived from first principles and presented in full detail. 

The beam-tracing method (or paraxial WKB) is an asymptotic representation of the electric field in anisotropic media for finite size wave packets with characteristic scale $\sim W$. Beam tracing extends ray tracing by including diffraction effects in a set of ODEs. As in ray tracing, the wavevector of the wave $\mathbf{K}$ is ordered inversely proportional to the wavelength $\sim 1/\lambda$, while the background equilibrium is assumed to vary on a scale $L \gg \lambda$. In beam tracing, the scale $W$ is intermediate and obeys the following orderings $L \gg W \gg \lambda$, with $W \sim (\lambda L)^\frac{1}{2}$. In this manuscript, we denote the wavevector associated to the wave with the capital letter $\mathbf{K}$. The same notation will apply to its components $K_g$, $K_x$, etc. while the wavevector components associated to the turbulence will be in lower cases, e.g. $k_x$ and $k_y$. 

In Gaussian beam tracing, the variation of the wave electric field perpendicular to a central ray is assumed to be of Gaussian shape. The central ray propagation obeys the traditional ray-tracing equations for the ray position vector $\mathbf{q}(\tau) \sim L$ and wavevector $\mathbf{K}(\tau) \sim 1/\lambda$, where $\tau$ is a parameter along the central ray. For reference, the ray-tracing equations are 

\begin{equation}
	\begin{alignedat}{2}
	& \frac{\text{d}\mathbf{q}}{\text{d}\tau} = \nabla_K H, \\
	& \frac{\text{d}\mathbf{K}}{\text{d}\tau} = - \nabla H, 
	\end{alignedat}
	\label{raytracing_eqs}
\end{equation}
where $H$ is the cold-plasma dispersion relation and $\mathbf{g} = \nabla_K H$ is the effective group velocity. Note that there are infinite ways to define $H$. We use a specific definition that ensures that the wave electric field scales as $\sim 1/g^\frac{1}{2}$ \citep{valerian_ppcf_2022}. Equations (\ref{raytracing_eqs}) are routinely solved by codes such as GENRAY \citep{genray_smirnov_2009}. 

\begin{figure}
	\begin{center}
		\includegraphics[height=9cm]{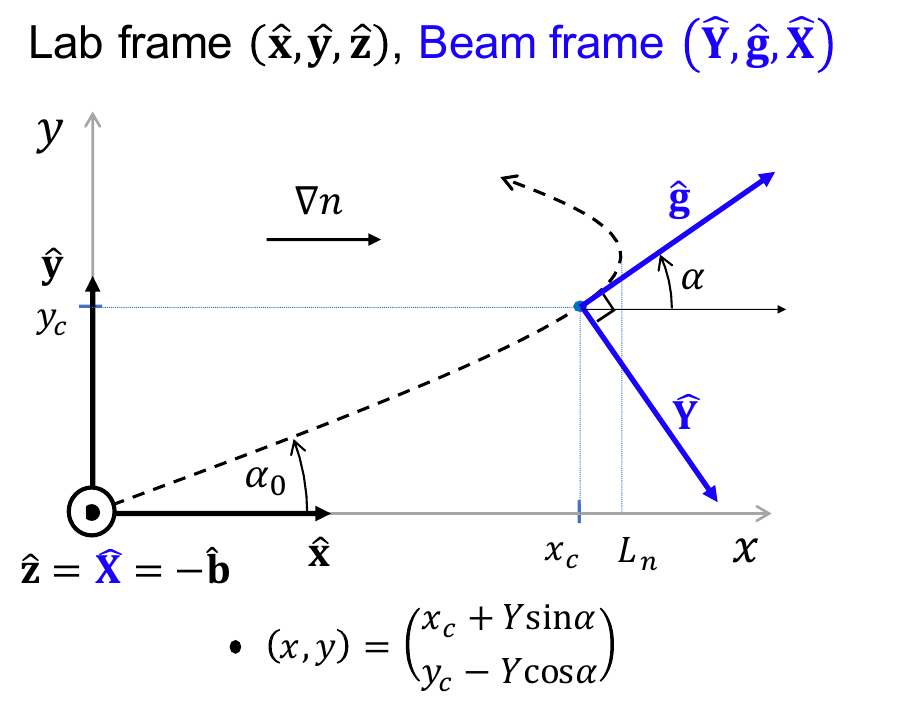}
	\end{center}
	\caption{Lab frame and beam frame for coordinates with oblique beam incidence (finite $\alpha_0$) used throughout this manuscript.}   
	\label{frames}
\end{figure}

In the Gaussian beam-tracing approximation, the beam electric field $\mathbf{E}_b$ is written in amplitude and phase as $\mathbf{E}_b(\mathbf{r}) = \mathbf{A}(\mathbf{r}) \exp[i \psi(\mathbf{r})] + $ c.c., where the phase is expanded to second order about the central ray in the perpendicular direction to $\mathbf{g} $, $\psi(\mathbf{r}) = s(\tau) + \mathbf{K}_w \cdot \mathbf{w} + \frac{1}{2} \mathbf{w} \cdot \boldsymbol{\Psi}_w \cdot \mathbf{w}$. A general point in space is described by $\mathbf{r} = \mathbf{q}(\tau) + \mathbf{w} = \mathbf{q}(\tau) + X \mathbf{\hat{X}} + Y \mathbf{\hat{Y}}$. We define the (orthonormal) beam-frame coordinate system $(\mathbf{\hat{Y}}, \mathbf{\hat{g}}, \mathbf{\hat{X}})$ as follows

\begin{equation}
	\begin{alignedat}{2}
	& \mathbf{\hat{Y}} = \frac{\mathbf{\hat{b}} \times \mathbf{\hat{g}}}{| \mathbf{\hat{b}} \times \mathbf{\hat{g}} |}, \qquad \mathbf{\hat{g}} = \frac{\mathbf{g}}{g}, \qquad \mathbf{\hat{X}} = \frac{\mathbf{\hat{Y}} \times \mathbf{\hat{g}}}{| \mathbf{\hat{Y}} \times \mathbf{\hat{g}} |},
	\end{alignedat}
	\label{beamframe_system}
\end{equation}
where $\mathbf{\hat{b}} = \mathbf{B}/B$ the unit vector along the background magnetic field. This is not to be confused with the Cartesian lab-frame system of coordinates $(\mathbf{\hat{x}}, \mathbf{\hat{y}}, \mathbf{\hat{z}})$ (see figure \ref{frames} for a comparison between beam frame and lab frame in 2D). The vector $\mathbf{K}_w = (\mathbf{I} - \mathbf{\hat{g}} \mathbf{\hat{g}} ) \cdot \mathbf{K}$ in the phase is the component of $\mathbf{K}$ perpendicular to $\mathbf{\hat{g}}$. The matrix $\boldsymbol{\Psi}_w = (\mathbf{I} - \mathbf{\hat{g}} \mathbf{\hat{g}} ) \cdot \boldsymbol{\Psi} \cdot (\mathbf{I} - \mathbf{\hat{g}} \mathbf{\hat{g}} )$ is $2 \times 2$ and only has components along the perpendicular directions $\mathbf{\hat{X}}$ and $\mathbf{\hat{Y}}$. The matrix $\boldsymbol{\Psi}_w$ contains information about the phase-front curvature of the Gaussian beam (through the eigenvalues of its real part = ${K^3}/{(K_g^2 R_{b,i})} $, where $K = |\mathbf{K}|$ and $K_g = \mathbf{K} \cdot \mathbf{\hat{g}}$) and beam width $W_i$ (through the eigenvalues of its imaginary part $ = {2}/{W_i^2}$). 

In the beam-tracing formulation which we use in this manuscript, the beam-tracing matrix $\boldsymbol{\Psi} \sim 1/W^2$ is a $3 \times 3$ matrix that is convenient to evolve along the central ray. The beam matrix ${\boldsymbol{\Psi}}$ follows the beam-tracing evolution equations, a set of ordinary differential equations parametrised by $\tau$, see \cite{valerian_ppcf_2022} and appendix \ref{app1}. Given $\boldsymbol{\Psi}$, the electric field $\mathbf{E}_b$ in Gaussian beam tracing is written as 

\begin{equation}
	\begin{alignedat}{2}
	& \mathbf{E}_b = &&A_{ant}\exp[{i(\phi_G+\phi_P)}] \bigg[ \frac{ \det(\Im[\boldsymbol{\Psi}_w]) }{ \det(\Im[\boldsymbol{\Psi}_{w,ant}]) } \bigg]^{\frac{1}{4}} \bigg( {\frac{g_{ant} }{g}} \bigg)^\frac{1}{2} \\
	& && \times \exp \Big( is + i \mathbf{K}_w\cdot \mathbf{w} + \frac{i}{2} \mathbf{w} \cdot \boldsymbol{\Psi}_w \cdot \mathbf{w}  \Big) \mathbf{\hat{e}},
	\end{alignedat}
	\label{valerian_beam}
\end{equation}
where $\mathbf{E}_b = E_b \mathbf{\hat{e}}$, $s = s(\tau) = \int^\tau{ \mathbf{K}(\tau') \cdot \mathbf{g} (\tau')} \ \text{d}\tau' = \int^\tau{ {K}_g  g  } \ \text{d}\tau' $ is the large phase Eikonal term measuring the variation of the electric field along the central ray, $s \sim L/\lambda \gg 1$ (note that $\tau \sim L/\lambda$ for $H \sim 1$). The unit vector $\mathbf{\hat{e}}$ is the polarization vector, $\phi_P$ and $\phi_G$ are respectively the polarization and Gouy phases, which are of limited interest in this manuscript (see \cite{valerian_ppcf_2022} for their evolution equations). The subscript $(.)_{ant}$ means that quantities are evaluated at the antenna launch location. As previously discussed, the prescription in equation (\ref{valerian_beam}) for the electric field shows how $E_b \sim 1/(gW)^\frac{1}{2} \sim 1/(KW)^\frac{1}{2}$. This is due to the term $\det(\Im[\boldsymbol{\Psi}_w])^\frac{1}{4} \propto 1/W^\frac{1}{2}$ and $1/g^\frac{1}{2} \propto 1/K^\frac{1}{2}$. 

In the next section, we show numerical solutions to the beam-tracing equations (appendix \ref{app1}) in real tokamak geometry, which exhibit the phenomenon of beam focusing. These will be compared to analytical solutions to beam tracing for the 2D linear-layer problem. We will see how the analytical solution qualitatively captures the phenomenon of beam focusing.

\subsection{Beam focusing in experiments}
\label{focusing_expts}

In this section, we will show that the beam-focusing phenomenon appears in numerical calculations of beam-tracing modelling of DBS experiments. We will see how the beam width has a tendency to focus as the beam moves towards higher density, a condition that is routinely encountered in DBS measurements. 

We describe the phenomenon of beam focusing, that is, the decrease or compression of the beam width as the beam propagates near the turning point that characterises a cutoff. This phenomenon was previously observed in numerical simulations \citep{poli_pop_1999, poli_fed_2001, kravtsov_berczynski_2007, conway_irw_2007, conway_irw_2015, conway_irw_2019} and in analytic solutions \citep{maj_pop_2009, maj_ppcf_2010, weber_jpcs_2018}. In order to confirm the existence of the beam-focusing phenomenon and its relevance to real-life experimental conditions, in figure \ref{jet97080_w1w2_analyt_} we show the result of a numerical calculation with the Scotty code \citep{valerian_ppcf_2022} for the DBS beam propagation in the core of the tokamak JET for L-mode discharge 97080 in an NBI-only phase. Figure \ref{jet97080_w1w2_analyt_}.a) shows the trajectory of the central ray projected on the 2D $(R,Z)$ plane overlaid by contour lines of the poloidal flux function $\Psi_p$. The turning point is at $\Psi_p/\Psi_{p0} \approx 0.7$, where $\Psi_{p0}$ is the separatrix value. Figure \ref{jet97080_w1w2_analyt_}.b) shows the perpendicular widths $W_1$ and $W_2$ that correspond to the eigenvalues of the beam-tracing matrix $\Im[ \boldsymbol{\Psi}_w ]$. We approximate the overmoded waveguide launch by a beam with circular cross section and initial conditions $W_0 \approx 1$ cm, $R_{Y0} = \infty$ (launch at the waist). The widths $W_1$ and $W_2$ are plotted against the path length $l = \int \text{d}\tau g$. The beam propagates in vacuum from $l \approx 0$ to $l \approx 0.4$ m, which roughly corresponds to the plasma edge. In vacuum, both principal beam widths follow the same behaviour: expansion following a launch from the waist. In the plasma, both widths initially continue expanding, but one of them (blue) starts contracting and reaching a minimum around $l \approx 0.7$ m (beam focusing). Following the beam focusing, the contracted width starts to expand again. The behaviour observed in figure \ref{jet97080_w1w2_analyt_}.b) consistently appears in beam-tracing numerical calculations near a turning point, and is confirmed by the beam-tracing code TORBEAM \citep{poli_cpc_2001, poli_cpc_2018} (not shown). This behaviour was also observed in other tokamaks \citep{conway_irw_2007, conway_irw_2015, conway_irw_2019}. This motivates studying the beam-focusing phenomenon using a reduced model that is analytically tractable, the 2D linear-layer model \citep{maj_pop_2009, maj_ppcf_2010}, which we present next.

\subsection{Beam focusing in the 2D linear layer}
\label{beam_foc_2d_sec}

\begin{figure}
	\begin{center}
		\includegraphics[height=5.4cm]{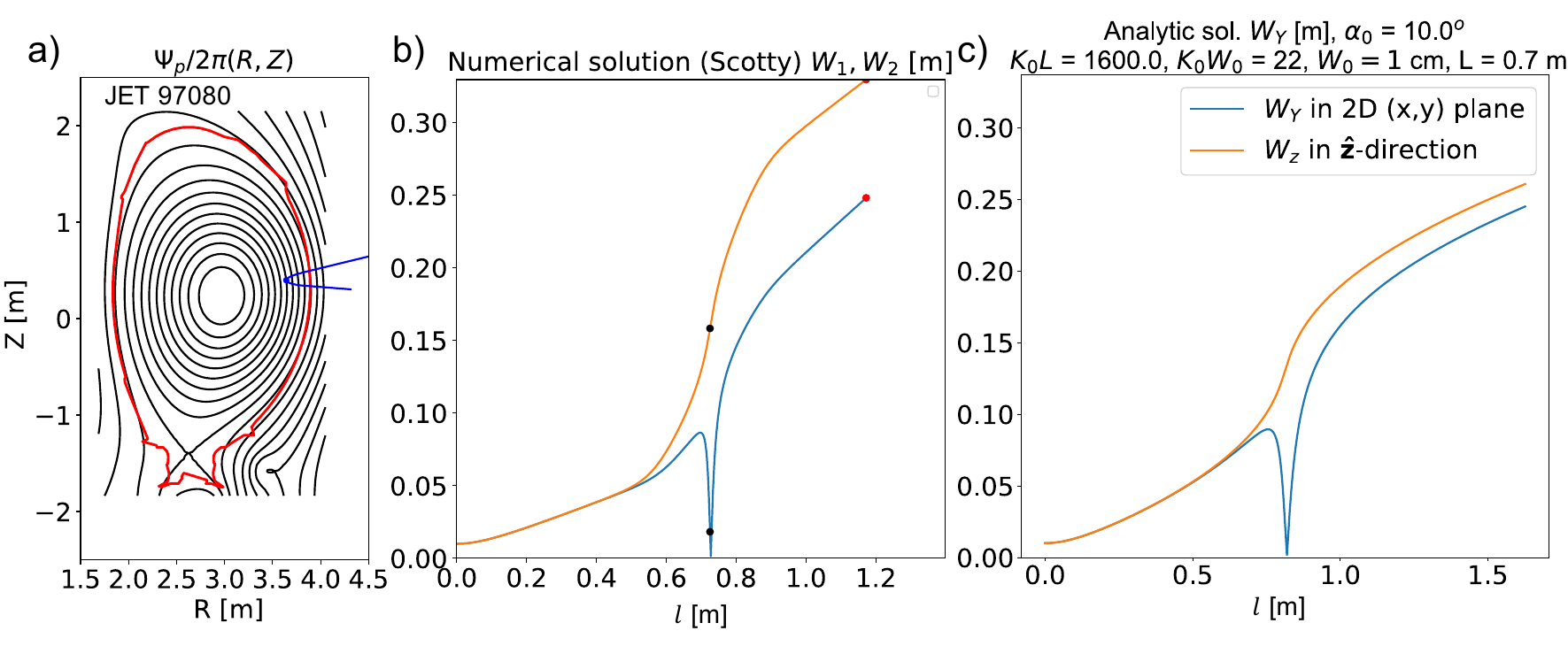}
	\end{center}
	\caption{a) Trajectory of the central ray of a DBS beam projected on the poloidal $(R,Z)$ plane overlaid by contour lines of the poloidal flux function $\Psi_p$ for JET discharge 97080 (NBI-heated L-mode). b) Numerical solution of the two principal widths perpendicular to the central ray propagation, noted here $W_1$ and $W_2$, using Scotty \citep{valerian_ppcf_2022}. The black dots correspond to the turning point in the trajectory (vanishing wavenumber component normal to the flux surface), while the red dots correspond to the plasma exit. c) Analytic solution of the beam-tracing equations for the 2D linear layer in slab geometry using experimental parameters corresponding to the case shown in a) and b): $K_0L\approx 1600, K_0W_0\approx 22$, $\alpha_0 \approx 10^\circ$, $R_{Y0}=\infty$ (launch at the waist). There is good qualitative agreement between the numerical and analytic solutions.}   
	\label{jet97080_w1w2_analyt_}
\end{figure}

In order to understand the beam-focusing phenomenon observed in the Scotty numerical solution in toroidal geometry in figure \ref{jet97080_w1w2_analyt_}.b), we first solve the equations for ray tracing (equation (\ref{raytracing_eqs})) using simplified Cartesian-slab geometry in 2D, and we refer to it as the 2D linear-layer model \citep{maj_pop_2009, maj_ppcf_2010}. Following the ray-tracing solution, we will subsequently solve the beam-tracing equations.

We solve the 2D linear-layer model in the lab frame, given by $\{ \mathbf{\hat{x}}, \mathbf{\hat{y}}, \mathbf{\hat{z}} \}$ (figure \ref{frames}). We will ignore the coordinate $\mathbf{\hat{z}}$, which points in the direction opposite to the magnetic field. These definitions will be needed in order to subsequently solve the beam-tracing equations. The 2D linear-layer problem in slab geometry has O-mode polarisation, uniform $\mathbf{B}(\mathbf{r}) = {B}_0 \mathbf{\hat{b}} = - B_0 \mathbf{ \hat{z} }$ and linear density profile $\omega_{pe}^2(x) = \Omega^2 x/L$. Here $x$ is our 'radial' coordinate, also Cartesian $x$ in figure \ref{frames}, $\Omega$ is the launch frequency, $\omega_{pe0}$ is the electron plasma frequency at the turning point location, $L_n = L \cos^2\alpha_0$ is the turning point location, and $L$ is the turning point location for zero incident angle ($\alpha_0=0$, see figure \ref{frames}). The ray-tracing equations determine the trajectory of the central ray $\mathbf{q}(\tau') = \big(x_c(\tau'), y_c (\tau') \big) $ and the wavevector of the central ray $\mathbf{K}(\tau') = \big(K_x(\tau'), K_y(\tau')\big) $, where $\tau' = \tau/K_0L$ is the normalised parameter along the path, $K_0$ is the wavenumber magnitude at launch. In this manuscript, we define the turning point, or cutoff, as the location where $K_x=0$ ($K_x$ is the Cartesian $x$-component of the beam wavevector $\boldsymbol{K}$, see equations (\ref{raytracing_sols})). We will make use of the dispersion relation $H=0$, where $H$ takes the form 

\begin{equation}
 	\begin{aligned}
	H = \frac{K^2}{K_0^2} - 1 + \frac{\omega_{pe}^2}{\Omega^2} 
	\end{aligned}
	\label{disp_rel_omode}
\end{equation}
for the O-mode. The form of $H$ given by equation (\ref{disp_rel_omode}) ensures that the electric field is given by equation (\ref{valerian_beam}). Using equation (\ref{disp_rel_omode}), we have $\nabla_K H = \mathbf{g} = 2 \mathbf{K}/K_0^2$ (initial condition $g_{ant} = 2/K_0$), and $\nabla H = \mathbf{\hat{x}}/L$. As can be seen from figure \ref{frames}, the magnetic field is into the page, and the angle $\alpha_0$ measures the vertical incidence angle of the central ray at launch. In what follows, $\alpha$ will measure the vertical incidence angle along the central ray. The initial conditions are $(x_{c}, y_{c})_0 = (0, y_0) $, $(K_{x}, K_{y})_0 = K_0 ( \cos\alpha_0, \sin\alpha_0)$. Straightforward calculations give the following solution for the ray-tracing equations (\ref{raytracing_eqs}) in the 2D linear layer

\begin{equation}
 	\begin{aligned}
	& x_c(\tau') = L \Big( \cos^2\alpha_0 - (\cos\alpha_0- {\tau'} )^2 \Big) = L \big( \cos^2\alpha_0 - {K_x^2}/{K_0^2}\big) , \\
	& y_c(\tau') = y_0 + 2 L \sin\alpha_0 \ \tau' , \\
	& K_x(\tau') = K_0 (\cos\alpha_0 - \tau') = K_0 (\cos^2\alpha_0 - x_c/L)^\frac{1}{2} , \\
	& K_y(\tau') = K_0\sin\alpha_0.
	& \end{aligned}
	\label{raytracing_sols}
\end{equation}
Equations (\ref{raytracing_sols}) describe a parabolic trajectory in $(x,y)$. We will find useful to parametrise the ray trajectory by the radial wavenumber of the central ray $K_x$, instead of $\tau$. We will also parametrise the beam-tracing solution by $K_x$. 

The 2D linear-layer problem with finite incident angle $\alpha_0$ is also analytically solvable in Gaussian beam tracing, as was previously shown by \cite{maj_pop_2009, maj_ppcf_2010}. For details of the derivation, we refer the reader to appendix \ref{app1}. The beam-tracing equations for the linear-layer problem reduce to ${\text{d} \boldsymbol{\Psi}}/{ \text{d} \tau } = -({2}/{K_0^2}) \boldsymbol{\Psi}^2$, which is a nonlinear matrix equation for $\boldsymbol{\Psi}$. This equation belongs to a class of ODEs known as Ricatti equations. The main component of the beam matrix $\boldsymbol{\Psi}$ of interest in this manuscript is the $\boldsymbol{\hat{Y}} \boldsymbol{\hat{Y}}$ component $\Psi_{YY}$, which captures the beam-focusing phenomenon through the perpendicular beam width $W_Y = (2/\Im[\Psi_{YY}])^\frac{1}{2}$. The solution for $\Psi_{YY}$ is

\begin{equation}
	\begin{aligned}
	& \Psi_{YY}(K_x) &&= \frac{ - \frac{1}{2} \sin^2\alpha_0 + \Psi_{yy0}' \Big[ \frac{K_x^3}{K_0^3} + 3 \sin^2\alpha_0 \frac{K_x}{K_0} - \frac{\sin^2\alpha_0}{\cos\alpha_0}(\cos^2\alpha_0 - \sin^2\alpha_0) \Big] }{ \Big[ \sin^2\alpha_0 + \frac{K_x^2}{K_0^2} \Big] \Big[ \frac{K_x}{K_0}  + 2 \Psi_{yy0}' \Big( \frac{\cos^2\alpha_0 - \sin^2\alpha_0 }{\cos\alpha_0} \frac{K_x}{K_0} + \sin^2\alpha_0 - \frac{K_x^2}{K_0^2}  \Big) \Big] } \frac{K_0}{L} ,
	\end{aligned}
	\label{psiYYkx}
\end{equation}
where $\Psi_{yy0}' = \Psi_{yy0} ({L}/{K_0}) $ is the initial condition for the beam $\Psi_{yy}$ component in the lab-frame direction $\mathbf{\hat{y}}$ (vertical direction in figure \ref{frames}) that contains information about the initial beam width and radius of curvature of the phase front (normalised to $K_0/L$, see appendix \ref{app1} for details). Equation (\ref{psiYYkx}) is the basis for most of the analysis in this manuscript. The reader is referred to appendix \ref{app1} for details in the calculation of $\Psi_{YY}$. The analytic solution of $\Psi_{YY}$ contains the beam-focusing phenomenon for the 2D linear-layer problem, and is used to characterise the Doppler backscattering power contributions along the beam trajectory through the filter function $|F_{x\mu}|^2$ (section \ref{gusakov_section}). 

We wish to compare the analytic beam-tracing solution for $W_{Y}$, computed using equation (\ref{psiYYkx}), to the numerical solution in toroidal geometry shown in figures \ref{jet97080_w1w2_analyt_}.a) and \ref{jet97080_w1w2_analyt_}.b). To do so, we extract the physical values of $K_0$, $L$ and the initial width $W_0$ from experimental conditions in the JET discharge 97080: $K_0 \approx 2200 $ m$^{-1}$, $W_0 \approx 1$ cm and $L\approx 0.7$ m. Knowing that in reality the beam propagates in a plasma with a varying density gradient, we choose the value of $L$ to be the average value of the density gradient experienced by the beam as it propagates through the plasma, yielding $L \approx 0.7$ m. We use the experimental values of $K_0$, $L$ and $W_0$ to normalise the initial conditions in the analytic beam-tracing solution, giving $K_0W_0\approx 22$, $K_0 L \approx 1600$. The width $W_Y$ that is perpendicular to the central ray propagation and in the $(x,y)$ plane is plotted in blue in figure (\ref{jet97080_w1w2_analyt_}).c), as a function of the beam path length $l$. The vacuum solution is plotted in orange, which is also the same solution as the $\hat{\mathbf{z}}$-component of the width for the linear layer in the $\hat{\mathbf{z}}$-direction $W_z$. Despite the limitations of the model (linear density gradient, 2D slab), a qualitative comparison between figures \ref{jet97080_w1w2_analyt_}.c) and \ref{jet97080_w1w2_analyt_}.b) clearly shows that the analytic solution for the 2D linear layer is able to recover the focusing behaviour observed in the numerical solution in toroidal geometry. This confirms that the focusing behaviour in the vicinity of the turning point is physical and not a numerical artifact, and motivates studying the beam-focusing phenomenon using the analytic solution to the beam-tracing equations in the 2D linear-layer problem. In what follows, we describe the dependence of the analytic beam-tracing solution for $\Psi_{YY}$ on the initial conditions for the width $W_0$, the radius of curvature $R_{Y0}$ and the vertical launch angle $\alpha_0$. The initial condition is at $\tau=0$, which corresponds to the plasma edge $x=0$. 


\begin{figure}
	\begin{center}
		\includegraphics[height=12cm]{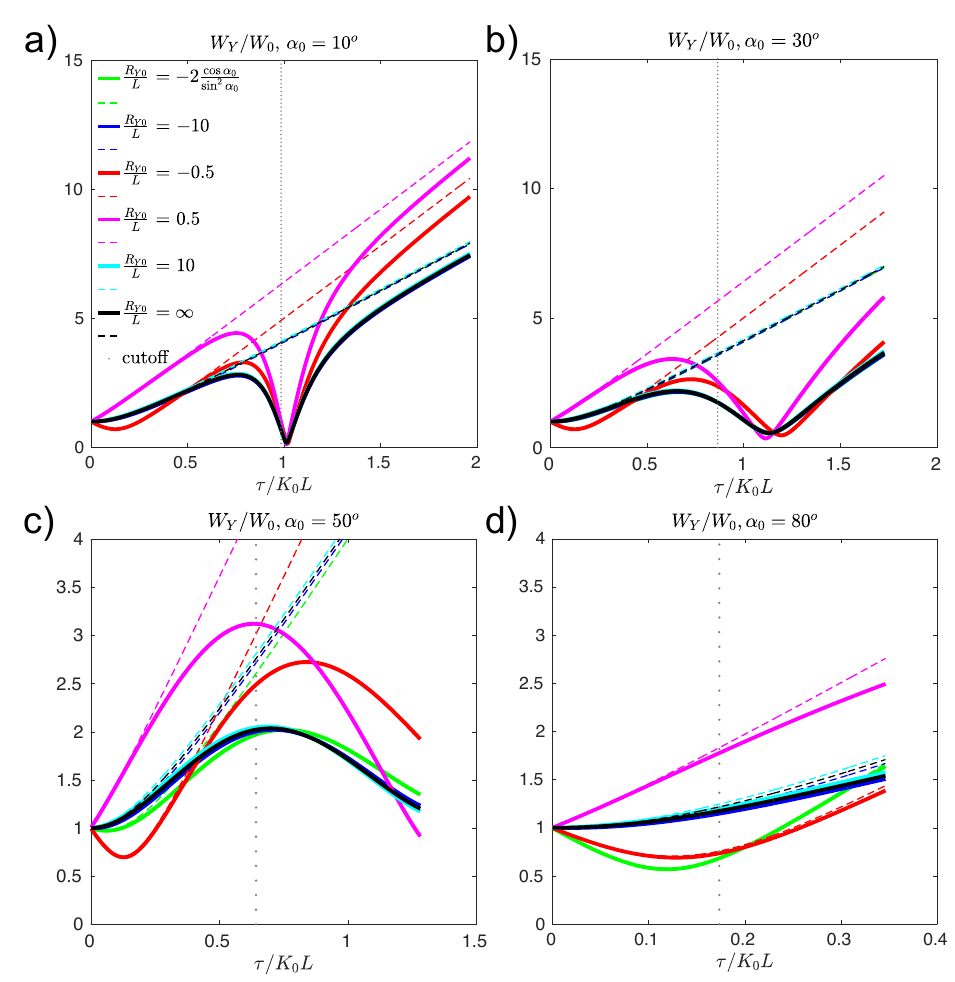}
	\end{center}
	\caption{Thick colored curves indicate values of the width $W_Y$ for varying initial incident angle $\alpha_0 = 10^\circ, 30^\circ, 50^\circ$, and $80^\circ$, and different initial $R_{Y0}$ (see legend), and fixed initial $W_0 = 0.40(\lambda L)^\frac{1}{2}$. Corresponding dashed lines of same color indicate vacuum values of $W_Y$ for same initial conditions as the thick colored curves. 
	The initial value of $R_{Y0}/L=-2 {\cos\alpha_0}/{\sin^2\alpha_0}$ (green) corresponds to the particular initial condition employed by \cite{gusakov_ppcf_2014, gusakov_pop_2017}. The vertical dotted points in grey indicate the location of the turning point, or cutoff (location where $K_x=0$). }   
	\label{wyry_fx_tau_alpha_ry0scan}
\end{figure}

We start by considering the analytic beam-tracing solution for $\Psi_{YY} = {K}/{R_Y} + i {2}/{W_Y^2}$ (appendix \ref{app1}). In this beam-tracing solution, the dimensional quantities $K_0 = 2\pi/\lambda$, $W_0$, and $L$ enter the equations through the beam parameter $W_0/(\lambda L)^\frac{1}{2}$. Therefore, in the rest of the manuscript, only the beam parameter $W_0/(\lambda L)^\frac{1}{2}$ will be specified since it is necessary  to determine the \emph{normalised} solution $\Psi_{YY} ({L}/{K_0})$. The beam width $W_Y$ is shown in figure \ref{wyry_fx_tau_alpha_ry0scan} as a function of $\tau$ for different initial conditions of the incident angle $\alpha_0 = 10^\circ, 30^\circ, 50^\circ$ and $80^\circ$, and the initial radius of curvature $R_{Y0}$ (colored curves). In this slab model, we assume that plasma is only present for $x>0$ and the vacuum exists for $x<0$. Therefore, one needs to calculate the range of $\tau$ for which the central ray trajectory has positive $x_c$. In appendix \ref{app1}, we show that for initial conditions at $\tau'=0$, the plasma exit corresponds to $\tau' = 2\cos\alpha_0$ (both corresponding to $x_c=0$, see equations (\ref{raytracing_sols})). The initial width $W_0$ is kept fixed to $W_0 = 0.40(\lambda L)^\frac{1}{2}$. The colored dashed lines are the vacuum solutions (absence of plasma) with initial conditions corresponding to the respective $W_Y$ of the same color in the 2D linear-layer problem. The vacuum solution initially follows the plasma solution until the beam approaches the turning point. The initial value of $R_{Y0}/L=-2 {\cos\alpha_0}/{\sin^2\alpha_0}$ corresponds to the particular initial condition employed by \cite{gusakov_ppcf_2014, gusakov_pop_2017} (see discussion in appendix \ref{app1}). For that particular case, the initial radius of curvature takes the values $R_{Y0}/L = -65.32, -6.93, -2.19$ and $-0.3581$ respectively for $\alpha_0=10^\circ, 30^\circ, 50^\circ$ and $80^\circ$. $R_{Y0}$ has little effect on the beam focusing around $\tau' \approx 1$ for small angles, but has a non-negligible effect in the vicinity of the initial launch and after the turning point for the outgoing beam. These differences appear for small initial $|R_{Y0}|/L \lesssim 1$, that is, for strongly focusing or diverging beams (see red and magenta curves in figure \ref{wyry_fx_tau_alpha_ry0scan}). The initial radius of curvature $R_{Y0}$ becomes more important in the vicinity of the turning point for increasing incident angles, where the beam focusing location is shifted (see $\alpha_0 = 30^\circ$). For $\alpha_0 = 50^\circ, 80^\circ$, the beam focusing does not take place inside the plasma. The only focusing region corresponds to the beam waist that is captured by the vacuum solution. Note how the initial condition of $R_{Y0}/L = -2 {\cos\alpha_0}/{\sin^2\alpha_0} $ from \citep{gusakov_ppcf_2014} and \citep{gusakov_pop_2017} strongly depends on the launch angle $\alpha_0$: it defines a beam that is closer to a launch at the waist ($R_{Y0}/L = \infty$) for small incident angles, but becomes strongly focusing (smaller, negative $R_{Y0}$) for larger $\alpha_0$.


\begin{figure}
	\begin{center}
		\includegraphics[height=4.8cm]{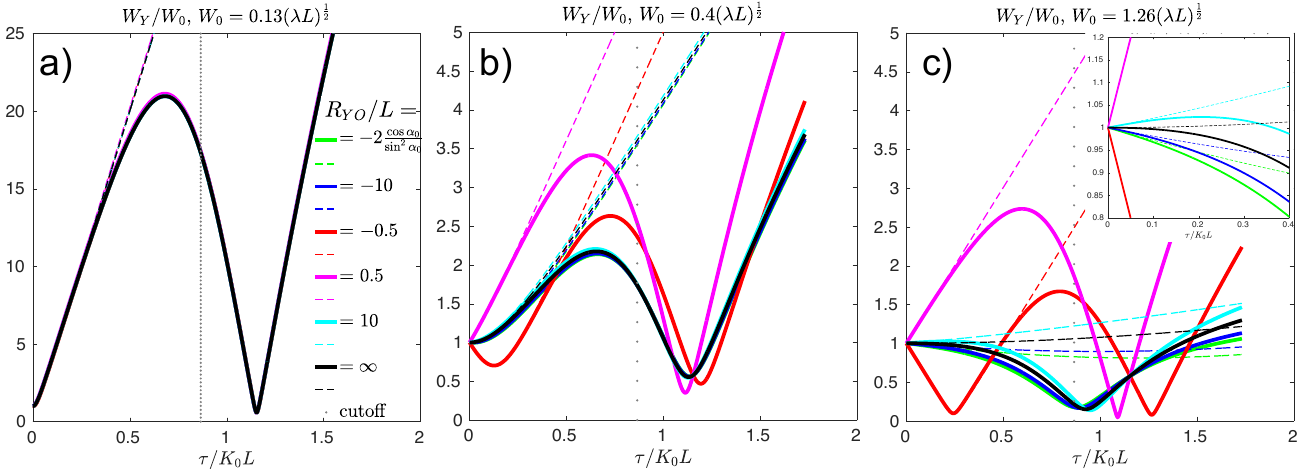}
	\end{center}
	\caption{Similar figure to figure \ref{wyry_fx_tau_alpha_ry0scan} for varying initial width $W_0 / (\lambda L)^\frac{1}{2} = 0.13, 0.40, 1.26$ and fixed $\alpha_0=30^\circ$. For small $W_0$ the beam is initially strongly focused and experiences a large expansion (due to diffraction) that follows the vacuum solution before focusing slightly past the turning point (for all of $R_{Y0}$). For increasing $W_0$ the initial growth is less severe and the beam focusing depends on the initial $R_{Y0}$. The beam even focuses twice along the path for particular initial conditions $R_{Y0}/L=-0.5, \alpha_0=30^\circ$, $W_0 / (\lambda L)^\frac{1}{2} = 0.4 $ (red in b)), and $1.26(\lambda L)^\frac{1}{2}$ (for the same $R_{Y0}/L, \alpha_0$, red in c)), where the first focusing region is the waist captured by the vacuum solution. Varying the initial width $W_0$ affects the initial expansion of the beam, from a pronounced initial expansion in a)-b) to no initial expansion for some $R_{Y0}/L$ in c). The inset in c) focuses on the initial propagation region, which exhibits converging or diverging beams depending on the initial conditions.
	}   
	\label{wyry_fx_tau_gammaalpha_ry0scan}
\end{figure}

Figure \ref{wyry_fx_tau_gammaalpha_ry0scan} shows the beam width for different initial conditions $W_0/(\lambda L)^\frac{1}{2} = 0.13, 0.4, 1.26$, and the same scan in $R_{Y0}/L$ as in figure \ref{wyry_fx_tau_alpha_ry0scan}, all while keeping a fixed initial incident angle $\alpha_0=30^\circ$. For small $W_0=0.13(\lambda L)^\frac{1}{2}$, the beam width $W_Y$ initially follows closely the vacuum solution (dashed line), independent of the initial conditions $W_0$ and $R_{Y0}$. Small $W_0/(\lambda L)^\frac{1}{2}$ corresponds to a highly focused beam. Diffraction produces a strong initial expansion of the beam width $W_Y$, which is followed by strong focusing slightly past the turning point (for all of $R_{Y0}$). For $W_0=0.40(\lambda L)^\frac{1}{2}$, the initial growth is less severe and the beam focusing (value and location) exhibits a noticeable dependence on the initial $R_{Y0}$. The dependence on the initial $R_{Y0}$ is even more noticeable for $W_0=1.26(\lambda L)^\frac{1}{2}$, especially for the focus location: the beam even focuses twice along the beam for $R_{Y0}/L=-0.5$ (red curve), which is an initially focusing beam at the plasma edge (the first focusing region is the waist captured by the vacuum solution). The double focusing happens for both $W_0 = 0.4 (\lambda L)^\frac{1}{2}$ and $W_0 = 1.26 (\lambda L)^\frac{1}{2}$. Comparing the same initial condition $R_{Y0}=\infty$ (launch at waist, black curve) between the different initial $W_0$, we see that the beam focusing location takes place after the turning point for small initial $W_0 = 0.13, 0.40 (\lambda L)^\frac{1}{2}$, and approaches the turning point for large $W_0 = 1.26 (\lambda L)^\frac{1}{2}$ (the turning point location is given by the vertical dotted lines). This discussion has important consequences for interpreting the regions in the plasma with predominant contributions to backscattering in DBS measurements, as will be discussed in section \ref{conseq_dbs}. 

In this section, we have seen how the 2D linear-layer model exhibits focusing of the beam width around the cutoff. This focusing is separate from the focusing caused at the beam waist in vacuum. The beam focusing depends on the initial conditions for the incident angle $\alpha_0$, normalised beam parameter $W_0/(\lambda L)^\frac{1}{2}$ and radius of curvature $R_{Y0}/L$. Beam focusing tends to be enhanced for small $\alpha_0$, while it tends to disappear for large $\alpha_0$. For small $W_0/(\lambda L)^\frac{1}{2}$, strong initial growth precedes strong focusing. For large $W_0/(\lambda L)^\frac{1}{2}$, the beam focuses from the initial condition towards the focusing region, and the initial growth disappears for some initial conditions. These phenomena are not captured by the vacuum solution.  

Having gained an understanding of the phenomenon of beam focusing, in what follows, we will use the analytical 2D model solution to assess the impact of beam focusing on Doppler Backscattering measurements.

\section{Beam-tracing model for DBS in the 2D linear layer} 
\label{gusakov_section}

In the previous section, we have shown the behaviour of the width $W_Y$ of a microwave beam propagating in a 2D linear layer. In this section, we will use that solution to find the backscattered signal amplitude and power in DBS measurements. This will allow us to study how the phenomenon of beam focusing impacts the scattered power measured by DBS. We will see how the scattered power can be written as the integral over $k$-space of a filter function $|F_{xy}|^2$ multiplied by the Fourier-transform of the correlation function, which corresponds to the density-fluctuation power spectrum. We will also see how the dependence of the filter function on $k_x$ and $k_y$ is not trivial, and we will show how $|F_{xy}|^2$ can be enhanced in the vicinity of the focusing region, where $W_Y$ reaches a minimum. The formulas the we obtain using a beam-tracing model for DBS recover previous analytic work based on a 2D Cartesian slab by \cite{gusakov_ppcf_2014, gusakov_pop_2017}, and connect it with more recent work based on a beam-tracing model of DBS in full toroidal geometry \citep{valerian_ppcf_2022}. In order to do so, we use a representation of the density fluctuations in 2D Cartesian coordinates, as in \citep{gusakov_ppcf_2014, gusakov_pop_2017}. A representation aligned along the beam trajectory, or beam-aligned representation \citep{valerian_ppcf_2022}, is not used in the main body of this manuscript, and will be the object of a future publication. In appendices \ref{mapping_density} and \ref{app_amplitude_2d}, we use the beam-aligned representation of the density to show the equivalence between the beam model by \cite{valerian_ppcf_2022} and the 2D DBS model by \cite{gusakov_ppcf_2014, gusakov_pop_2017}. For the purposes of calculating the scattered power contributions along the beam path, we show in appendices \ref{mapping_density} and \ref{app_amplitude_2d} that both representations of the density are related to each other by a rotation. With respect to the wavenumber resolution, the situation is more subtle, and one cannot simply assume that both representations of the density are related by a rotation. This will be the object of an upcoming publication. 

We start by defining the Cartesian representation of the density fluctuations, or lab-frame density-fluctuation amplitude. We expand the density fluctuations in $\mathbf{r} = (x,y,z)$ as follows

\begin{equation}
	\begin{alignedat}{2}
	& \delta n(x,y,z,t) = \int {\text{d}k_x \text{d}k_y \ \delta \hat{n}(k_x, k_y, z, t) \exp({i k_x x + i k_y y }}),
	\end{alignedat}
	\label{density_cartesian_2d}
\end{equation}
where $x$ and $y$ are the directions perpendicular to the magnetic field, and $z$ is along the magnetic field (see figure \ref{frames} for the 2D linear layer). In the 2D linear layer, the density fluctuations in equation (\ref{density_cartesian_2d}) are evaluated at $z=0$. A point $(x,y)$ in the 2D plane is given by 

\begin{equation}
	\begin{alignedat}{2}
	& x = x_c(\tau) + Y \sin\alpha(\tau) \\
	& y = y_c(\tau) - Y \cos\alpha(\tau),
	\end{alignedat}
	\label{xcyc_2d}
\end{equation}
where $\big( x_c(\tau), y_c(\tau) \big)$ in equation (\ref{xcyc_2d}) are the coordinates along the central ray (figure \ref{frames} and appendix \ref{app1}), and $\alpha$ is the angle between the central ray tangent vector $\mathbf{\hat{g}}$ and the horizontal ($\sin\alpha$ and $\cos\alpha$ are given by equation (\ref{sincos})). The wavevector of the \emph{turbulence} is written in its Cartesian components $\mathbf{k}_\perp = k_x \mathbf{\hat{{x}}} + k_y \mathbf{\hat{{y}}}$ (figure \ref{frames}). Here, the Cartesian $k_x$ naturally corresponds to the radial direction, or normal to the flux surface, and $k_y$ corresponds to the binormal component, that is, the component in the flux surface and perpendicular to the magnetic field. This is a useful frame in which to express turbulent fluctuations perpendicular to the background magnetic field \citep{catto_pp_1978, frieman_chen_physflu_1982}. 

Having introduced the representation of the density fluctuations, we calculate the backscattered amplitude and power. To do so, we use a theoretical beam-tracing model of DBS \citep{valerian_ppcf_2022} that gives an analytical relationship between the beam electric field and the scattered amplitude. As shown in appendix \ref{app_power}, the scattered amplitude takes an analogous form to equation (153) from \citep{valerian_ppcf_2022}

\begin{equation}
	\begin{alignedat}{2}
	& {A}_r(t) &&= A_{ant} \int \text{d}k_x \text{d}k_y \ F_{xy,\mu}(k_x, k_y) \ \delta \hat{n} (k_x, k_y) \exp\big[i\big(2 {s}_\mu + k_x {x}_{c\mu} + k_y {y}_{c\mu} \big) \big] ,
	\end{alignedat}
	\label{as_kxky_dnxy_cart}
\end{equation}
where $({x}_{c\mu}, {y}_{c\mu}) = \big( x_c({\tau}_\mu), y_c({\tau}_\mu) \big)$ denote the point along the central ray trajectory evaluated at a particular $\tau = {\tau}_\mu$, and $x_c$ and $y_c$ are given by equations \eqref{raytracing_sols}.  The subscript $(.)_\mu$ means that functions are evaluated at a point along the trajectory $\tau = {\tau}_\mu$ that satisfies the Bragg condition for backscattering (equation \eqref{bragg_dfdt_xy}). In equation \eqref{as_kxky_dnxy_cart}, the phase in the argument of the exponent is equivalent to the phase $2s_\mu + k_1 l_\mu$ in \citep{valerian_ppcf_2022}, and it is ordered large $\sim L/\lambda$. The filter function $F_{xy,\mu}(k_x, k_y)$ is the Cartesian equivalent to $F_\mu(k_1,k_2)$ in \citep{valerian_ppcf_2022}. The expression for $F_{xy,\mu}$ is given in appendix \ref{app_power}. For the rest of the manuscript, we will only need the magnitude of $F_{xy,\mu}$, given in equation \eqref{fkxky_2_cartesian_}. Note that the expression for $A_r(t)$ is related to the spectral amplitude $\tilde{A}_r(\omega)$ by the Fourier transform $\mathscr{F}$ in time, $\tilde{A}_r(\omega) = \mathscr{F} \big[A_r(t)\big] (\omega) = (2\pi)^{-1} \int{A_r(t) \exp(i\omega t) \text{d}t}$ (only $\delta n$ is assumed to depend on time). In this manuscript, we will preferentially work with $A_r(t)$. The location ${\tau}_\mu$ is defined by the condition of stationary phase \citep{bender_orzag_78}, given to lowest order by 

\begin{equation}
	\begin{alignedat}{2}
	& 2Kg + k_x\frac{\text{d}x_c}{\text{d}\tau} + k_y\frac{\text{d}y_c}{\text{d}\tau} = 0 . \\
	\end{alignedat}
	\label{bragg_dfdt_xy}
\end{equation}
Since $k_y + 2K_0\sin\alpha_0 \sim 1/W \ll 1/\lambda$, the equation for ${\tau}_\mu$ is

\begin{equation}
	\begin{alignedat}{2}
	& K_x({\tau}_\mu) \approx - \frac{k_x}{2} + \sin\alpha_0 (k_y + 2 K_0 \sin\alpha_0) \frac{K_0}{k_x} \approx - \frac{k_x}{2}.
	\end{alignedat}
	\label{taumu_bar}
\end{equation}
Note that equation (\ref{taumu_bar}) defines a ${\tau}_\mu$ that fails near the turning point $k_x \approx 0$. More details are given in appendix \ref{app_power}. 

From equation (\ref{as_kxky_dnxy_cart}) we calculate the scattered power, which takes an analogous form to equation (177) in \citep{valerian_ppcf_2022},

\begin{equation}
	\begin{alignedat}{2}
 		& \frac{p_r}{P_{ant}} &&= \int{ \text{d}k_x \text{d}k_y |F_{xy,\mu}|^2 \langle |\delta \hat{n} (k_x, k_y)|^2 \rangle_T},
	\end{alignedat}
	\label{prkxky_cart}
\end{equation}
where the slowly varying filter $|F_{xy,\mu}|^2(k_x, k_y)$ is
\begin{equation}
	\begin{alignedat}{3}
	& |F_{xy,\mu}|^2(k_x, k_y) &&= 2 \pi \frac{e^4}{m_e^2 \epsilon_0^2 \Omega^4} \frac{ \Im[{{\Psi}}_{YY}]_\mu }{| {\Psi}_{YY}|_\mu} \frac{ \exp \left[ - \frac{ \big(2 \frac{\text{d} {K}_\mu}{\text{d}\tau} \big)^2 {g}_\mu^2 }{ \big| 2 \frac{\text{d} {K}_\mu}{\text{d}\tau} {g}_\mu - \frac{2 {K}_\mu^2}{{\Psi}_{YY\mu}} \big( \frac{\text{d}\alpha}{\text{d}\tau} \big )^2 ( {\tau}_\mu ) \big|^2 } \frac{2 {q}_{2\mu}^2}{\Delta {k}_{\mu2}^2} \right] }{  \big| 2 \frac{\text{d} {K}_\mu}{\text{d}\tau} {g}_\mu - \frac{2 {K}_\mu^2}{{\Psi}_{YY\mu} } \big( \frac{\text{d}\alpha}{\text{d}\tau} \big)^2 ({\tau}_\mu) \big| } ,
	\end{alignedat}
	\label{fkxky_2_cartesian_}
\end{equation}
where $m_e$ is the electron mass, $e$ the electron charge, and $\epsilon_0$ the vacuum permittivity. In equation (\ref{fkxky_2_cartesian_}), we have introduced the shorthand notation 

\begin{equation}
	\begin{alignedat}{2}
	& q_{1\mu} = k_x \cos\alpha_\mu + k_y \sin\alpha_\mu , \\ 
	& q_{2\mu} = -k_y \cos\alpha_\mu + k_x \sin\alpha_\mu , 
	\label{rotation_q_k}
	\end{alignedat}
\end{equation}
which defines a rotation in the 2D $(x,y)$ plane for every point along the beam path denoted by $\tau$ (note that $\alpha$ depends on $\tau$ through the trajectory in equation \eqref{raytracing_sols}). The Gaussian exponential term in ${q}_{2\mu}$ in equation (\ref{prkxky_cart}), entering through $|F_{xy,\mu}|^2$ in equation (\ref{fkxky_2_cartesian_}), is to be considered as a Gaussian exponential in $k_y$, where ${q}_{2\mu}$ is defined in terms of $k_x$ and $k_y$ by equation \eqref{rotation_q_k}. Note that all functions of $\tau$ in equation (\ref{fkxky_2_cartesian_}) are evaluated at $\tau = {\tau}_\mu$. Here, ${\alpha}_\mu = \alpha({\tau}_\mu)$ is the vertical incidence angle at $\tau = {\tau}_\mu$, figure \ref{frames}. The function $|F_{xy,\mu}(k_x, k_y)|^2$ in equation (\ref{fkxky_2_cartesian_}) is a function of $(k_x, k_y)$ through the Bragg condition relating ${\tau}_{\mu}$ to $k_x$ and $k_y$ (equations (\ref{bragg_dfdt_xy}), (\ref{taumu_bar})). 

In equation \eqref{fkxky_2_cartesian_} we have introduced the wavenumber resolution $\Delta k_{\mu2} \sim 1/W$ \citep{valerian_ppcf_2022}. In the context of the 2D linear layer, the quantity $\Delta {k}_{\mu2}$ is a measure of the resolution in the wavenumber component that is perpendicular to $\hat{\mathbf{g}}$. This can be seen by noting that $q_{2\mu}$ is in fact the component of $\mathbf{k}_\perp$ in the perpendicular direction to $\hat{\mathbf{g}}$. The dependence on ${q}_{2\mu}$ of equation \eqref{fkxky_2_cartesian_} therefore implies ${q}_{2\mu} \sim 1/W$, that is $ k_x \approx k_y \cos{\alpha}_\mu/\sin{\alpha}_\mu $, because $\Delta {k}_{\mu2} \sim 1/W \ll 1/\lambda$. The resolution $\Delta k_2$ is of limited interest in this manuscript, beyond the fact that it is of order $\sim 1/W$. 

The Gaussian exponential term in $q_{2\mu}$ in equation \eqref{fkxky_2_cartesian_} should be regarded as a Gaussian in $k_y$. This allows us to calculate the wavenumber resolution in the lab frame, or DBS $k_y$-resolution $\Delta k_y$ (see appendix \ref{app_power}). The final expression for the scattered power, expressed in terms of $k_x$ and $k_y$, is 

\begin{equation}
	\begin{alignedat}{2}
	&\frac{p_r}{P_{ant}} && \approx 2^\frac{1}{2} \pi \frac{e^4}{m_e^2 \epsilon_0^2 \Omega^4} \frac{K_0^2L}{\Delta k_y} \int{\text{d}k_x \text{d}k_y } \frac{ \exp \big[-{2(k_y + 2K_{0}\sin\alpha_0 )^2/\Delta k_y^2} \big] }{ K_\mu W_{Y\mu} } \langle |\delta \hat{n} (k_x, k_y)|^2 \rangle_T  \\
	\label{prkxky2_cart}
	& && \approx \pi^\frac{3}{2} K_0^2L \frac{e^4}{m_e^2 \epsilon_0^2 \Omega^4} \int{\text{d}k_x } \frac{ \langle |\delta \hat{n} (k_x, - 2 K_{0}\sin\alpha_0)|^2 \rangle_T  }{ K_\mu W_{Y\mu} },  \\
	\end{alignedat}
\end{equation}
where we have defined the $k_y$ resolution of the DBS diagnostic as

\begin{equation}
	\begin{alignedat}{2}
 	& \Delta k_y^2 = 4 \frac{ | \Psi_{yy0}|^2 }{ \Im[\Psi_{yy0}] } .
	\end{alignedat}
	\label{ky_resolution}
\end{equation}
Equations \eqref{prkxky2_cart} define a one-dimensional filter $|F_{x\mu}|^2$, given by, 
\begin{equation}
	\begin{alignedat}{2}
 	& |F_{x\mu}|^2 = \pi^\frac{3}{2} K_0^2L \frac{e^4}{m_e^2 \epsilon_0^2 \Omega^4} \frac{1}{K_\mu W_{Y\mu}} ,
	\end{alignedat}
	\label{fxmu2_def}
\end{equation}
where the product $K_\mu W_{Y\mu}$ is given by
\begin{equation}
	\begin{alignedat}{2}
	& {K_\mu} W_{Y\mu} &&=  \bigg| - \frac{k_{x}}{K_0} + \Psi_{yy0}' \Big( - 2\frac{\cos^2\alpha_0 - \sin^2\alpha_0}{\cos\alpha_0} \frac{k_{x}}{K_0} + 4\sin^2\alpha_0 - \frac{k_{x}^2}{K_0^2} \Big) \bigg|  . \\ 
	\end{alignedat}
	\label{kw_equation}
\end{equation}
In equation (\ref{kw_equation}), we used the beam-tracing analytic solution for $W_Y = W_{Y}(\tau)$ (equation (\ref{psiYYkx}), appendix \ref{app1}) as well as the expression for $K = K(\tau)$ in the 2D linear-layer problem. 

The first approximately equal sign in equation (\ref{prkxky2_cart}) recovers the expected Gaussian exponential dependence of the power with $k_y$. Note how the Gaussian term in $k_y$ from equation (\ref{fkxky_2_cartesian_}) has explicit dependence on $k_x$, while that dependence on $k_x$ is hidden in $\tau_\mu$ in equation (\ref{prkxky2_cart}). Equation (\ref{prkxky2_cart}) is the lowest-order contribution to the Gaussian exponential term in equation (\ref{fkxky_2_cartesian_}), and shows that the selected wavenumber $k_y$ is given approximately by $k_y \approx -2K_0\sin\alpha_0 \sim 1/\lambda$, and the correction to that is ordered $\sim 1/W$. This justifies the approximately equal signs in equation (\ref{prkxky2_cart}). More details can be found in appendix \ref{app_power}. 

Importantly, equations \eqref{prkxky2_cart}, \eqref{ky_resolution} and \eqref{kw_equation} recover equation (16) in \citep{gusakov_ppcf_2014}, which was extended to equations (14) and (15) in \citep{gusakov_pop_2017} for the cross-correlation function CCF. Gusakov \emph{et al}. carry their analysis for radial correlation Doppler reflectometry (RCDR), and not for standard DBS, as done in this manuscript. In RCDR, one is interested in the cross-correlation function CCF$(\Delta x)$, which depends on the radial separation $\Delta x$ between DBS scattering locations. The scattered power calculated in this work can be recovered by setting $\Delta x = 0$ in \citep{gusakov_ppcf_2014, gusakov_pop_2017} (power is auto correlation). \cite{gusakov_ppcf_2014, gusakov_pop_2017} also use a particular initial condition $\Psi_{yy0}' = i {K_0L}/{(K_0 \rho)^2} $, where $\rho$ is related to the launched beam width. One can see that setting the particular initial condition for $\Psi_{yy0}' = i {K_0L}/{(K_0 \rho)^2}$ in equation (\ref{kw_equation}) recovers equations (14) and (15) in \citep{gusakov_pop_2017} for $\Delta x = 0$. 

The backscattered power can be written as an integral along the path $l$ (equation (196) in \citep{valerian_ppcf_2022}). This can be achieved by realising that the Bragg condition and the fact that $q_2 \approx 0$ imply $k_x \approx - 2 {K}_{x\mu}$ (equation \eqref{rotation_q_k}). Having integrated in $k_y$, the scattered $k_x$ along the Cartesian $x$-direction can parametrise the location along the path. To express the $k_x$-integral (\ref{prkxky2_cart}) in terms of $l$, change variables using $\text{d}k_x = 2 K_\mu |\text{d}K_\mu/\text{d}\tau| \text{d}l / |g_\mu K_{x\mu}|$. Then, the denominator under the integral sign $\int \text{d}k_x $ can be explicitly expressed as a function of the path length $l$. Additional details on the calculation are given in appendices \ref{app1} and \ref{app_power}. We find
   
\begin{equation}
	\begin{alignedat}{2}
	&\frac{p_r}{P_{ant}} &&\approx 4 \pi^\frac{3}{2} \frac{e^4}{m_e^2 \epsilon_0^2 \Omega^4} \int \frac{ \text{d}l }{g^2 W_{Y}} \big\langle \big| \delta \hat{n} \big(k_x(l), -2K_0\sin\alpha_0 \big) \big|^2 \big\rangle_T.
	\end{alignedat}
	\label{pr_l_beam_}
\end{equation}
Equation \eqref{pr_l_beam_} closely connects to the expression for the scattered power in equation (196) of \citep{valerian_ppcf_2022}. In that case, $1/g^2$ is the \emph{ray} piece, while the \emph{beam} piece can still be written as ${1}/{W_{Y}}$. We see that $W_Y$ plays an explicit role in the $k_x$ wavenumber resolution (equation (\ref{prkxky2_cart})) and equivalently in the spatial localisation of the power along the beam path (equation (\ref{pr_l_beam_})). 
  
Next, we discuss the filter function $|F_{x\mu}|^2 \propto 1/K_\mu W_{Y\mu}$ in equation (\ref{fxmu2_def}). \cite{gusakov_ppcf_2014, gusakov_pop_2017} argue that non-local forward scattering events produced by large-radial-scale fluctuations cause the denominator of $|F_{xy,\mu}|^2$ to approach zero. This happens for specific $k_x$, and should cause an enhancement in the DBS signal. \cite{gusakov_ppcf_2014, gusakov_pop_2017} state that this enhancement is due to forward scattering events taking place all along the beam path through the plasma, making the DBS measurement spatially non local. They argue that this effect should preferentially enhance the DBS signal for small incidence angles $\alpha_0$, which they observe as an enhancement in the CCF. In our work, forward scattering is absent by design. Our model only includes contributions from backscattering events, which are selected through the Bragg condition in equations (\ref{bragg_dfdt_xy}) and (\ref{taumu_bar}). Equation (\ref{prkxky2_cart}) also shows that all the contributions to the so-called forward scattering can be explained by the filter function, which is the term $ \propto 1/KW_Y$ in equation \eqref{prkxky2_cart}, or equivalently $\propto 1/g^2 W_Y$ when written as an integral in $l$. Importantly, in section \ref{conseq_dbs} we will see how the focusing effect due to $W_Y$ can be far greater than the decrease of $K$ near the turning point. It is therefore crucial to understand the behaviour of $W_Y$ in order to understand its effect on the backscattered signal. In this work, we interpret equation (\ref{prkxky2_cart}) and the signal enhancement produced for specific $k_x$ as a consequence of beam focusing in the vicinity of a turning point, and not as due to forward scattering. The more the beam is focused (small $W_Y$), the more the signal is localised around the focusing region due to the local increase in the wave intensity, which takes place at a finite $k_x$ (in the vicinity of, but not exactly at the turning point). This effect preferentially happens for small incident angles $\alpha_0$, as was shown in section \ref{section_beammodel}. 

With respect to the wavenumber resolution, equation (\ref{prkxky2_cart}) recovers previous calculations of the DBS wavenumber resolution \citep{lin_ppcf_2001, hirsch_rsi_2001, hillesheim_rsi_2012}, which is derived here in the context of a beam-tracing model. Note the difference between $\Delta k_y$ in the lab frame, which is constant along the path in the particular case of the 2D linear layer, and $\Delta k_2$ in \citep{valerian_ppcf_2022}, which is perpendicular to the central ray propagation and depends strongly on the distance along the path. Equation (\ref{ky_resolution}) simplifies to $\Delta k_y = 2/\rho $ for the choice of $\Psi_{yy0}$ made by \cite{gusakov_ppcf_2014, gusakov_pop_2017}. Note how the value $2/\rho$ is a lower limit to the diagnostic $k_y$-wavenumber resolution. Interestingly, the wavenumber resolution $\Delta k_y$ depends on the incidence angle $\alpha_0$ for a given initial $W_0$, $R_{Y0}$ and a given $L$. Inspecting equation \eqref{ky_resolution} and using the expression for $\Psi_{yy0}$ in terms of $W_0$ and $R_{Y0}$ (appendix \ref{app1}), we find that 
\begin{equation}
	\begin{alignedat}{2}
 	& \Delta k_y^2 = 2 \cos^2\alpha_0 \frac{K_0^2W_0^2}{K_0L} \left[ \left( \frac{L}{R_{Y0}} + \frac{\sin^2\alpha_0}{2\cos\alpha_0} \right)^2 + 4 \left( \frac{K_0L}{K_0^2W_0^2} \right)^2 \right] \frac{K_0}{L}.
	\end{alignedat}
	\label{ky_resolution_alpha0}
\end{equation}
Equation \eqref{ky_resolution_alpha0} shows how $\Delta k_y$ can have a non-trivial dependence on $\alpha_0$, which should be taken into consideration when comparing the scattered power in DBS measurements and in numerical turbulence simulations via synthetic diagnostics.

 
Equations (\ref{prkxky2_cart}), (\ref{ky_resolution}), and \eqref{kw_equation} are the basis for synthetic DBS diagnostics that can be readily applied to direct nonlinear gyrokinetic simulations of plasma turbulence. In that context, and borrowing notation from previous work \citep{ruizruiz_ppcf_2022}, $k_x$ and $k_y$ can be directly identified to the physical normal and binormal wavenumbers $k_n$ and $k_b$. The Cartesian $k_x$ employed here corresponds to the normal wavenumber component of the gyrokinetic simulation $k_x = k_n = \mathbf{k_\perp} \cdot \mathbf{\hat{e}_n}$, where $\mathbf{\hat{e}_n} = \nabla r/|\nabla r|$ is the unit vector normal to the flux surface identified by minor radius $r$. The Cartesian $k_y$ corresponds to the binormal wavenumber of gyrokinetic simulations $k_y = k_b = \mathbf{k_\perp} \cdot \mathbf{\hat{e}_b}$, where $\mathbf{\hat{e}_b} = \mathbf{\hat{e}_n} \times \mathbf{\hat{b}}$ is the binormal wavevector, perpendicular to the unit vector of the magnetic field $\mathbf{\hat{b}}$ and to $\mathbf{\hat{e}_n}$. Depending on the gyrokinetic code used, $k_n$ and $k_b$ might need to be mapped from internal code wavenumber definitions, examples of which are given by \cite{ruizruiz_ppcf_2022}.

\section{Consequences of beam focusing for DBS measurements}
\label{conseq_dbs}

The phenomenon of beam focusing affects the DBS signal localisation and wavenumber resolution, and it enhances the DBS signal in the vicinity of the beam focusing region, as shown by the filter function $|F_{x\mu}|^2 \propto 1/K_\mu W_{Y\mu}$ for the 2D linear-layer problem in equation \eqref{fxmu2_def}. This challenges the interpretation of 'forward scattering' in references \citep{gusakov_ppcf_2004, gusakov_ppcf_2014} and \citep{gusakov_pop_2017} as the responsible mechanism for the DBS signal enhancement for small launch angle $\alpha_0$. In this section, we characterise the consequences of beam focusing on the DBS signal through the filter function $|F_{x\mu}|^2$ for the 2D linear-layer problem. We scan the possible initial conditions: incident angle $\alpha_0$, initial width $W_{0}$ and initial radius of curvature $R_{Y0}$. In what follows, $K_\mu$ and $W_{Y\mu}$ are normalised to the initial conditions $K_0$ and $W_0$, respectively. 

\subsection{The 1D filter function $|F_{x\mu}|^2$}
\label{filter_section}

The filter function $|F_{x\mu}|^2$ in equation \eqref{fxmu2_def} provides the $k_x$-selectivity of the DBS power. This is tied to the localisation along the path: the scattering turbulent $k_x$ can be thought of as a parameter determining the location of scattering along the path $\tau_\mu$, as $\tau_\mu$ is related to the turbulent scattering $k_x$ via the Bragg condition for backscattering in equation \eqref{taumu_bar}, which is used to arrive at equation \eqref{pr_l_beam_}. The radial component $K_{x\mu} = K_0(\cos\alpha_0 - \tau'_\mu)$ from equation (\ref{raytracing_sols}) is a decreasing function of $\tau'_\mu$ from the launch, becoming negative after the turning point ($K_{x\mu}=0$, i.e. $\tau'_\mu = \cos\alpha_0$). The vertical component $K_y$ is constant, since the system is homogeneous in $y$. This means that $K_\mu$ reaches its minimum at the turning point, which should enhance the DBS signal contribution in that location. As we saw in the previous section, the perpendicular beam width $W_{Y\mu}$ has a tendency to focus, which should further enhance the DBS signal power. In this section we separate the enhancement due to ray and the beam contributions in the filter function. 

\begin{figure}
	\begin{center}
		\includegraphics[height=12cm]{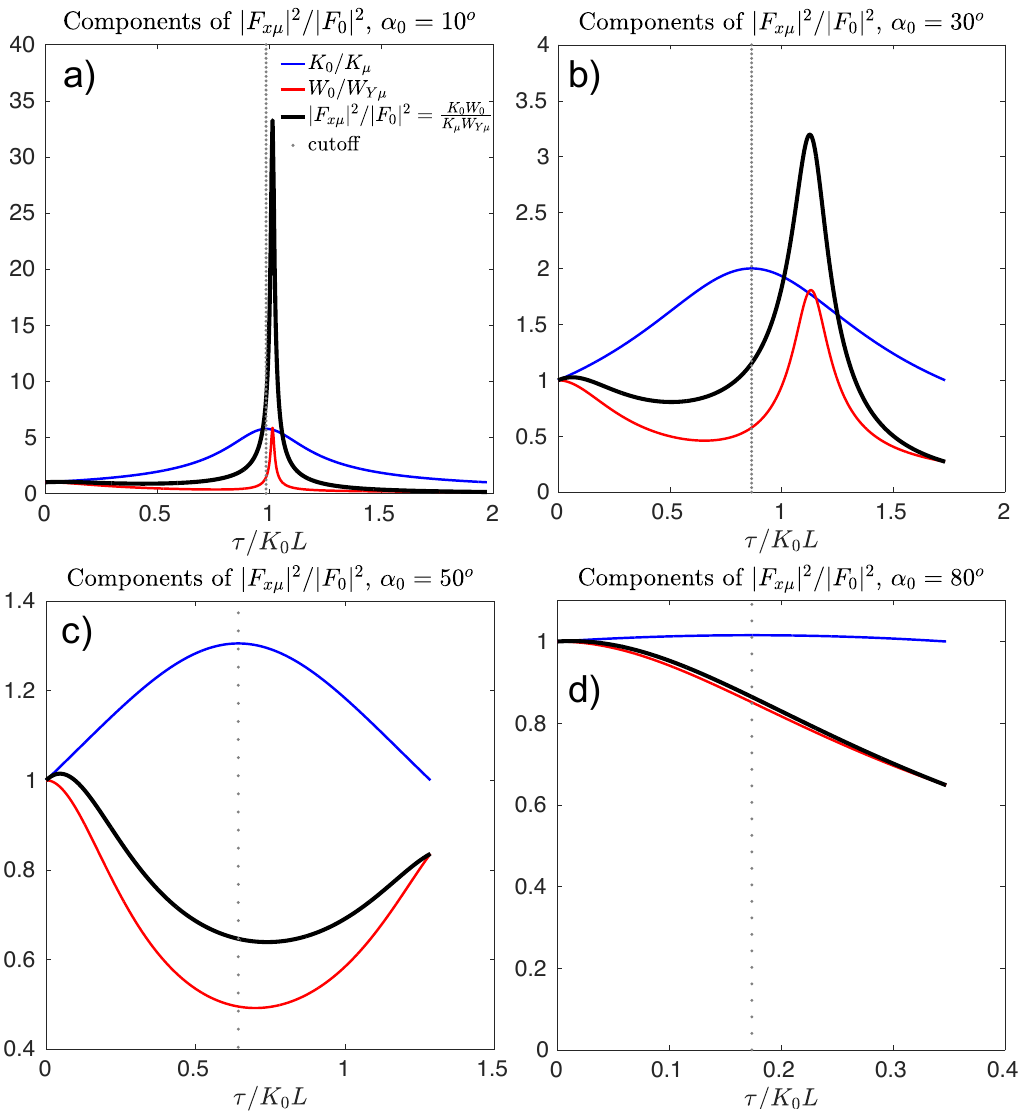}
	\end{center}
	\caption{Filter function $|F_{x\mu}|^2(\tau)/|F_0|^2 = K_0 W_0/K_\mu W_{Y\mu}$ and the corresponding ray component $K_0/K_\mu$ and beam component $W_0/W_{Y\mu}$ for varying values of the incident launch angle $\alpha_0$, and fixed $R_{Y0} = \infty$ and $W_0 = 0.40(\lambda L)^\frac{1}{2}$. Note that in all cases the filter function is predominantly affected by the beam term $1/W_{Y\mu}$ which represents the focusing. }   
	\label{fxkmuwymu_alphascan}
\end{figure}

Figure \ref{fxkmuwymu_alphascan} shows the ray (blue) and beam (red) components of $|F_{x\mu}|^2$ for varying incident angles $\alpha_0 = 10^\circ, 30^\circ, 50^\circ$ and $ 80^\circ$, and fixed $W_0 = 0.40(\lambda L)^\frac{1}{2}$ and $R_{Y0} = \infty$ (launch at the waist). For convenience, we normalise the filter function to its value at $\tau=0$, $|F_0|^2$. This filter function corresponds to the same launch conditions used in figure \ref{wyry_fx_tau_alpha_ry0scan}. For $\alpha_0=10^\circ$, the filter function $|F_{x\mu}|^2$ is strongly localised at the focus location, which almost matches the location of the turning point, or cutoff (maximum of the ray term in blue). The turning point, or cutoff, is represented in each figure by a vertical grey dotted line. Careful inspection shows that the focus location takes place in the vicinity but \emph{after} the turning point $\tau'_\mu = \cos\alpha_0$ for $\alpha_0=10^\circ$, consistent with beam focusing taking place after the turning point in figure \ref{wyry_fx_tau_alpha_ry0scan}. This shows that the contributions to the DBS power are predominantly from the vicinity (but after) the turning point for these initial conditions. 

\begin{figure}
	\begin{center}
		\includegraphics[height=12cm]{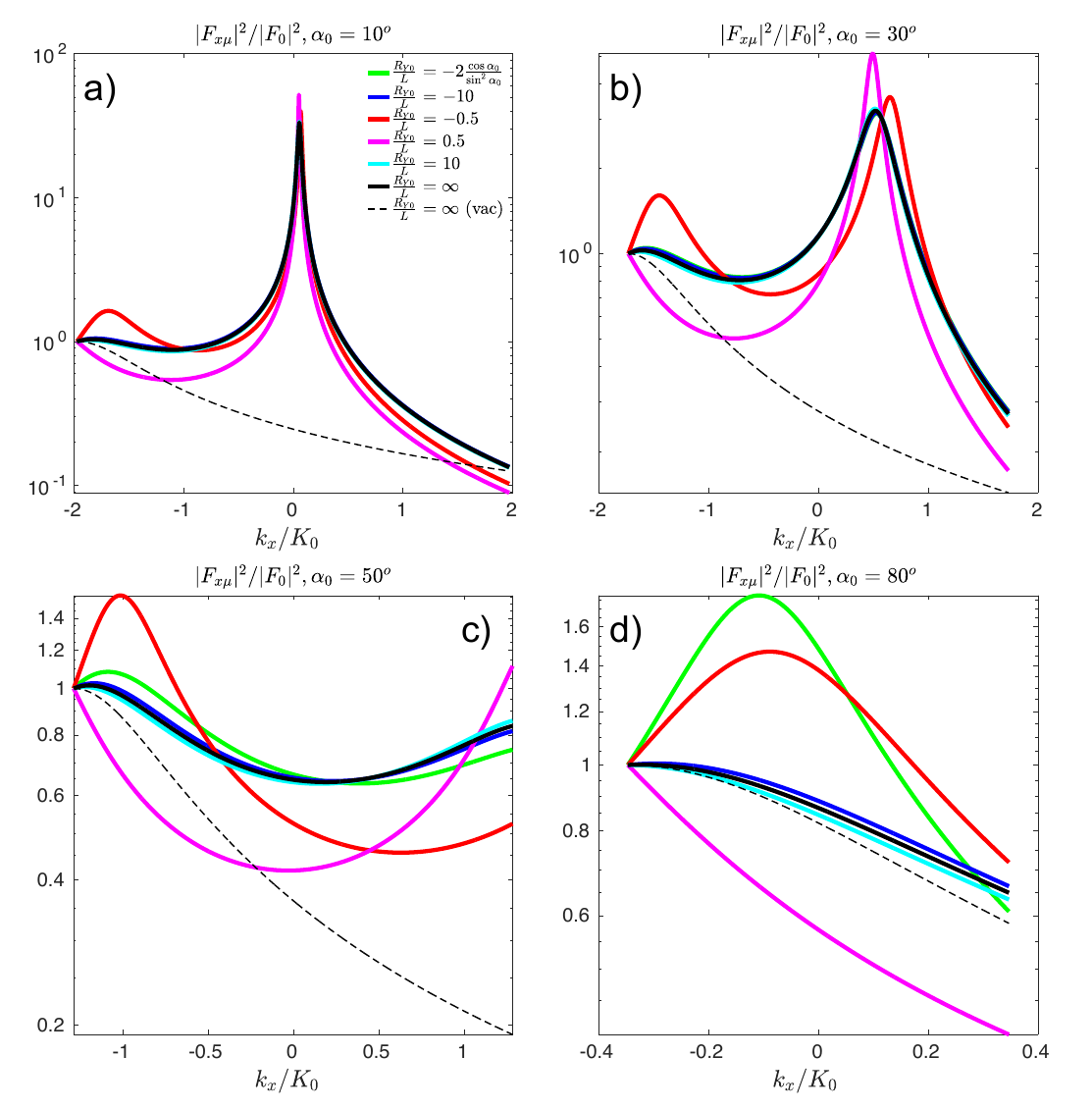}
	\end{center}
	\caption{Values of the filter function $|F_{x\mu}|^2/|F_0|^2$ for varying incident angles $\alpha_0$ and fixed initial $W_0 = 0.40 (\lambda L)^\frac{1}{2}$ as a function of the turbulent, selected $k_x$ component. For small incident angle $\alpha_0=10^\circ$, the filter is strongly peaked near $k_x = 0$, consistent with the signal being strongly localised near the turning point region. The signal is in fact sensitive to slightly positive $k_x>0$, which corresponds to a focusing slightly after the turning point. For $\alpha_0=30^\circ$, the peak near $k_x=0$ decreases and shifts to larger $k_x$'s, meaning that the signal starts getting important contributions from finite $k_x$ turbulent fluctuations away from the turning point. For $\alpha_0=50^\circ$ the peak has almost disappeared and the signal receives close-to-uniform contributions from $-1 \lesssim k_x/K_0 \lesssim 2$, corresponding to a highly delocalised signal along the beam path. See $|F_{x\mu}|^2(\tau)$ for $\alpha_0=50^\circ$ in figure \ref{wyry_fx_tau_alpha_ry0scan}. For $\alpha_0 = 80^\circ$ the beam is very close to fully expanding and the filter function approaches the vacuum solution. }   
	\label{fx_kx_alpha_ry0scan}
\end{figure}

For larger incident angles, the turning point takes place for smaller $\tau' = \cos\alpha_0$, while the beam term peaks at larger $\tau'$ (beam focus takes place at larger $\tau'$): the ray and beam terms compete to yield a filter function that is less peaked and less localised around the focus location for increasing $\alpha_0$. The ray term (blue) decreases in amplitude for larger $\alpha_0$, but less than the beam term (red), which decreases more. The beam term becomes broader in $\tau'$ for larger angles. These effects make the signal more delocalised along the path. This can be clearly seen in figure \ref{fxkmuwymu_alphascan}.c). The localisation becomes broad for $\alpha_0=50^\circ$, where beam focusing becomes less important in favour of contributions from $\tau'_\mu = 0$. In figure \ref{fxkmuwymu_alphascan}, the initial condition is chosen to be at the waist $W_{Y}(\tau=0)=W_0$. In figure \ref{fx_kx_alpha_ry0scan}, we show the effect of the initial radius of curvature $R_{Y0}$ on the filter function. An initially expanding beam (positive and finite $R_{Y0}/L$) is shown in figure \ref{fx_kx_alpha_ry0scan} not to make a dramatic difference on $|F_{x\mu}|^2$. For larger incident angles ($\alpha_0 = 80^\circ$), the filter peaks around $\tau = 0$, corresponding to the waist initial condition of figure \ref{fxkmuwymu_alphascan} (figure \ref{fx_kx_alpha_ry0scan} shows that the behaviour is qualitatively similar for different initial $R_{Y0}$). Therefore, for large incident angles, the DBS signal becomes delocalised with predominant contributions before and after the turning point (figures \ref{fxkmuwymu_alphascan} and \ref{fx_kx_alpha_ry0scan}). 

\begin{figure}
	\begin{center}
		\includegraphics[height=5cm]{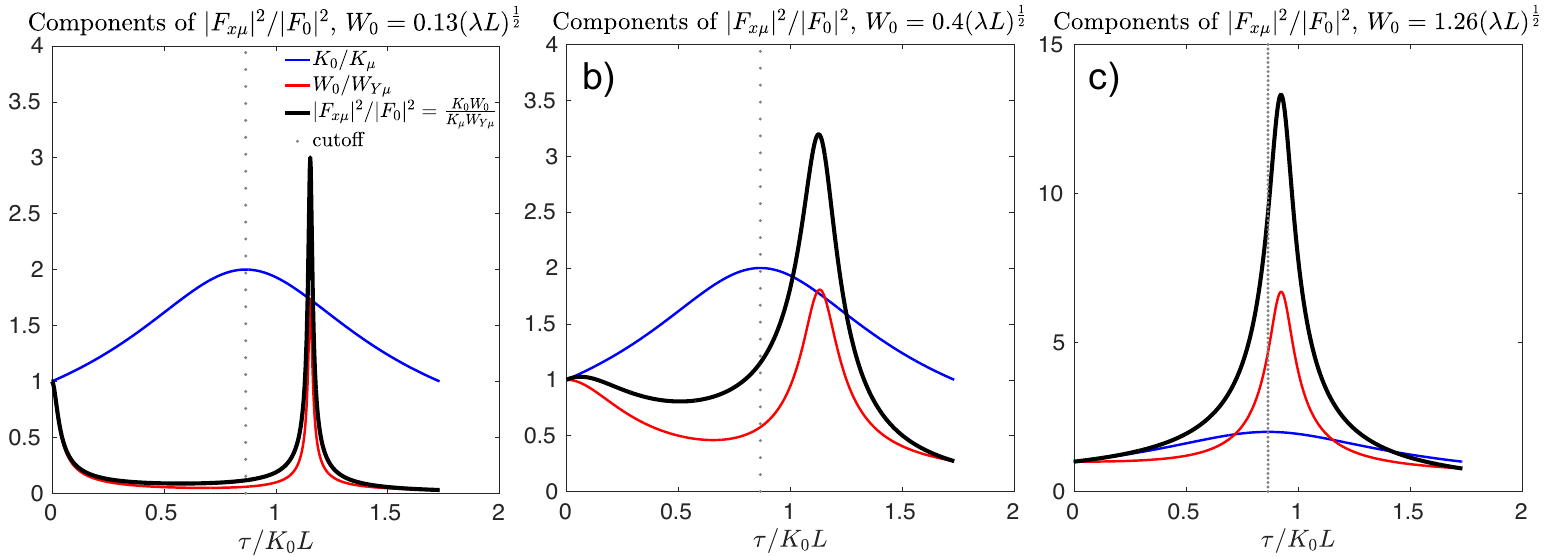}
	\end{center}
	\caption{Filter function $|F_{x\mu}|^2(\tau)/|F_0|^2 = K_0 W_0/K_\mu W_{Y\mu}$ and the corresponding ray component $K_0/K_\mu$ and beam component $W_0/W_{Y\mu}$ for varying values of the initial width $W_0 = 0.13, 0.40, 1.26(\lambda L)^\frac{1}{2}$, and fixed $R_{Y0} = \infty$ and $\alpha_0 = 30^\circ$. Note that in all cases the filter function is predominantly affected by the beam term $1/W_{Y\mu}$ which represents the focusing. }   
	\label{fxkmuwymu_gammascan}
\end{figure}

Figure \ref{fxkmuwymu_gammascan} shows the ray (blue) and beam (red) components of $|F_{x\mu}|^2$ for different values of the initial width $W_0 / (\lambda L)^\frac{1}{2} = 0.13, 0.40$ and $1.26 $, and fixed incident angle $\alpha_0 = 30^\circ$ and $R_{Y0} = \infty$ (launch at the waist). This filter function corresponds to the same launch conditions used in figure \ref{wyry_fx_tau_gammaalpha_ry0scan}. In this case, the ray term remains constant. The beam term is narrowest around the initial launch (waist) and around the focusing point for small initial width $W_0=0.13 (\lambda L)^\frac{1}{2}$. This is because a small initial width produces rapid expansion of the beam (see figure \ref{wyry_fx_tau_gammaalpha_ry0scan}) which gives negligible contributions to the beam term outside the waist and focus point. This suggests a very localised contribution to the DBS power from the vicinity of the focus point as well as the plasma edge (if enough fluctuations are present). For initial width $W_0=0.40(\lambda L)^\frac{1}{2}$, the location of the peak in $|F_{x\mu}|^2$ is similar to the one for $W_0 = 0.13 (\lambda L)^\frac{1}{2}$, but the filter function becomes broader, suggesting that the beam focusing is slower and takes place over a larger region, as shown in figure \ref{wyry_fx_tau_gammaalpha_ry0scan}. For $W_0=1.26(\lambda L)^\frac{1}{2}$, the filter function maximum has increased in value (beam focuses more) and has a peak as broad as the filter function for $W_0 = 0.4 (\lambda L)^\frac{1}{2}$. Interestingly, in this case the filter function $|F_{x\mu}|^2$ does not exhibit an initial decrease after the launch. This is because, in this condition, the beam does not initially expand following the launch, but only contracts from launch to the focus location (see black line in figure \ref{wyry_fx_tau_gammaalpha_ry0scan}.c)). Moreover, the beam focuses closer to the turning point than for $W_0/(\lambda L)^\frac{1}{2} = 0.13, 0.4$.  

Figures \ref{wyry_fx_tau_alpha_ry0scan}, \ref{wyry_fx_tau_gammaalpha_ry0scan}, \ref{fxkmuwymu_alphascan} and \ref{fxkmuwymu_gammascan} show that the beam focusing in $W_Y$ tends to take place close to but after the turning point ($K_{x\mu}=0, \tau'_\mu=\cos\alpha_0$) depending on the initial conditions for the incident angle $\alpha_0$, width $W_0/(\lambda L)^\frac{1}{2}$, and especially on the phase front radius of curvature $R_{Y0}$. This challenges the common belief that the DBS signal is always most sensitive at the turning point, and has important consequences for interpreting the DBS scattered power. 

It is equally instructive to characterise the filter function in terms of the turbulent scattered $k_x$. Note that negative $k_x $ corresponds to scattering from the beam in its first pass from launch towards the turning point, and positive $k_x$ to scattering from the beam in its return journey away from the turning point. Figure \ref{fx_kx_alpha_ry0scan} shows the variation of the filter function $|F_{x\mu}|^2$ for different incident angles $\alpha_0$, different initial radii of curvature $R_{Y0}$ and $W_0 = 0.40 (\lambda L)^\frac{1}{2}$ as a function of the turbulent scattered $k_x$. For $\alpha_0=10^\circ$, $|F_{x\mu}|^2$ is strongly peaked at $k_x=0$. This means that the DBS signal is strongly localised near the turning point region, and predominantly sensitive to fluctuations with $k_x \approx 0$. For $\alpha_0=30^\circ$, the filter peak shifts towards larger $k_x$. For even larger $\alpha_0=50^\circ$, the filter is sensitive almost uniformly to $-1 \lesssim k_x/K_0 \lesssim 2$, with peaks of the filter function at both positive and negative $k_x$, and surprisingly a dip in the vicinity of $k_x=0$. This suggests that the DBS power is sensitive predominantly to specific values of $k_x$, with a subdominant contribution from $k_x=0$. For $\alpha_0=80^\circ$, the beam follows closely vacuum propagation and the filter has contributions from small $k_x/K_0$ near $0$, but this time originating near $\tau=0$ (near the vacuum beam-waist, or edge of the plasma in a real experiment) and decays for larger $k_x$. The beam is very close to fully expanding and the filter function approaches the vacuum solution (black dashed line). 

\begin{figure}
	\begin{center}
		\includegraphics[height=5cm]{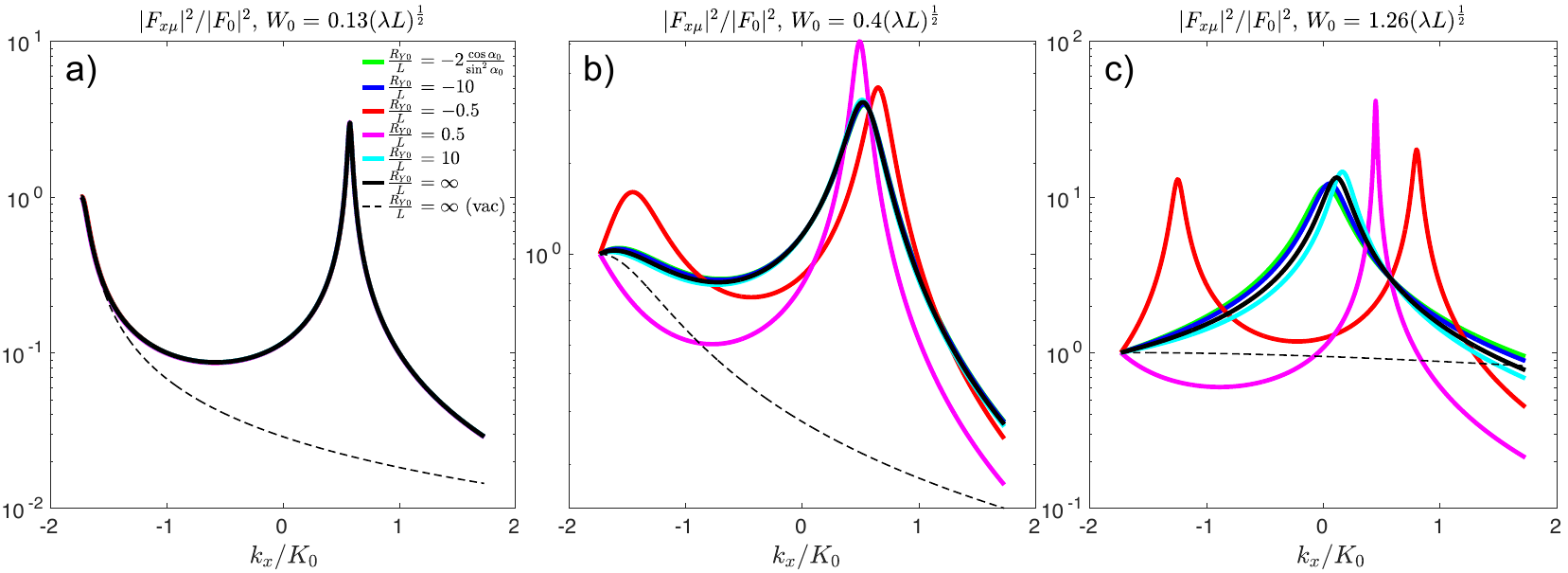}
	\end{center}
	\caption{Values of the filter function $|F_{x\mu}|^2/|F_0|^2$ for varying initial width $W_0/(\lambda L)^\frac{1}{2} = 0.13, 0.40, 1.26$ and fixed $\alpha_0 = 30^\circ$ as a function of the turbulent, selected $k_x$. For small initial $W_0$, the filter is strongly peaked at two values of $k_x$: one negative that corrresponds to the entrance to the plasma, and one positive from the beam focusing. For small $W_0$, the filter is also independent of the initial condition $R_{Y0}$. For $W_0 = 0.40 (\lambda L)^\frac{1}{2}$, the filter is dominated by a peak of amplitude and location similar to those of the peak due to focusing for small $W_0$, but broader. For $W_0 = 1.26 (\lambda L)^\frac{1}{2}$, the peak in $k_x$ stays broad and shifts towards $k_x=0$ for launch near the waist (large $|R_{Y0}/L|$).}   
	\label{fx_kx_gamma_ry0scan}
\end{figure}

Figure \ref{fx_kx_gamma_ry0scan} shows the corresponding filter function for varying $W_0 / (\lambda L)^\frac{1}{2} = 0.13, 0.40$ and $1.26$ and radii of curvature and fixed initial angle $\alpha_0=30^\circ$. The behaviour of the filter function for small $W_0$ can be understood from figure \ref{wyry_fx_tau_gammaalpha_ry0scan}. For small initial $W_0$, the filter is strongly peaked at two different values of $k_x$ and is independent of the initial condition $R_{Y0}$, consistent with the signal being strongly localised at the plasma entrance and near the turning point region. For $W_0 = 0.40 (\lambda L)^\frac{1}{2}$, the filter is dominated by a peak with amplitude and location similar to those of the peak due to focusing for small $W_0$, but broader, and hence the signal gets enhanced contributions from turbulent fluctuations away from the focusing region, depending on the initial condition $R_{Y0}$. For $W_0 = 1.26 (\lambda L)^\frac{1}{2}$ the peak in $k_x$ stays broad and shifts towards $k_x=0$ for launch near the waist (large $|R_{Y0}/L|$). The small $|R_{Y0}/L|$ cases are different, exhibiting distinctive narrow peaks that even appear twice for $R_{Y0}/L=-0.5$, consistent with two consecutive focusing regions, as in figure (\ref{wyry_fx_tau_gammaalpha_ry0scan}).

\subsection{Predictions of the measured turbulent spectra}

Throughout this manuscript, we have focused on describing the dependence of the filter function $|F_{x\mu}|^2$ on experimentally relevant initial values for $\alpha_0$, $W_0$ and $R_{Y0}$. We should note, however, that the intrinsic power spectrum of the turbulence itself has a direct impact on the measurement. The turbulent spectrum is generally a decreasing function of $k_x$, as demonstrated by direct numerical gyrokinetic turbulence simulations. In conditions where the filter function predominantly selects large $k_x$'s (finite to large $\alpha_0$), the turbulence spectrum in $k_x$ should be taken into account in order to understand the dominant $k_x$ contributing to the backscattering signal. In this section, we apply what we have learned about the filter function $|F_{x\mu}|^2$ in the beam-tracing model, and use it to understand its effect on the DBS backscattered power from a realistic density fluctuation spectrum. 

We use the electron density fluctuation wavenumber power spectrum $ \langle | \delta \hat{n}(k_x, k_y)|^2 \rangle_T$ obtained from nonlinear gyrokinetic simulations resolving strongly developed ETG turbulence in NSTX, which was examined by \cite{ruizruiz_pop_2015, ruizruiz_ppcf_2019, ruizruiz_ppcf_2020, ruizruiz_pop_2020, ren_nf_2020, ruizruiz_ppcf_2022}. The turbulent spectrum can be approximated by the following shape 

\begin{equation}
	\frac{\langle |\delta \hat{n} (k_x, k_y) |^2 \rangle_T}{n^2} \sim \frac{A}{1 + \big| \frac{k_x}{w_{k_x} }\big|^\zeta + \big| \frac{k_y - k_{y*}}{w_{k_y} } \big|^\eta }.
	\label{spec_analytic}
\end{equation}
This expression is fitted to the specific strongly-driven ETG simulations that we have mentioned above \citep{ruizruiz_ppcf_2022} to find: $\zeta \approx 3.14, \eta \approx 3.22, w_{k_x} \rho_s \approx 0.95, w_{k_y}\rho_s \approx 4.49, k_{y*}\rho_s \approx 6.31, A \approx 9.21 \ 10^{-7}$. \cite{ruizruiz_ppcf_2022} represented the fluctuation spectra in terms of the normal and binormal wavenumber components $k_n$ and $k_b$ perpendicular to the magnetic field, which can be easily calculated from the internal wavenumber components in a gyrokinetic code. The normal and binormal components correspond to the Cartesian $k_x$ and $k_y$ employed throughout this manuscript, as we have explained at the end of section \ref{gusakov_section}. Importantly, from here on we assume that the turbulent spectrum is uniform in space, that is, the same fit parameters from equation (\ref{spec_analytic}) are assumed throughout the whole plasma volume through which the beam propagates.

 
\begin{figure}
	\begin{center}
		\includegraphics[height=7cm]{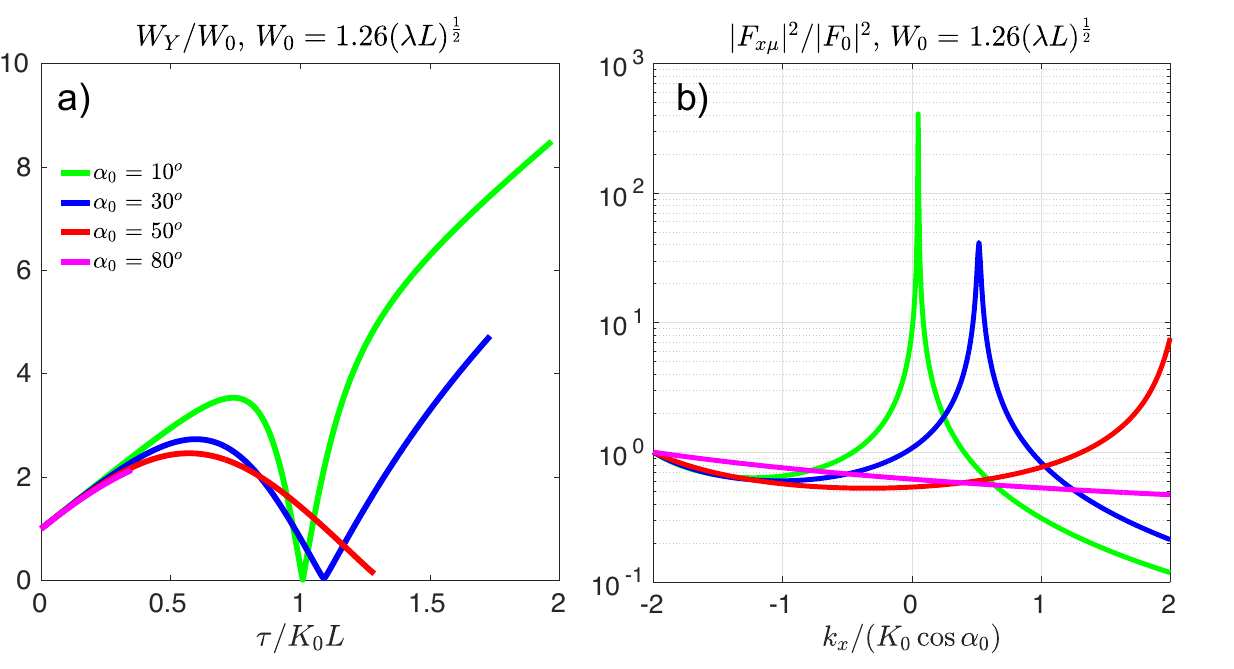}
	\end{center}
	\caption{a) Values of the normalised beam width $W_Y/W_0$ along the path $\tau$ for varying values of the incident angle $\alpha_0$, $W_0 \approx 1.26(\lambda L)^\frac{1}{2}$ and $R_{Y0}/L = 0.5 $. The curves are plotted for the values of $\tau$ in which the beam is traversing the plasma slab, $x>0$. We assume vacuum for $x<0$ and the beam is launched at $x=0$. b) Corresponding filter function $|F_{x\mu}|^2/|F_0|^2 = {K_0 W_0}/{ K_\mu W_{Y\mu} }$ plotted as a function of the scattered turbulent $k_x$ within the plasma slab, that is, for $|k_x/K_0 \cos\alpha_0| < 2 $.}   
	\label{wYtau_fx2kxk0_alpha0_plot}
\end{figure}

Figure \ref{wYtau_fx2kxk0_alpha0_plot}.a) shows the normalised beam width $W_Y/W_0$ from the beam-tracing equations as a function of $\tau$ for a range of incident angles $\alpha_0 = (10^\circ, 30^\circ, 50^\circ, 80^\circ)$, $W_0 = 1.26(\lambda L)^\frac{1}{2}$ and $R_{Y0}/L = 0.5 $. The beams in figure \ref{wYtau_fx2kxk0_alpha0_plot} have the same initial frequency $\Omega$, and reach a turning point location at $x=L_n = L\cos^2\alpha_0$ (see figure \ref{frames}). Figure \ref{wYtau_fx2kxk0_alpha0_plot}.a) exhibits beam focusing for $\alpha_0 = 10^\circ, 30^\circ$ within the plasma while the beam focusing takes place outside the plasma for $\alpha_0 = 50^\circ$ (vacuum propagation). Both the value of the minimum beam width and the focusing location increase with $\alpha_0$ in this particular case. 

Figure \ref{wYtau_fx2kxk0_alpha0_plot}.b) shows the DBS filter function $|F_{x\mu}|^2$ corresponding to figure \ref{wYtau_fx2kxk0_alpha0_plot}.a) plotted as a function of the turbulent scattered $k_x$ normalised to $K_0\cos\alpha_0$. The scattered $k_x$ at every $\tau$ location is computed using the Bragg condition for backscattering written in Cartesian coordinates, $k_x = - 2 K_{x\mu} = - 2 K_0 (\cos\alpha_0 - \tau')$ (see section \ref{gusakov_section} and appendix \ref{app_power}). The turning point takes place where $K_{x\mu}=0$, i.e. $\tau'=\cos\alpha_0$. Note how the filter function peaks closer to $k_x=0$ for decreasing $\alpha_0$, corresponding to the beam focusing location approaching the turning point for small angles. The intensity of the enhancement decreases for increasing $\alpha_0$. The dominant $k_x$ contributing to scattering is always positive $k_x>0$ in this situation, which means that the signal is predominantly originating from plasma locations $\emph{after}$ the turning point. Note how this depends on the initial condition: figure \ref{wyry_fx_tau_gammaalpha_ry0scan}.c) showed how the beam focusing can take place arbitrarily close to the turning point for the same $\alpha_0=30^\circ$ and $W_0 = 1.26 (\lambda L)^\frac{1}{2}$ values but different $R_{Y0}$. Increasing values of $\alpha_0$ will move the filter function localisation of the DBS signal further and further away from the turning point, reaching the plasma exit for $\alpha_0 \approx 50^\circ$ in these conditions (see red line in figure \ref{wYtau_fx2kxk0_alpha0_plot}.a)). 

\begin{figure}
	\begin{center}
		\includegraphics[height=12cm]{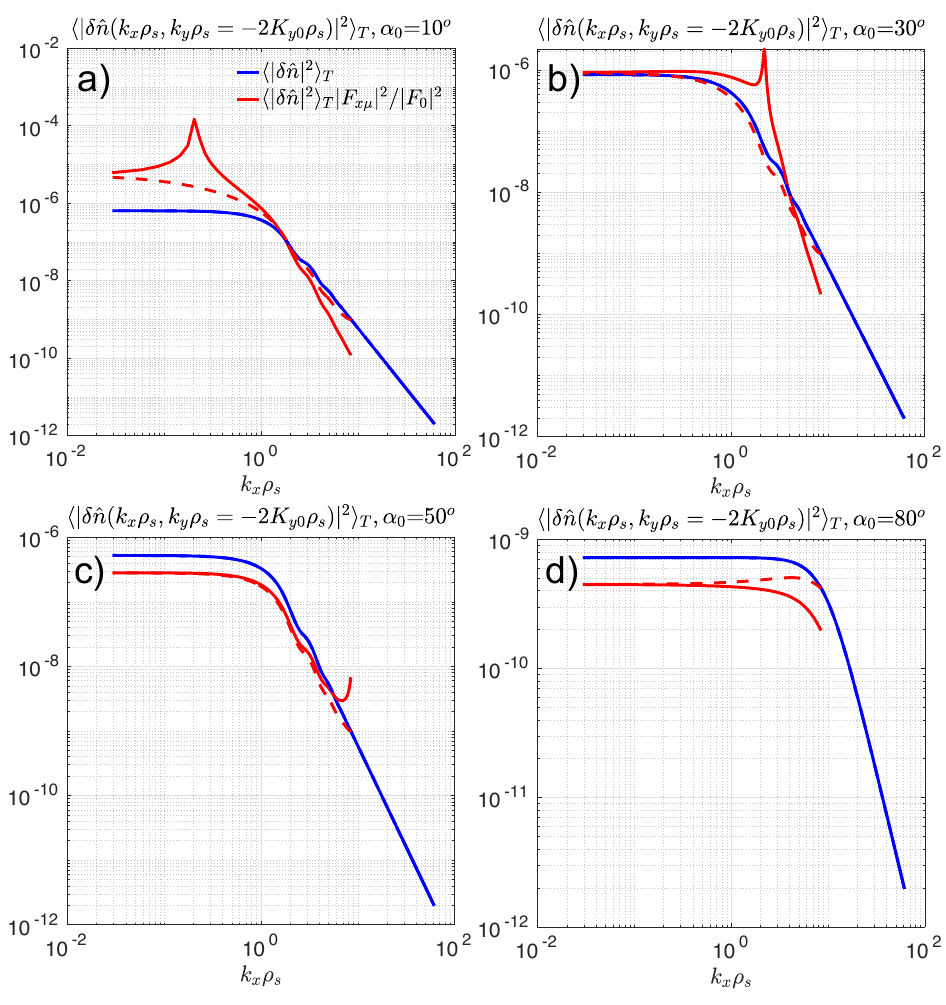}
	\end{center}
	\caption{Density fluctuation power $\langle |\delta \hat{n}|^2 \rangle_T$ and the product of the density fluctuation power and the filter function $\langle |\delta \hat{n}|^2 \rangle_T |F_{x\mu}|^2/|F_0|^2$ as a function of the normalised scattered turbulent radial wavenumber $k_x\rho_s$, for different values of the scattered turbulence $k_{y0}\rho_s = -2K_{0}\rho_s \sin\alpha_0$. The different $\alpha_0 = (10^\circ, 30^\circ, 50^\circ, 80^\circ)$ correspond to scattered wavenumbers $k_{y0}\rho_s = -(1.50, 4.92, 10.15, 48.29)$. Solid lines correspond to $k_x > 0$ while dashed lines correspond to $k_x < 0$. The red curve shows the combination of $k_x$ that the DBS signal power is sensitive to, showing that the beam focusing has a strong effect in the measurement $k_x$ selection for $\alpha_0 \lesssim 30-40^\circ$ while the effect becomes negligible for $\alpha_0 \gtrsim 40^\circ$.}   
	\label{dn2fx2_kxrhos_alpha0_plot}
\end{figure}

\begin{figure}
	\begin{center}
		\includegraphics[height=7cm]{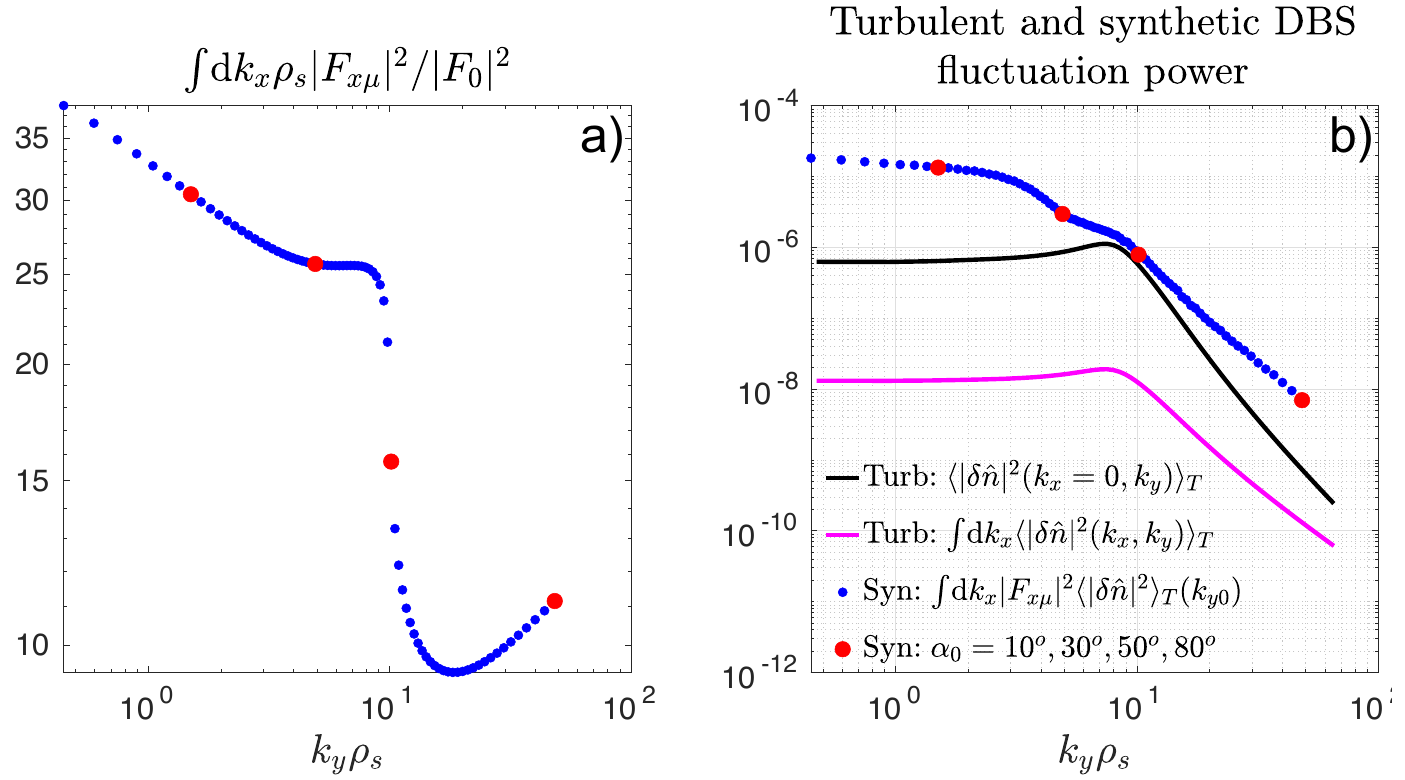}
	\end{center}
	\caption{a) Integrated filter function $\int \text{d}k_x |F_{x\mu}|^2$ for the scattered $k_x$ within the plasma ($|k_x/K_0| < 2\cos\alpha_0$). Note how the integrated weight $\int \text{d}k_x |F_{x\mu}|^2$ is not constant, meaning that the filter function can affect the measured DBS $k_y$ spectrum. b) Turbulent spectrum $\langle |\delta \hat{n}|^2(k_x=0, k_y) \rangle_T$ (black), $k_x$-integrated spectrum $\int \text{d}k_x \langle |\delta \hat{n}(k_x, k_y)|^2 \rangle_T$ (magenta) and synthetic DBS spectrum $\int \text{d}k_x |F_{x\mu}|^2 \langle | \delta \hat{n} |^2 \rangle_T (k_{y})$ (blue) as a function of $k_y$ of the turbulence. The turbulence spectrum has been fitted to a gyrokinetic simulation \citep{ruizruiz_ppcf_2022}. Note how the predicted DBS measurement is not able to capture the spectral peak of the turbulent spectrum (injection scale), but it is able to quantitatively reproduce the spectral exponent of the $k_x$-integrated spectrum. The red dots indicate the specific angles $\alpha_0 = 10^\circ, 30^\circ, 50^\circ, 80^\circ$ given for reference.}   
	\label{intfxmu2dn2_ky0_3}
\end{figure}

Next, we quantify the combined effect of the filter function, and its dependence on $k_x$, in conjunction with a realistic, power law turbulence spectrum. In figures \ref{dn2fx2_kxrhos_alpha0_plot} and \ref{intfxmu2dn2_ky0_3} we normalise $k_x$, $k_y$ and the initial wavenumber magnitude $K_0$ to the local ion sound gyroradius at the cutoff $\rho_s = c_s/\Omega_D$, where $c_s = (T_e/m_D)^{1/2}$ is the local ion sound speed, $m_D$ the deuterium mass, and $\Omega_D = eB/m_D$ is the deuterium gyro-frequency. Using the relation $\Omega = \omega_{pe0}/\cos\alpha_0$ between $\Omega$ and the local value of the electron plasma frequency at the cutoff $\omega_{pe0}$ (see section \ref{beam_foc_2d_sec}), we have $K_0\rho_s = (\beta_e m_D/2m_e)^{1/2}/\cos\alpha_0 \approx 4.26/\cos\alpha_0$, where $\beta_e \approx 1 \%$ is the electron beta using the local magnetic field, electron density and temperature, and $m_e$ is the electron mass. 

Figure \ref{dn2fx2_kxrhos_alpha0_plot} shows the radial wavenumber dependence of the density fluctuation power spectrum (blue lines) that would be sampled as the microwave beam propagates through the plasma. The product of the filter function $|F_{x\mu}|^2$ and the density fluctuation power is shown in red. These quantities are plotted as a function of the turbulent $k_x$ normalised by $\rho_s$. The solid lines correspond to $k_x$ scattered after the turning point ($k_x > 0$ in figure \ref{wYtau_fx2kxk0_alpha0_plot}.b)), while the dashed lines correspond to $k_x$ scattered before the turning point ($k_x<0$ in figure \ref{wYtau_fx2kxk0_alpha0_plot}.b)). As expected, the filter function has an important effect on the DBS signal power for smaller incidence angles $\alpha_0 = 10^\circ, 30^\circ$, while it becomes unimportant for larger $\alpha_0$, at which point the spectral falloff of the density spectrum becomes the dominant effect determining the $k_x$ selection in the DBS measurement. Note, for example, that the peaking of the filter function near the plasma exit for $\alpha_0=50^\circ$ becomes negligible because it occurs at large $k_x$. Importantly, figure \ref{dn2fx2_kxrhos_alpha0_plot} contains both information about the radial localisation of the DBS power (if plotted as a function of $\tau$), as well as the $k_x$-selectivity of the power spectrum in the DBS measurement (plotted as a function of $k_x$) through the combined effect of the filter function and the turbulent spectrum. We stress again that, in this discussion, we are neglecting the spatial dependence of the turbulence spectrum, $\langle |\delta \hat{n}|^2 \rangle_T$, which is assumed not to vary along the path.

Ultimately, one of the main purposes of DBS diagnostics as a turbulence measurement is to obtain the density fluctuation power spectrum in the binormal wavenumber of the turbulence, known as $k_\perp$ in the DBS jargon. In the context of this manuscript, $k_\perp$ is $k_y$. In order to understand the impact of the filter function in such a measurement, in figure \ref{intfxmu2dn2_ky0_3}.a) we plot the integrated filter function $|F_{x\mu}|^2$ over the relevant $k_x$ which would be sampled as the beam would propagate through the plasma as a function of the selected wavenumber $k_{y} = 2K_0 \sin\alpha_0$. Since $|F_{x\mu}|^2$ multiplies the density fluctuation spectrum $\langle |\delta \hat{n}|^2 \rangle_T$, a constant $\int \text{d}k_x |F_{x\mu}|^2$ as a function of $k_{y}$ would suggest that the DBS measurement is able to reproduce the true shape of the density fluctuation wavenumber power spectrum $\langle |\delta \hat{n}|^2(k_{y}) \rangle_T$. Figure \ref{intfxmu2dn2_ky0_3}.a) shows that the $k_{y}$ dependence of $\int \text{d}k_x |F_{x\mu}|^2$ varies by a factor $\sim 3$ for $k_{y}\rho_s \sim 0.4-40 $ ($\alpha_0 \sim 3^\circ-80^\circ$). The red dots correspond to the values of the integral of the filter function for $\alpha_0 = 10^\circ, 30^\circ, 50^\circ$ and $80^\circ$. For small $k_{y}$ (small $\alpha_0$), the filter function is a decreasing function of $k_{y}$. This is due to the fact that for small $\alpha_0$, the beam width focuses and produces an overall enhancement upon integration over the sampled $k_x$. In fact, $W_Y \sim \alpha_0$ for small $\alpha_0$ \citep{belrhali_jpp_2024}, that is $|F_{x\mu}|^2 \sim 1/\alpha_0^2$ ($K_\mu \sim \alpha_0$). This means that for small enough angles, beam tracing is bound to break down: the prediction of a signal enhancement for small angles might overestimate the true enhancement in reality, as discussed in detail by \cite{maj_pop_2009, maj_ppcf_2010}. The precise angles for which this might happen and corrections to the beam-tracing model for small angles will be the object of future publications. 

For $k_{y}\rho_s \approx 4-5$ ($\alpha_0 \approx 30^\circ$), the integrated filter function plateaus before abruptly decreasing for $\alpha_0 \approx 50^\circ$. A similar plateau behaviour is observed in detailed measurements of the scattered power in DIII-D, reported in a recent publication by \cite{pratt_nf_2024}. In our case, this is due to the fact that for such large angles, beam focusing has started to take place close to the plasma exit (see figure \ref{wYtau_fx2kxk0_alpha0_plot}.b)). This eliminates the signal enhancement due to the beam focusing inside the plasma (the $k_x$ integration is only performed inside the plasma), therefore decreasing the value of the integrated filter function. For larger $k_{y}$ ($\alpha_0 \gtrsim 60^\circ$), the integrated filter increases again. This last increase is due to the fact that the launch condition at the plasma edge starts to become important. For such large incidence angles, the beam is practically glancing and can be treated as a high-k backscattering system \citep{rhodes_rsi_2006, hillesheim_nf_2015b}. Figure \ref{intfxmu2dn2_ky0_3}.a) shows that the range of sampled radial wavenumbers in the density fluctuation spectrum can have an important effect on the measured DBS power spectrum. 

Figure \ref{intfxmu2dn2_ky0_3}.b) shows the turbulence binormal wavenumber power spectrum $\langle |\delta \hat{n}|^2(k_x=0, k_y) \rangle_T$ (black) corresponding to $k_x=0$. This follows the traditional interpretation that dominant contributions to the DBS signal originate from the cutoff. The magenta line shows the $k_x$-integrated wavenumber spectrum $\int \text{d}k_x\rho_s \langle |\delta \hat{n}|^2 (k_x, k_y)\rangle_T$. This could be interpreted as a line-integrated measurement of the density-fluctuation spectrum. The DBS system can be interpreted as a line-integrated measurement for large angles $\alpha_0$, since for large $\alpha_0$, $|F_{x\mu}|^2$ varies slowly and is not strongly peaked (see figure \ref{fx_kx_alpha_ry0scan}.d)). The extent to which it can be interpreted as a line integral rather than a localised measurement is quantified by the filter function $|F_{x\mu}|^2$, that we discuss next.

The blue dots in figure \ref{intfxmu2dn2_ky0_3}.b) show expected \emph{synthetic} scattered power spectrum in a DBS measurement, given by the quantity $\int \text{d} k_x\rho_s |F_{x\mu}|^2/|F_0|^2 \langle |\delta \hat{n}|^2 \rangle_T (k_{y})$ in this beam-tracing model (see equations (\ref{prkxky2_cart})). For each incidence angle $\alpha_0$ (each $k_{y}$), the {synthetic} DBS scattered power includes the effect of the filter function and the radial wavenumber dependence of the density fluctuation spectrum. As expected from the non-monotonic nature of $\int \text{d}k_x |F_{x\mu}|^2$ from figure \ref{intfxmu2dn2_ky0_3}.a), the synthetic scattered power does not reproduce the true density fluctuation wavenumber spectrum for $k_x=0$ nor the $k_x$-integrated spectrum. The peak in the density fluctuation power spectrum ($k_y\rho_s \approx 6-8$), which is the driving scale of the turbulence, cannot be identified in the synthetic power spectrum. The peak should be visible for $\alpha_0 \approx 30^\circ-40^\circ$, where the beam model is predicted to be quantitatively accurate. The enhancement of the synthetic scattered power with respect to the true turbulence spectrum for low $k_{y}$ (small incidence angles $\alpha_0 \lesssim 20^\circ$) is due to the enhancement of the signal by beam focusing. For larger $k_{y}\rho_s \gtrsim 10$ ($\alpha_0 \gtrsim 40^\circ$), the synthetic spectrum exhibits a power-law spectral decay $\propto k_{y}^{-3.05}$, which is different from that of the true turbulent spectrum for $k_x=0$ ($\propto k_{y}^{-4.16}$). The spectral decay is shown to quantitatively agree with the $k_x$-integrated spectrum, for which $\propto k_{y}^{-2.84}$. This suggests that DBS scattered power measurements could accurately capture the spectral exponent of the $k_x$-integrated spectrum, contrary to the traditional belief that DBS measurements only select the $k_x=0$ component \citep{hillesheim_rsi_2012, holland_nf_2012}. For this particular case, the DBS scattered power measurements can quantitatively reproduce the true spectral exponent of the $k_x$-integrated, density-fluctuation power spectrum.

\section{Conclusions and discussion}

In this manuscript, we have discussed the phenomenon of beam focusing in the vicinity of a turning point. This phenomenon is not new and was already observed in numerical simulations for the 2D linear-layer problem \citep{poli_pop_1999, poli_fed_2001, kravtsov_berczynski_2007}, analytical calculations \citep{maj_pop_2009, maj_ppcf_2010}, as well as in numerical simulations in realistic tokamak geometry \citep{conway_irw_2007, conway_irw_2015, conway_irw_2019}. These results were confirmed in this work by numerical simulations using the Scotty code for an NBI-heated L-mode plasma in the JET tokamak. The analytic solution derived from a 2D linear-layer problem adjusted to fit the experimental conditions of JET discharge 97080 showed encouraging agreement with the Scotty simulations for realistic 3D tokamak geometry. Moreover, the beam-focusing phenomenon was quantitatively reproduced by the TORBEAM code (not shown). This motivated us to study the phenomenon of beam focusing in the linear-layer problem. We have characterised the beam focusing in terms of the initial incident angle $\alpha_0$, initial width $W_0$ and initial radius of curvature $R_{Y0}$. In particular, we have seen that beam focusing tends to be enhanced at small angles, independently of $R_{Y0}$, to the point that the beam-tracing approximation becomes quantitatively inaccurate near the turning point, as discussed by \cite{maj_pop_2009, maj_ppcf_2010}. Beam focusing also exhibits a clear dependence on the initial width $W_0$: initial widths small with respect to $(\lambda L)^\frac{1}{2}$ lead to pronounced initial expansion due to diffraction of the electromagnetic waves as captured by the beam-tracing equations. Following the initial beam expansion for small $W_0$, the beam exhibits focusing in the vicinity of the turning point. For initial widths large with respect to $(\lambda L)^\frac{1}{2}$, the initial expansion can completely disappear to yield only an initial contraction towards a maximum focusing location in the vicinity of the turning point. The value and location of the beam focusing is shown to depend on the initial conditions $\alpha_0$, $W_0$ and $R_{Y0}$. 

We have used the analytical beam-tracing solution and applied it to the beam-tracing model for DBS \citep{valerian_ppcf_2022}. We have found an analytic expression for the filter function $|F_{x\mu}|^2/ |F_0|^2 = K_0W_0/K_\mu W_{Y\mu}$ that characterises the scattering intensity along the beam path through the plasma, and we have studied it as a function of the beam initial conditions. This shows that $W_{Y}$ is a critical parameter affecting the DBS scattering intensity along the path. We apply the lessons learned from our beam focusing formula to study the DBS filter function $|F_{x\mu}|^2$. When the beam focuses most ($W_{Y}$ is minimum), the filter function is enhanced. This means that DBS is most sensitive to enhanced contributions from the vicinity of the focusing point, and as a consequence, it is most sensitive for small angles $\alpha_0$, for which the focusing is large. 

The filter function $|F_{xy,\mu}|^2$ in terms of $k_x$ and $k_y$ in Cartesian coordinates opens the door to implementing synthetic diagnostics for DBS as in \citep{white_pop_2008, holland_pop_2009, leerink_cpp_2010, hillesheim_rsi_2012, holland_nf_2012, leerink_prl_2012, gusakov_ppcf_2013, krutkin_nf_2019}. In the $k_x$ direction, we recover formulas reported in \citep{gusakov_pop_2017} when applied to the case of auto-correlation (power) spectrum. In previous work, \cite{gusakov_ppcf_2004, gusakov_ppcf_2014, gusakov_pop_2017} argued that a forward scattering contribution to the DBS amplitude is responsible for the signal enhancement observed for small incidence angle $\alpha_0$, in DBS as well as in radial correlation Doppler reflectometry. The calculations by \cite{gusakov_ppcf_2004, gusakov_ppcf_2014, gusakov_pop_2017} were based on the full-wave analytic solution to the Helmholtz equation, including the Airy function behaviour near the turning point that is absent in our present description. Our work demonstrates that the enhancement of the DBS scattered power for small incidence angles observed in \citep{gusakov_ppcf_2014, gusakov_pop_2017} can be completely explained via beam tracing, without the need of capturing the Airy behaviour in the vicinity of the turning point. We show that the underlying principle behind the enhancement of the DBS power is the beam focusing near the turning point. The formulas from section \ref{gusakov_section} recover the $k_y$-resolution, or diagnostic wavenumber resolution $\Delta k_y$, that was already reported in previous work \citep{lin_ppcf_2001, hirsch_rsi_2001, hillesheim_rsi_2012}. We find that $\Delta k_y$ is dependent on the initial conditions $\alpha_0$, $W_0$ and $R_{Y0}$ and on $L$, and is constant along the path. 


In addition to discussing the effect of the filter function $|F_{x\mu}|^2$ on the total DBS scattered power measurement, we have also studied the effect of the turbulent wavenumber spectrum on the measured total scattered power. We have used a realistic representation of the turbulence spectrum based on gyrokinetic simulations, as done by \cite{ruizruiz_ppcf_2022}. We show the combined effect of the $k_x$ dependence of the filter function and the $k_x$ dependence of the turbulence spectrum. The filter function is shown to have a most dominant effect for small angles $\alpha_0 = 10^\circ, 30^\circ$ since for these specific conditions DBS is sensitive to low-to-intermediate $k_y$ and the spectrum is large. For larger angles, the filter function selects larger $k_{y} = - 2 K_0 \sin\alpha_0$. In the specific condition analysed, larger angles $\alpha_0 > 50^\circ$ corresponded to wavenumbers $k_{y}$ with small turbulent power in the spectrum. In those conditions, the filter function was shown to have little effect on the measurement and the dominant contributions to the scattered power could directly be assessed from the intrinsic turbulent spectrum itself. We also show that the filter function integrated over $k_x$ is far from being constant and is strongly dependent on the $k_{y}$ selected. The integrated filter function can be affected by the beam focusing inside or outside the plasma and the launch condition. This shows that one expects the total scattered power to depend on the incident angle of beam injection in the plasma, $\alpha_0$. For the particular NSTX turbulent spectrum studied in this manuscript, we show that the measured DBS scattered power spectrum could not recover the peak in the true density fluctuation spectrum. The power law decay exponent of the synthetic DBS spectrum can reproduce the $k_x$-integrated turbulence spectrum, but not the spectrum for $k_x=0$. 

This work helps characterise and interpret the total scattered power measured by DBS. Little attention has been paid to the spectral frequency power spectrum and specific experimental analysis techniques that are usually employed to study it. For example, of particular importance is the dependence of the Doppler shifted DBS frequency spectrum on the diagnostic wavenumber resolution, as well as the effect of the $k_x = 0$ component (which is directly related to the beam focusing). The impact on the measured frequency spectrum remains an open question, but could be easily tackled using the reduced filter function developed in this work. This will require further analysis of the model and comparisons to experiments, which could be the object of future publications. 

Lastly, the model presented here has limitations for small enough incident angles $\alpha_0$, that is, for close to normal incidence beams. One can show that the filter function $|F_{x\mu}|^2 \propto 1/\alpha_0^2$ for small angles. That explains the highly peaked behaviour of the filter function in figures \ref{fxkmuwymu_alphascan}.a) and \ref{fx_kx_alpha_ry0scan}.a) near the focusing region. This behaviour has its origin in the beam-tracing solution for the electric field, for which one can show that $E_b \propto 1/\alpha_0$ for small angles \citep{belrhali_jpp_2024}. This was studied by \cite{maj_pop_2009, maj_ppcf_2010}. However, as pointed out in Maj's work, the validity of the beam-tracing approximation is questionable for small enough incident angles. The beam-tracing solution for $E_b$ might be overpredicting the electric-field enhancement due to focusing. In the exact solution for the electric field, one expects Airy behaviour to become important through interference between the incident and returning beams in the vicinity of the turning point, especially in conditions in which the beam width $W_Y$ approaches the Airy length $l_{Ai} = (L/K_0^2)^\frac{1}{3}$. This behaviour is absent in the beam-tracing model, which expands the Airy behaviour asymptotically far from the turning point. Although the beam-tracing solution might be overpredicting the electric-field enhancement in the vicinity of the focusing region, we are confident that the phenomenon of beam focusing is physical, as confirmed by \cite{maj_pop_2009, maj_ppcf_2010} and by our recent work. It is therefore important to understand the reasons driving the beam focusing as the beam approaches the turning point, as well as the exact beam focusing amplitude near the turning point. Although our work suggests that the electric field is overly enhanced by beam tracing in the vicinity of the focusing region for small incidence angles, nothing is said about the total integrated power along the path that DBS measures. Future work will seek to employ more sophisticated models than beam tracing that capture Airy behaviour (as done in recent publications, such as \cite{lopez_pre_2023}) to quantify how accurate the total integrated power enhancement predicted by beam tracing is for small incidence angles.

\section*{Acknowledgements}

Discussions with T. Rhodes have been insightful for the preparation of this manuscript. This work has been supported by the Engineering and Physical Sciences Research Council (EPSRC), with grant numbers EP/R034737/1 and EP/W026341/1. V. H. Hall-Chen was partly funded by an A*STAR Green Seed Fund, C231718014, and a YIRG, M23M7c0127. This work was supported by the U.S. Department of Energy under contract number DE-AC02-09CH11466. The United States Government retains a non-exclusive, paid-up, irrevocable, world-wide license to publish or reproduce the published form of this manuscript, or allow others to do so, for United States Government purposes. This work has been carried out within the framework of the EUROfusion Consortium, funded by the European Union via the Euratom Research and Training Programme (Grant Agreement No 101052200 — EUROfusion). Views and opinions expressed are however those of the author(s) only and do not necessarily reflect those of the European Union or the European Commission. Neither the European Union nor the European Commission can be held responsible for them.

\makeatletter
\def\fps@table{h}
\def\fps@figure{h}
\makeatother

\appendix

\section{Analytic beam-tracing solution for the linear-layer problem.}
\label{app1}

We solve the beam-tracing equations in a 2D slab for the O-mode with a constant density gradient ($\omega_{pe}^2(x) = \Omega^2 x/L$, where $x$ is our Cartesian coordinate in figure \ref{frames}) and launch frequency $\Omega$. We will make use of the dispersion relation $H=0$, which takes the form $H = {K^2}/{K_0^2} - 1 + {\omega_{pe}^2}/{\Omega^2} = 0$ for the O-mode. The general beam-tracing evolution equation for $\boldsymbol{\Psi}$ is calculated along the central ray path $(x_c, y_c)$, and reads \citep{valerian_ppcf_2022}

\begin{equation}
	\frac{\text{d} \boldsymbol{\Psi}}{\text{d} \tau} = - \big( \boldsymbol{\Psi} \cdot \nabla_K \nabla_K H \cdot \boldsymbol{\Psi} + \boldsymbol{\Psi} \cdot \nabla_K \nabla H + \nabla \nabla_K H \cdot \boldsymbol{\Psi} + \nabla \nabla H \big) .
	\label{bt_eq_general}
\end{equation}

The analytic solution to the Gaussian beam-tracing equations was already obtained by \cite{pereverzev_pop_1998} and \cite{ poli_pop_1999} for the 2D linear-layer problem and normal incidence ($\alpha_0 = 0$). The solution for the oblique incidence angle was obtained by \cite{maj_pop_2009, maj_ppcf_2010}. The beam-tracing equation (\ref{bt_eq_general}) in the linear-layer problem reduces to ${\text{d} \boldsymbol{\Psi}}/{ \text{d} \tau } = -({2}/{K_0^2}) \boldsymbol{\Psi}^2$ since we have $\nabla \nabla H = 0, \nabla_K \nabla H = \nabla \nabla_K H= 0$ and $\nabla_K \nabla_K H = ({2}/{K_0^2}) \mathbf{I}$, where $\mathbf{I}$ is the identity matrix. Normalizing $\boldsymbol{\Psi}$ by $K_0/L$ and $\tau$ by $K_0L$, the corresponding solution for $\boldsymbol{\Psi}' = \boldsymbol{\Psi} ({L}/{K_0})$ is

\begin{equation}
	\begin{aligned}
		& \boldsymbol{\Psi'^{-1}}(\tau') = 2 \tau' \mathbf{I} + \boldsymbol{\Psi'^{-1}}(0).
		\end{aligned}
	\label{beam_solution_mat}
\end{equation}
The matrix $\boldsymbol{\Psi}'$ must satisfy the constraint 

\begin{equation}
	\nabla_K H \cdot \boldsymbol{\Psi}' + \nabla H \frac{L}{K_0} = 0.
	\label{constraint}
\end{equation}
Taken at the initial condition, equation (\ref{constraint}) gives 

\begin{equation}
	\boldsymbol{\Psi}'(0) = 
	\begin{pmatrix}
	 	\Psi_{xx0}' & \Psi_{xy0}' & 0 \\
		\Psi_{xy0}' & \Psi_{yy0}' & 0 \\
		0 & 0 & \Psi_{zz0}' \\
	\end{pmatrix}
	= 
	\begin{pmatrix}
	 	\Psi_{yy0}' \tan^2\alpha_0  - \frac{1}{2\cos\alpha_0} & - \Psi_{yy0}' \tan\alpha_0 & 0 \\
		- \Psi_{yy0}' \tan\alpha_0 & \Psi_{yy0}' & 0 \\
		0 & 0 & \Psi_{zz0}' \\
	\end{pmatrix}. \\
	\label{psi_lab_t0}
\end{equation}
Note that one only needs $\Psi_{yy0}'$ and $\alpha_0$ to specify the initial conditions for the beam in the 2D linear-layer problem ($\Psi_{zz0}'$ is along $\mathbf{\hat{b}}$). Equation (\ref{constraint}) also gives the beam components $\Psi_{Yg}'$ and $\Psi_{gg}'$

\begin{equation}
	\begin{aligned}
	& \Psi_{Yg}'(\tau) &&= - \frac{1}{g} \big( \nabla H \cdot \mathbf{\hat{Y}} \big) \frac{L}{K_0} = - \frac{1}{2} \sin\alpha_0 \frac{K_0^2}{K^2} , \\
	& \Psi_{gg}'(\tau) &&= - \frac{1}{g} \big( \nabla H \cdot \mathbf{\hat{g}} \big) \frac{L}{K_0} = - \frac{1}{2} \frac{K_x}{K_0} \frac{K_0^2}{K^2} . \\
	\end{aligned}
	\label{psiyg_psigg}
\end{equation}
Taken at the initial condition, equation (\ref{psiyg_psigg}) gives 

\begin{equation}
	\begin{pmatrix}
	 	\Psi_{YY0}' & \Psi_{Yg0}' & 0 \\
		\Psi_{Yg0}' & \Psi_{gg0}' & 0 \\
		0 & 0 & \Psi_{XX0}' \\
	\end{pmatrix}
	= 
	\begin{pmatrix}
	 	\Psi_{YY0}' & - \frac{1}{2} \sin\alpha_0 & 0 \\
		- \frac{1}{2} \sin\alpha_0 & - \frac{1}{2} \cos\alpha_0 & 0 \\
		0 & 0 & \Psi_{zz0}' \\
	\end{pmatrix}, \\
	\label{psi_beam_t0}
\end{equation}
where $\Psi_{YY0}'$ is the beam-frame initial condition. The components $\Psi_{YY0}'$ and $\Psi_{yy0}'$ are related to each other. We proceed to discuss how. 

The beam-tracing equations are more easily solved in the lab frame $\{ \mathbf{\hat{x}}, \mathbf{\hat{y}}, \mathbf{\hat{z}} \}$ (figure \ref{frames}). The spatial coordinates in the lab frame are related to those in the beam frame by

\begin{equation}
	\begin{pmatrix}
		x - x_c \\
		y - y_c \\
		z \\		
	\end{pmatrix}
	= \mathbf{R}_\alpha 
		\begin{pmatrix}
		Y \\
		q \\
		X \\		
	\end{pmatrix},
	\label{change_basis_vec}
\end{equation}
where $q$ is the coordinate that is tangent to $\hat{\mathbf{g}}$, $Y$ is along $\hat{\mathbf{Y}}$, $X$ is along $\mathbf{\hat{X}}$ and $\mathbf{R_\alpha}$ is a rotation matrix given by 

\begin{equation}
	\mathbf{R_\alpha}(\tau) = 
	\begin{pmatrix}
		\sin\alpha (\tau) & \cos\alpha (\tau) & 0 \\
		-\cos\alpha (\tau) & \sin\alpha (\tau) & 0 \\
		0 & 0 & 1 \\		
	\end{pmatrix}
	.
	\label{rotation}
\end{equation}
The beam-frame solution can be computed from the lab-frame solution via a simple rotation of $\pi/2 - \alpha$ along $\mathbf{\hat{z}} = \mathbf{\hat{X}}$ (figure \ref{frames}, equation (\ref{change_basis_vec})). The change of basis is 




\begin{equation}
	\begin{pmatrix}
		\Psi_{YY} & \Psi_{Yg} & \Psi_{YX} \\
		\Psi_{Yg} & \Psi_{gg} & \Psi_{Xg} \\
		\Psi_{YX} & \Psi_{Xg} & \Psi_{XX} \\		
	\end{pmatrix}
	= \mathbf{R}^\text{T}_\alpha(\tau) \\
	\begin{pmatrix}
	 	\Psi_{xx} & \Psi_{xy} & \Psi_{xz} \\
		\Psi_{xy} & \Psi_{yy} & \Psi_{yz} \\
		\Psi_{xz} & \Psi_{yz} & \Psi_{zz} \\
	\end{pmatrix} \\
	\mathbf{R}_\alpha(\tau).
	\label{change_basis_mat}
\end{equation}

The evolution of the angle $\alpha$ along the path can be easily computed from the trajectory of the central ray (see equations (\ref{raytracing_sols})). It is given by 

\begin{equation}
	\begin{aligned}
		& \tan\alpha = \frac{\tan\alpha_0}{1-\tau' / \cos\alpha_0} = \sin\alpha_0 \frac{K_0}{K_x} , \\
		& \sin\alpha = \frac{\sin\alpha_0}{(1 - 2\tau'\cos \alpha_0 + \tau'^2)^{1/2}} = \sin\alpha_0 \frac{K_0}{K} , \\
		& \cos\alpha = \frac{\cos\alpha_0 - \tau' }{(1-2\tau'\cos\alpha_0+\tau'^2)^{1/2}} = \frac{K_x}{K} . \\
	\end{aligned}
	\label{sincos}
\end{equation}
Using equations (\ref{change_basis_mat}) and (\ref{rotation}), one can relate the beam-frame initial condition $\Psi_{YY0} = {L}/{R_{Y0}} + i {2}/{W_{0}^2}$ to the lab-frame initial condition $\Psi_{yy0}'$, yielding $\Psi_{YY0}' = - {\sin^2\alpha_0}/{(2\cos\alpha_0)} + {\Psi_{yy0}'}/{\cos^2\alpha_0} $. The change of basis in equation (\ref{change_basis_mat}) also gives the following relations between all the beam-frame and lab-frame components


\begin{equation}
 	\begin{aligned}
	& \Psi_{YY}(\tau') = \Psi_{xx} \sin^2\alpha - 2 \Psi_{x y}\sin\alpha \cos\alpha + \Psi_{yy}\cos^2\alpha , \\ 
	& \Psi_{Yg}(\tau') = \big(\Psi_{xx}-\Psi_{yy} \big)\sin\alpha \cos\alpha + \Psi_{x y} \big(\sin^2\alpha - \cos^2\alpha \big) , \\ 
	& \Psi_{gg}(\tau') = \Psi_{xx}\cos^2\alpha + 2 \Psi_{x y}\sin\alpha \cos\alpha + \Psi_{yy}\sin^2\alpha , \\
	& \Psi_{XX}(\tau') = \Psi_{zz} . \\	
	\end{aligned}
	\label{frame_to_beam}
\end{equation}

We now provide the lab-frame solution, derived from equation (\ref{beam_solution_mat})

\begin{equation}
	\begin{aligned}
		& \Psi_{xx}'(\tau') &&= \frac{ \Psi'_{xx0} + 2 \Delta_0 \tau'  }{ D(\tau')  } , \\
		& \Psi_{x y}'(\tau') &&= \frac{ \Psi'_{x y 0} }{ D(\tau')  } , \\
		& \Psi_{yy}'(\tau') &&= \frac{ \Psi'_{yy0} + 2 \Delta_0 \tau'  }{ D(\tau')  } . \\
	\end{aligned}
	\label{beam_solution}
\end{equation}
Here the denominator $D(\tau')$ can be readily expressed in terms of the lab-frame $\mathbf{\hat{x}}$ component of the central ray $\mathbf{K}$, $K_x$, yielding 

\begin{equation}
	\begin{aligned}
	& D(\tau') &&= 1 + 2\tau' \Tr_0 + 4\Delta_0\tau'^2 \\
	& &&= \frac{1}{\cos\alpha_0} \frac{K_x}{K_0} + 2 \frac{\Psi_{yy0}'}{\cos\alpha_0} \bigg( \frac{\cos^2\alpha_0 - \sin^2\alpha_0}{\cos\alpha_0} \frac{K_x}{K_0} + \sin^2\alpha_0 - \frac{K_x^2}{K_0^2} \bigg) , \\
	\end{aligned}
	\label{d_denominator}
\end{equation}
with
\begin{equation}
	\begin{aligned}
	& \Delta_0 &&= (\Psi_{yy}'\Psi_{xx}' - \Psi_{x y}'^2)_0 = (\Psi_{YY}'\Psi_{gg}' - \Psi_{Yg}'^2)_0 = - \frac{\Psi_{yy0}'}{2\cos\alpha_0} ,  \\
	& \Tr_0 &&= (\Psi_{yy}' + \Psi_{xx}')_0 = (\Psi_{gg}' + \Psi_{YY}')_0 = - \frac{1}{2\cos\alpha_0} + \frac{ \Psi_{yy0}' }{\cos^2\alpha_0} , \\
	\end{aligned}
	\label{delta0_tr0}
\end{equation}
where we used $\tau' = \cos\alpha_0 - K_x/K_0$. It is convenient to express equation (\ref{beam_solution}) in terms of $K_x$ and the initial conditions $\alpha_0$, $\Psi_{yy0}'$ and $K_0$. We have

\begin{equation}
	\begin{aligned}
		& \Psi_{xx}' && = \sin^2\alpha_0 \frac{K_0^2}{K_x^2} \Psi_{yy}' - \frac{1}{2} \frac{K_0}{K_x} , \\
		& \Psi_{x y}' && = - \sin\alpha_0 \frac{K_0}{K_x} \Psi'_{yy} , \\
		& \Psi_{yy}' &&= \frac{ \Psi'_{yy0} \frac{K_x}{K_0}  }{ \cos\alpha_0 D(K_x)  } . \\
	\end{aligned}
	\label{beam_solution_2}
\end{equation}
The denominator $D$ in equation \eqref{beam_solution_2} is a function of $K_x$, and is evaluated using the second line of equation \eqref{d_denominator}.

Using the evolution of the angle $\alpha$ along the path in equation (\ref{sincos}), and $\Psi_{YY}(\tau') = \Psi_{xx} \sin^2\alpha - 2 \Psi_{x y}\sin\alpha \cos\alpha + \Psi_{yy}\cos^2\alpha$ (equation (\ref{frame_to_beam})), we arrive at the final solution for $\Psi_{YY}'$ given in equation (\ref{psiYYkx}). Equation (\ref{psiYYkx}) contains the beam-focusing phenomenon for the 2D linear-layer problem through $W_Y = \big( 2/\Im[\Psi_{YY}]\big)^\frac{1}{2}$. Expressing $\Psi_{YY}$ as a function of the central ray $K_x$ is convenient, since it readily allows us to express $W_Y$ as a function of the turbulent scattered $k_x$, as done in this manuscript. We simply substitute $k_x = - 2 K_x$ (Bragg condition).

At this point, it is worth relating the beam-tracing solution to previous work. \cite{gusakov_ppcf_2014, gusakov_pop_2017} employ particular initial conditions for the beam-tracing equations. \cite{gusakov_ppcf_2014, gusakov_pop_2017} perform their calculations in the lab frame, setting the initial conditions to 

\begin{equation}
	\begin{aligned}
		& \Psi_{yy0}' && \equiv i \gamma , \\
		& \Psi_{xx0}' && = i \gamma \tan^2\alpha_0 - \frac{1}{2\cos\alpha_0} ,  \qquad \\
		& \Psi_{xy0}' && = - {i \gamma \tan\alpha_0} , \\
	\end{aligned}
	\label{initial_conditions}
\end{equation}
where $\gamma = L/(K_0 \rho^2)$ is purely real. Equations (\ref{initial_conditions}) are very particular initial conditions since they define a purely imaginary $\Psi_{yy0}$ in the lab frame, but not in the beam frame. This means that this initial condition does not correspond to the beam waist at launch but has a specific initial value for the radius of curvature in the beam frame. Using the relation between $\Psi_{YY0}$ and $\Psi_{yy0}$ in equation \eqref{frame_to_beam}, we have $ R_{Y0}/L = - {2 \cos\alpha_0}/{\sin^2\alpha_0} $, $W_{0} = \sqrt{2}\rho \cos\alpha_0 $. Note that we can recover expressions by \cite{gusakov_pop_2017} by setting $\Psi_{yy0}' \rightarrow i \gamma$. In \citep{gusakov_ppcf_2014, gusakov_pop_2017}, equations (\ref{beam_solution}), (\ref{d_denominator}) and (\ref{delta0_tr0}) show that $\Psi_{yy}'$ can be explicitly written as 

\begin{equation}
	\begin{aligned}
		& \Psi_{yy}'(K_x)  &&= \frac{2 \gamma^2 \big( \frac{\cos^2\alpha_0 - \sin^2\alpha_0}{\cos\alpha_0} + \frac{K_0}{K_x} \sin^2\alpha_0 - \frac{K_x}{K_0} \big) }{1 + 4 \gamma^2 \big( \frac{\cos^2\alpha_0 - \sin^2\alpha_0}{\cos\alpha_0} + \frac{K_0}{K_x}\sin^2\alpha_0 - \frac{K_x}{K_0} \big)^2 } \\ 
		& && + \frac{i \gamma }{ 1 + 4 \gamma^2 \big( \frac{\cos^2\alpha_0 - \sin^2\alpha_0}{\cos\alpha_0} + \frac{K_0}{K_x} \sin^2\alpha_0 - \frac{K_x}{K_0} \big)^2 } ,
	\end{aligned}
	\label{beam_sol_explicit}
\end{equation}
The particular initial conditions from \citep{gusakov_ppcf_2014, gusakov_pop_2017} finally yield

\begin{equation}
	\begin{aligned}
	& \Psi_{YY}'(K_x) &&= \frac{ -\frac{1}{2} \sin^2\alpha_0 + i \gamma \Big[ \frac{K_x^3}{K_0^3} + 3\sin^2\alpha_0 \frac{K_x}{K_0} - \frac{\sin^2\alpha_0}{\cos\alpha_0}(\cos^2\alpha_0-\sin^2\alpha_0) \Big] }{ \Big[ \sin^2\alpha_0 + \frac{K_x^2}{K_0^2} \Big] \Big[ \frac{K_x}{K_0}  + 2 i \gamma \Big( \frac{\cos^2\alpha_0 - \sin^2\alpha_0 }{\cos\alpha_0} \frac{K_x}{K_0} + \sin^2\alpha_0 - \frac{K_x^2}{K_0^2} \Big) \Big] } .
	\end{aligned}
	\label{psiYYkx_gusakov}
\end{equation}
Using $W_Y$ computed from equation (\ref{psiYYkx_gusakov}), and $|F_{x\mu}|^2/|F_0|^2 = K_0W_0/K_\mu W_{Y\mu}$ recovers equations (14) and (15) from \citep{gusakov_pop_2017}.

\section{Details of the scattered power calculations}
\label{app_power}

In this appendix, we provide details of the scattered power calculations in the 2D linear-layer problem from section \ref{gusakov_section}. We follow a derivation analogous to the one in \citep{valerian_ppcf_2022}, but using Cartesian coordinates instead of beam-aligned coordinates. The position $(x,y)$ in Cartesian coordinates is related to the $(\tau, Y)$ coordinates by $x = x_c(\tau) + Y \sin\alpha(\tau), y = y_c(\tau) - Y\cos\alpha(\tau) $ (equation (\ref{xcyc_2d})), where $(x_c, y_c)$ are the coordinates along the central ray (figure \ref{frames}), and $\sin\alpha$ and $\cos\alpha$ are given by equation (\ref{sincos}). The density fluctuations are expressed in terms of $(x,y)$ and the conjugate turbulent wavevector is expressed in its Cartesian components $\mathbf{k}_\perp = k_x \mathbf{\hat{{x}}} + k_y \mathbf{\hat{{y}}}$, that is, we have $\delta n(x,y,t) = \int {\text{d}k_x \text{d}k_y \ \delta \hat{n}(k_x, k_y) \exp({i k_x x + i k_y y }})$ (equation \eqref{density_cartesian_2d}). Starting from equation (100) in \citep{valerian_ppcf_2022}, the scattered amplitude can be written as

\begin{equation}
	\begin{alignedat}{2}
	& {A}_r(t) = &&\frac{i \Omega A_{ant} }{2\pi c} \int{ \text{d}V \ \text{d}k_x \ \text{d}k_y} \ \Im[{\Psi}_{YY}]^{\frac{1}{2}} \Im[{\Psi}_{XX}]^{\frac{1}{2}} \frac{g_{ant}}{g} \exp(i2\phi_G) \frac{\delta \hat{n}}{n} \Big[ \bold{\hat{e}}^* \cdot ( \boldsymbol{\epsilon}_{eq} - \bold{1} ) \cdot \bold{\hat{e}} \Big]  \\ 
	& && \times \exp \Big( 2is + ik_x \big( x_c(\tau) + Y \sin\alpha \big) + i k_y \big( y_c(\tau) - Y \cos\alpha \big) + i X^2 \Psi_{XX} + i Y^2 \Psi_{YY} \Big) , \\
	\end{alignedat}
	\label{}
\end{equation}
where the differential volume element is $\text{d} V = g\text{d}\tau \ \text{d}X \ \text{d}Y$ and $\boldsymbol{\epsilon}_{eq} $ is the equilibrium part of the cold plasma dielectric tensor. The integrals in $X$ and $Y$ are integrals of a complex Gaussian, and give 

\begin{equation}
	\begin{alignedat}{2}
	& {A}_r(t) = && \frac{- \Omega A_{ant} g_{ant}}{2 c} \int {\text{d}k_x \text{d}k_y \frac{\delta \hat{n}}{n} } \int{\text{d}\tau}\bigg[ \frac{\Im[{\Psi}_{YY}]}{ \Psi_{YY} } \bigg]^{\frac{1}{2}} \Big[ \bold{\hat{e}}^* \cdot ( \boldsymbol{\epsilon}_{eq} - \bold{1} ) \cdot \bold{\hat{e}} \Big] \\
	& && \times \exp({i2\phi_G}) \exp[{if(\tau)}] , \\ 
	\end{alignedat}
	\label{as_kxkytau}
\end{equation}
where we set $\Im[\Psi_{XX}]/\Psi_{XX} = 1$ (to be able to ignore the variation along the field line direction $X$). Here, $f(\tau)$ is 

\begin{equation}
	\begin{alignedat}{2}
	& f(\tau) = 2s + k_xx_c(\tau) + k_yy_c(\tau) - (k_x\sin\alpha - k_y\cos\alpha )^2/4 \Psi_{YY} , \\ 
	\end{alignedat}
	\label{bragg_f_xy}
\end{equation}
where the largest piece in $f$ is given by 

\begin{equation}
	\begin{alignedat}{2}
	& f_0 = 2s + k_x x_c + k_y y_c , \\ 
	\end{alignedat}
	\label{f0}
\end{equation}
where $f_0 \sim L/\lambda$. The integral in $\tau$ from equation (\ref{as_kxkytau}) is the integral of an exponential with a complex argument, so it can be calculated via steepest descent \citep{bender_orzag_78}. The integral is dominated by the region around $\text{d} f / \text{d} \tau = 0$, where 

\begin{equation}
	\begin{alignedat}{2}
	& \frac{ \text{d} f }{ \text{d} \tau } = 2Kg + k_x\frac{\text{d}x_c}{\text{d}\tau} + k_y\frac{\text{d}y_c}{\text{d}\tau} - \frac{q_{1}q_{2}}{2\Psi_{YY}} \frac{\text{d}\alpha}{\text{d}\tau} , \\
	\end{alignedat}
	\label{bragg_dfdt_xy_app}
\end{equation}
where we have introduced the shorthand notation $q_1 = k_x \cos\alpha + k_y \sin\alpha$ and $q_2 = k_x \sin\alpha - k_y \cos\alpha $ in equation (\ref{bragg_dfdt_xy_app}). Note that setting $ \text{d} f / \text{d} \tau ({\tau}_\mu) = 0$ (stationary phase) will give a complex value for ${\tau}_\mu$. Moreover, equation (\ref{bragg_f_xy}) implies that $q_2 \sim 1/W$, and hence the term proportional to $q_1q_2 $ in equation (\ref{bragg_dfdt_xy_app}) is small in $\lambda/W$. This motivates splitting the root into two pieces ${\tau}_\mu + \Delta \tau$, with $\Delta \tau \sim ({\lambda}/{W}) {\tau}_\mu$ complex, and expanding the Bragg condition order by order in $\lambda/W$. We denote quantities evaluated at ${\tau}_\mu$ with the subscript $\mu$ (e.g. $K({\tau}_\mu) = {K}_\mu$). At the lowest order, we have 

\begin{equation}
	\begin{alignedat}{2}
	& \frac{ \text{d} f }{ \text{d} \tau } \Big|_{{\tau}_\mu} \approx  \frac{ \text{d} f_0 }{ \text{d} \tau }  \Big|_{{\tau}_\mu} = 2 {K}_\mu {g}_\mu + k_x\frac{\text{d}x_c}{\text{d}\tau} \Big|_{{\tau}_\mu} + k_y\frac{\text{d}y_c}{\text{d}\tau} \Big|_{{\tau}_\mu} = 0. \\
	\end{alignedat}
	\label{bragg_dfdt_xy_0}
\end{equation}
The $0^\text{th}$ order Bragg condition in equation (\ref{bragg_dfdt_xy_0}) can be explicitly written in terms of ${K}_{x\mu}$ as follows 

\begin{equation}
	\begin{alignedat}{2}
	& 2 \frac{{K}_{x\mu}^2}{K_0^2} + \frac{k_x}{K_0} \frac{{K}_{x\mu}}{K_0} + \sin\alpha_0 \bigg( \frac{k_y}{K_0} + 2 \sin\alpha_0 \bigg) = 0 , \\
	\end{alignedat}
	\label{bragg_kx2}
\end{equation}
where we used equation (\ref{raytracing_sols}) and the expression ${g}_\mu = 2 {{K}_\mu}/K_0^2$. Using the fact that $q_2 \sim 1/W$ and employing equation (\ref{sincos}) gives $k_x \approx k_y/\tan{\alpha}_\mu$ and $ k_y + 2 K_0 \sin\alpha_0 \sim {1}/{W}$. Thus, we have

\begin{equation}
	\begin{alignedat}{2}
	& k_y + 2 K_0 \sin\alpha_0 \approx 0.
	\end{alignedat}
	\label{bragg_k}
\end{equation}
Note that equation (\ref{bragg_kx2}) exhibits two solutions for ${K}_{x\mu}$. Following equation (\ref{bragg_k}), we expand the two solutions for $k_y + 2 K_0 \sin\alpha_0 \approx 0$. We find 

\begin{equation}
	\begin{alignedat}{2}
	& {K}_{x\mu} \approx 
		\begin{cases} 
		-\sin\alpha_0(k_y + 2 K_0 \sin\alpha_0)K_0/k_x  \\
		-k_x/2 + {\sin\alpha_0(k_y + 2 K_0 \sin\alpha_0)}K_0/{k_x} 
		\end{cases} , 
	\end{alignedat}
	\label{bragg_kx}
\end{equation}
where only the second solution $\sim 1/\lambda$ is physically relevant. This shows that ${K}_{x\mu} \approx - k_x/2$. 

Next, we use the fact that $q_1 \sim 1/\lambda$ and $q_2 \sim 1/W$ to find the Bragg condition to next order in $\lambda/W$,

\begin{equation}
	\begin{alignedat}{2}
	& \frac{ \text{d} f }{ \text{d} \tau } \Big|_{{\tau}_\mu + \Delta \tau} = 2K g \big|_{{\tau}_\mu + \Delta \tau} + k_x\frac{\text{d}x_c}{\text{d}\tau} \Big|_{{\tau}_\mu + \Delta \tau} + k_y\frac{\text{d}y_c}{\text{d}\tau} \Big|_{{\tau}_\mu + \Delta \tau} - \frac{q_{1}q_{2}}{2\Psi_{YY}} \frac{\text{d}\alpha}{\text{d}\tau} \bigg|_{{\tau}_\mu + \Delta \tau} = 0. \\
	\end{alignedat}
	\label{bragg_dfdt_xy_1}
\end{equation}
Equation (\ref{bragg_dfdt_xy_1}) determines the value of $\Delta \tau$ 

\begin{equation}
	\begin{alignedat}{2}
	& \Delta \tau \approx \frac{\frac{ {q}_{1\mu} {q}_{2\mu}}{2 {\Psi}_{YY\mu}} \big( \frac{\text{d}\alpha}{\text{d}\tau} \big)_{{\tau}_\mu} }{2 \frac{\text{d} {K}_\mu}{\text{d}\tau} {g}_\mu - \frac{ {q}_{1\mu}^2}{2 {\Psi}_{YY\mu}} \big( \frac{\text{d}\alpha}{\text{d}\tau} \big)^2_{{\tau}_\mu} } . \\
	\end{alignedat}
	\label{bragg_tau1}
\end{equation}
Having calculated the value of ${\tau}_\mu$ and $\Delta \tau$ that satisfy the Bragg condition order by order in equations (\ref{bragg_dfdt_xy_0}), (\ref{bragg_dfdt_xy_1}) and (\ref{bragg_tau1}), next we approximate $f(\tau) \approx f({\tau}_\mu + \Delta \tau) +  {\text{d}^2 f}/{ \text{d} \tau^2} \big|_{{\tau}_\mu + \Delta \tau} (\tau - {\tau}_\mu - \Delta \tau)^2 / 2$. We find  

\begin{equation}
	\begin{alignedat}{2}
	& f({\tau}_\mu + \Delta \tau) &&\approx f( {\tau}_\mu) + \Delta \tau \frac{\text{d} f}{ \text{d} \tau } \Big|_{{\tau}_\mu} + \frac{1}{2} \Delta \tau^2 \frac{\text{d}^2 f}{ \text{d} \tau^2 } \Big|_{ {\tau}_\mu } \approx f_0( {\tau}_\mu) - \frac{\frac{\text{d}{K}_\mu}{\text{d}\tau} {g}_\mu \frac{{q}_{2\mu}^2}{2 {\Psi}_{YY\mu}} }{ 2 \frac{\text{d} {K}_\mu}{\text{d}\tau} {g}_\mu - \frac{2 {K}_\mu^2}{{\Psi}_{YY\mu}} \big( \frac{\text{d}\alpha}{\text{d}\tau} \big)_{{\tau}_\mu}^2 } , \\
	\end{alignedat}
	\label{f_second_approx}
\end{equation}
where $f_0$ is given by equation (\ref{f0}), and the second derivative term in equation (\ref{f_second_approx}) is given by
\begin{equation}
	\begin{alignedat}{2}
	& \frac{\text{d}^2 f}{ \text{d} \tau^2 } \Big|_{ {\tau}_\mu + \Delta \tau} && \approx \frac{\text{d}^2 f}{ \text{d} \tau^2 }  \Big|_{{\tau}_\mu} = 2 \frac{\text{d} {K}_\mu}{\text{d}\tau} {g}_\mu - \frac{2 {K}_\mu^2}{{\Psi}_{YY\mu}} \Big( \frac{\text{d}\alpha}{\text{d}\tau} \Big)^2_{{\tau}_\mu} .
	\end{alignedat}
	\label{df2}
\end{equation}
We are now in a position to perform the steepest-descent integral in equation (\ref{as_kxkytau}). We find the following expression for the scattered amplitude in Cartesian coordinates

\begin{equation}
	\begin{alignedat}{2}
	& {A}_r(t) &&= A_{ant} \int \text{d}k_x \text{d}k_y \ F_{xy,\mu}(k_x, k_y) \delta \hat{n} (k_x, k_y) \exp\big[i\big(2 {s}_\mu(k_x, k_y) + k_x {x}_{c\mu} + k_y {y}_{c\mu} \big) \big], \\
	\end{alignedat}
	\label{as_kxky_dnxy}
\end{equation}
where the slowly varying function (filter) ${F}_{xy,\mu}(k_x, k_y)$ takes the form 

\begin{equation}
	\begin{alignedat}{2}
	&F_{xy,\mu}(k_x, k_y) = &&- \frac{\Omega g_{ant}}{2c} \left[ \frac{\Im[{ {\Psi}}_{YY\mu}] }{ {\Psi}_{YY\mu}} \right]^\frac{1}{2} \frac{\big[ \mathbf{\hat{e}^*} \cdot (\boldsymbol{\epsilon}_{eq} - \mathbf{1} ) \cdot \mathbf{\hat{e}} \big]_{{\tau}_\mu}}{{n}_\mu} \exp({i 2 {\phi}_{G\mu}}) \\ 
	& && \times \left( \frac{2 i\pi}{ 2 \frac{\text{d} {K}_\mu}{\text{d}\tau} {g}_\mu - \frac{2 {K}_\mu^2}{{\Psi}_{YY\mu}} \Big( \frac{\text{d}\alpha}{\text{d}\tau} \Big)^2_{{\tau}_\mu} } \right)^\frac{1}{2} \exp\left[  - i \frac{\frac{\text{d}{K}_\mu}{\text{d}\tau} {g}_\mu \frac{{q}_{2\mu}^2}{2 {\Psi}_{YY\mu}} }{ 2 \frac{\text{d} {K}_\mu}{\text{d}\tau} {g}_\mu - \frac{2 {K}_\mu^2}{{\Psi}_{YY\mu}} \big( \frac{\text{d}\alpha}{\text{d}\tau} \big)_{{\tau}_\mu}^2 } \right]  .
	\end{alignedat}
	\label{fkxky}
\end{equation}

Next, we calculate the scattered power following the same procedure as \cite{valerian_ppcf_2022}. For this calculation, we need the turbulence correlation function $C$, defined as the Fourier transform of the density fluctuation wavenumber power spectrum in Cartesian $(k_x, k_y)$ coordinates

\begin{equation}
	\begin{alignedat}{2}
	& C(\mathbf{r}, t, \Delta \mathbf{r}, \Delta t) &&= \frac{\langle \delta n (\mathbf{r} + \Delta \mathbf{r}, t + \Delta t) \delta n (\mathbf{r}, t)  \rangle_T}{ \langle \delta {n}^2 \rangle_T } \\
	& &&= \int \text{d}k_x \text{d}k_y \ \hat{C}(\mathbf{r}, t, k_x, k_y, z=0, \Delta t) \exp[{i(k_x \Delta x + k_y \Delta y )}].
	\end{alignedat}
	\label{corr_func}
\end{equation}
Here $\langle \delta {n}^2 \rangle_T$ is the r.m.s. value of the density fluctuation power. The function $\hat{C}$ is given by 

\begin{equation}
	\begin{alignedat}{2}
	& \hat{C}(\mathbf{r}, t, k_x, k_y) =&& \frac{1}{\langle \delta n^2 \rangle_T} \int{ \text{d}k_x' \text{d}k_y' \big\langle \delta \hat{n}(k_x, k_y) \delta \hat{n}^*(k_x', k_y') \big\rangle_T } \\
	& && \times \exp[i(k_x-k_x')x + i(k_y-k_y')y].
	\end{alignedat}
	\label{chat_cartesian}
\end{equation}
We order $k_x - k_x' \sim k_y - k_y' \sim 1/L$. We use equation \eqref{chat_cartesian} for $\hat{C}$ to calculate the scattered power. We start by calculating the product $\langle |A_r|^2 \rangle_T$, which is given by 

\begin{equation}
	\begin{alignedat}{2}
	& \big\langle |{A}_r(t)|^2 \big\rangle_T = &&|A_{ant}|^2 \int \text{d}k_x \text{d}k_y \ F_{xy,\mu}(k_x, k_y) \exp\big[i f_0(\tau_\mu) \big] \\
	& && \times \int \text{d}k_x' \text{d}k_y' \ F_{xy,\mu'}^*(k_x', k_y') \exp\big[-i f_0(\tau_{\mu'}) \big] \\
	& && \times \big\langle \delta \hat{n}(k_x, k_y) \delta \hat{n}^*(k_x', k_y') \big\rangle_T
	\end{alignedat}
	\label{ar2_power_cart}
\end{equation}
where $\mu'$ is the scattering position corresponding to $(k_x', k_y')$, different from $(k_x, k_y)$. Note that $F_{xy,\mu}$ only contains slow variations $\sim 1/L$. This means that we can neglect the differences $k_x - k_x' \sim k_y - k_y' \sim 1/L$ between $F_{xy,\mu}(k_x, k_y)$ and $F_{xy,\mu'}^*(k_x', k_y')$, so that $F_{xy,\mu'}^*(k_x', k_y') \approx F_{xy,\mu'}^*(k_x, k_y)$. The last piece to calculate $ \langle |{A}_r(t)|^2 \rangle_T $ is to compute the difference in the large phase $f_0(\tau_\mu) - f_0(\tau_{\mu'})$. Since $f_0 \sim L/\lambda$, it is important here keep the differences $k_x - k_x' \sim k_y - k_y' \sim 1/L$, since they will provide an order unity contribution to the phase. Straightforward calculations lead to 

\begin{equation}
	\begin{alignedat}{2}
	2{s}_\mu - 2 {s}_{\mu'} + k_x {x}_{c\mu} - k_x' {x}_{c\mu'} + k_y {y}_{c\mu} - k_y' {y}_{c\mu'} \approx (k_x - k_x') {x}_{c\mu} + (k_y - k_y') {y}_{c\mu} ,
	\end{alignedat}
	\label{f0f0_cartesian}
\end{equation}
where we used the expressions for $x_c$ and $y_c$ from equations \eqref{raytracing_sols}, the Bragg condition \eqref{bragg_kx2} and the definition of $s = \int K \text{d}l $ in Cartesian coordinates (see equation \eqref{valerian_beam}). Using equation \eqref{f0f0_cartesian}, $\mu=\mu'$ and $F_{xy,\mu'}^*(k_x', k_y') \approx F_{xy,\mu'}^*(k_x, k_y)$, we can relate the expression for $\langle |{A}_r(t)|^2 \rangle_T$ to the correlation function in equation \eqref{chat_cartesian}. Defining $p_r = \langle |{A}_r(t)|^2 \rangle_T$, we find 

\begin{equation}
	\begin{alignedat}{2}
	& \frac{p_r}{ P_{ant} } = && \int \text{d}k_x \text{d}k_y \  | F_{xy,\mu} |^2 \langle \delta n^2 \rangle_{T} \hat{C}( {\mathbf{r}}_{c\mu}, t, k_x, k_y) , \\
	\end{alignedat}
	\label{prkxky_2d}
\end{equation}
where ${\mathbf{r}}_{c\mu} = ({x}_{c\mu}, {y}_{c\mu}, z=0)$. The correlation function is related to the 2D wavenumber power spectrum via $\langle |\delta \hat{n} (k_x, k_y)|^2 \rangle_T = \langle \delta {n}^2 \rangle_T \hat{C} (\mathbf{r}, t, k_x, k_y, z=0, \Delta t=0)$. The Cartesian filter function $|F_{xy,\mu}|^2$ in equation \eqref{prkxky_2d} can be calculated directly using equation \eqref{fkxky}, and is given by

\begin{equation}
	\begin{alignedat}{3}
	& |F_{xy,\mu}|^2 && = && 2\pi \Big( \frac{\Omega g_{ant}}{2c} \Big)^2 \frac{ \big| \mathbf{\hat{e}^*} \cdot (\boldsymbol{\epsilon}_{eq} - \mathbf{1} ) \cdot \mathbf{\hat{e}} \big|^2_{{\tau}_\mu} }{ {n}_\mu^2} \\
	& && && \times \frac{ \Im[{{\Psi}}_{YY}]_\mu }{| {\Psi}_{YY}|_\mu} \frac{ \exp \Bigg[ - \frac{ \big(2 \frac{\text{d}{K}_\mu}{\text{d}\tau} \big)^2 {g}_\mu^2 }{ \big| 2 \frac{\text{d} {K}_\mu}{\text{d}\tau} {g}_\mu - \frac{2 {K}_\mu^2}{{\Psi}_{YY\mu}} \big( \frac{\text{d}\alpha}{\text{d}\tau} \big )^2_{{\tau}_\mu} \big|^2 } \frac{2 {q}_{2\mu}^2}{\Delta {k}_{\mu2}^2} \Bigg] }{  \big| 2 \frac{\text{d} {K}_\mu}{\text{d}\tau} {g}_\mu - \frac{2 {K}_\mu^2}{ {\Psi}_{YY\mu} } \big( \frac{\text{d}\alpha}{\text{d}\tau} \big)^2_{{\tau}_\mu} \big| } ,
	\end{alignedat}
	\label{fkxky_2d}
\end{equation}
where ${q}_{2\mu} = {q}_{2\mu}(k_x, k_y) = k_x \sin {\alpha}_\mu - k_y\cos {\alpha}_\mu$, and we have introduced the wavenumber resolution $\Delta {k}_{\mu2} = 2 \big( {-1}/{\Im[{1}/{{\Psi}_{YY\mu} } ]} \big)^\frac{1}{2} = 2 {| {\Psi}_{YY\mu}|}/{ \Im[{\Psi}_{YY\mu} ]^\frac{1}{2} } \sim 1/W$ as defined in \citep{valerian_ppcf_2022}. 

Equation (\ref{prkxky_2d}) for $|F_{xy,\mu}|^2$ is equation (\ref{fkxky_2_cartesian_}) in the main text. We note that equation (\ref{prkxky_2d}) does not exactly recover the expressions in \cite{gusakov_ppcf_2014, gusakov_pop_2017}. Equation (\ref{prkxky_2d}) can be further simplified by using the Bragg condition and the analytic beam-tracing solution, which we describe next. We will recover equations (\ref{prkxky2_cart}) in the main text. This result is valid for general beam initial conditions. By using Gusakov's initial conditions in equation (\ref{initial_conditions}), we will recover the expressions for the scattered power (auto-correlation) derived by \cite{gusakov_ppcf_2014, gusakov_pop_2017}. 

We want to express $|F_{xy,\mu}|^2$ as a function of its variables $k_x$ and $k_y$. The function $|F_{xy,\mu}(k_x, k_y)|^2$ in equation (\ref{prkxky_2d}) is a function of $(k_x, k_y)$ through the Bragg condition relating ${\tau}_{\mu}$ to $k_x$ and $k_y$ (equation (\ref{bragg_dfdt_xy_0})). We proceed to calculate the different terms in equation (\ref{prkxky_2d}). Using the Bragg condition, the ray-tracing solution for the central ray, the analytic expression for $\Psi_{YY}$ (equation (\ref{psiYYkx})) and $k_x \approx -2 K_{x\mu}$, we proceed to simplify the terms outside of the exponential in equation (\ref{prkxky_2d}). Using the formulas

\begin{equation}
	\begin{alignedat}{2}	
	& \frac{ \text{d}K_\mu}{ \text{d} \tau} g_\mu &&= - \frac{2}{K_0L} \frac{K_{x\mu}}{K_0} , \\
	\end{alignedat}
	\label{dkdtau}
\end{equation}
and
\begin{equation}
	\begin{alignedat}{2}	
	& \bigg( \frac{ \text{d}\alpha}{ \text{d}\tau} \bigg)_\mu &&= \frac{\sin\alpha_0}{K_0L} \frac{K_0^2 }{K_\mu^2} , \\
	\end{alignedat}
	\label{dalphadtau}
\end{equation}
we find
\begin{equation}
	\begin{alignedat}{2}	
	& 2 \frac{\text{d} {K}_\mu}{\text{d}\tau} {g}_\mu - 2 \frac{ {K}_\mu^2}{ {\Psi}_{YY\mu} } \bigg( \frac{\text{d}\alpha}{\text{d}\tau}\bigg)_{{\tau}_\mu}^2 = \\
	& - \frac{4 \Psi_{yy0}'}{K_0L} \frac{ {K}_\mu^4 / K_0^4 }{ - \frac{1}{2} \sin^2\alpha_0 + \Psi_{yy0}' \Big( \frac{ {K}_x^3}{K_0^3} + 3\sin^2\alpha_0 \frac{ {K}_x}{K_0} - \frac{\sin^2\alpha_0}{\cos\alpha_0} (\cos^2\alpha_0 - \sin^2\alpha_0) \Big) } . \\
	\end{alignedat}
	\label{denominator_app}
\end{equation}
We then obtain 
\begin{equation}
	\begin{alignedat}{2}	
	& \frac{\Im[ {\Psi}_{YY} ]_\mu }{| {\Psi}_{YY}|_\mu \bigg| 2 \frac{ \text{d} {K}_\mu}{\text{d}\tau} {g}_\mu - 2 \frac{ {K}_\mu^2}{ {\Psi}_{YY\mu} } \Big( \frac{\text{d}\alpha}{\text{d}\tau}\Big)_{{\tau}_\mu}^2 \bigg|} =  \frac{K_0^2 L }{ 4 {K}_\mu }\frac{ \sqrt{ 2 \Im[\Psi_{yy0}]} }{ \big| \Psi_{yy0} \big| {W}_{Y\mu} } , \\
	\end{alignedat}
	\label{fxymu_nonexp}
\end{equation}
where we have used ${W}_{Y\mu} = ({2/ \Im[ {\Psi}_{YY\mu}]})^\frac{1}{2}$ and
\begin{equation}
	\begin{alignedat}{2}	
	& \Im[ {\Psi}_{YY\mu}] = \frac{ \Im[\Psi_{yy0}] \frac{ {K}_\mu^2}{K_0^2}  }{ \Big| \frac{ {K}_{x\mu}}{K_0} + 2 \Psi_{yy0}'  \Big( \frac{\cos^2\alpha_0 - \sin^2\alpha_0}{\cos\alpha_0} \frac{ {K}_{x\mu}}{K_0} + \sin^2\alpha_0 - \frac{ {K}_{x\mu}^2}{K_0^2} \Big)  \Big|^2 } . \\
\end{alignedat}
	\label{impsiYY}
\end{equation}

With respect to the terms inside the exponential in equation (\ref{prkxky_2d}), the $k_2$-wavenumber resolution $\Delta {k}_{\mu2}$ takes the following form 

\begin{equation}
	\begin{alignedat}{2}	
	& \Delta {k}_{\mu2}^2 = \frac{ 4 | {\Psi}_{YY\mu}|^2 }{ \Im[ {\Psi}_{YY\mu}] } = \\
	& \frac{ 4 }{ \Im[ \Psi_{yy0}' ] } \frac{ \Big| - \frac{1}{2} \sin^2\alpha_0 + \Psi_{yy0}' \Big(  \frac{ {K}_x^3}{K_0^3} + 3\sin^2\alpha_0 \frac{ {K}_x}{K_0} - \frac{\sin^2\alpha_0}{\cos\alpha_0} (\cos^2\alpha_0 - \sin^2\alpha_0) \Big) \Big|^2 }{ {K}_\mu^6 / K_0^6 } \frac{K_0}{L} . \\
	\end{alignedat}
	\label{dk2_expression}
\end{equation} 
Using equation (\ref{dk2_expression}) for $\Delta {k}_{\mu2}$ and employing expressions \eqref{dkdtau} and \eqref{denominator_app}, the argument of the exponential in $|F_{xy,\mu}|^2$ (equation (\ref{prkxky_2d})) can be written as $\exp[-h(k_x, k_y)]$, where

\begin{equation}
	\begin{alignedat}{2}	
	& h(k_x, k_y) = \frac{  \Big( 2 \frac{ \text{d} {K}_\mu}{ \text{d} \tau} {g}_\mu \Big)^2}{ \bigg| 2 \frac{ \text{d} {K}_\mu}{ \text{d} \tau} {g}_\mu - 2 \frac{ {K}_\mu^2}{ {\Psi}_{YY\mu} } \Big( \frac{ \text{d} \alpha}{ \text{d} \tau}\Big)_{{\tau}_\mu}^2 \bigg|^2} \frac{2 {k}_{2\mu}^2}{\Delta {k}_{\mu2}^2} = K_0 L \frac{\Im[\Psi_{yy0}']}{2 |\Psi_{yy0}'|^2} \frac{ {K}_{x\mu}^4}{ {K}_\mu^4} \bigg( \frac{k_y}{K_0} - \frac{k_x}{K_0} \frac{\sin {\alpha}_\mu}{\cos {\alpha}_\mu}  \bigg)^2 . \\
	\end{alignedat}
	\label{hxy_exp}
\end{equation}
We express $h$ in normalised quantities to emphasise that $h$ is large, $h \sim K_0L \gg 1$, unless $k_y \approx k_x \tan{\alpha}_\mu$. The $k_y$ dependence of the integrand in equation (\ref{prkxky_2d}) is through the exponential term (via $h$) and through the turbulence spectrum $\langle |\delta \hat{n} (k_x, k_y)|^2 \rangle_T$. Given the assumption of large $h$, we can further simplify the expression for $|F_{xy,\mu}|^2$ by performing the $k_y$ integral using Laplace's method \citep{bender_orzag_78}. This leads to equation (\ref{prkxky2_cart}), which can be proven as follows. The Gaussian integral in $k_y$ is dominated by contributions where $\partial h / \partial k_y \approx 0$, which takes place for a given $k_{y*}$. This suggests expanding $h$ around $k_{y*}$, giving $h(k_x, k_y) \approx h(k_x, k_{y*}) + ({1}/{2}) (k_y - k_{y*})^2 { \partial^2 h }/{ \partial k_y^2 }\big|_{k_{y*}}$. To calculate $\partial h / \partial k_y$ and $\partial ^2 h / \partial k_y^2$, we first use equation (\ref{sincos}) to express $\sin{\alpha}_\mu / \cos{\alpha}_\mu$ as a function of ${K}_{x\mu}$, giving $\sin{\alpha}_\mu/\cos{\alpha}_\mu = K_0 \sin\alpha_0/{K}_{x\mu}$. Then, we employ the expression ${K}_{x\mu} \approx -k_x/2 + \sin\alpha_0(k_y + 2 K_0 \sin\alpha_0)K_0/k_x$ in equation (\ref{bragg_kx}), giving ${\partial {K}_{x\mu}}/{\partial k_y} \approx \sin\alpha_0 K_0/k_x$. Using this, we find

\begin{equation}
	\begin{alignedat}{2}
	& \frac{\partial h}{ \partial k_y } \bigg|_{k_x, k_y} \approx L \frac{\Im[\Psi_{yy0}']}{ \big| \Psi_{yy0}' \big|^2} \frac{ {K}_{x\mu}^4}{ {K}_\mu^4} \bigg( \frac{k_y}{K_0} - k_x \frac{\sin\alpha_0}{ {K}_{x\mu}} \bigg) \bigg( 1 + \frac{K_0^2 \sin^2\alpha_0}{ {K}_{x\mu}^2 } \bigg) , \\
	\end{alignedat}
	\label{dhdky}
\end{equation}
where we have neglected terms small in $(k_y - k_x \tan{\alpha}_\mu)$. The dominant contribution to the integral comes from $k_{y}$ satisfying the extrema condition $\partial h / \partial k_y = 0$, which gives
\begin{equation}
	\begin{alignedat}{2}
	& k_{y*} = k_x \sin\alpha_0 \frac{K_0}{ {K}_{x\mu*}} \approx - 2 K_0 \sin\alpha_0 , \\
	\end{alignedat}
	\label{kystar}
\end{equation}
where we used once more equation (\ref{bragg_kx}). Using equation (\ref{dhdky}), we calculate $\partial ^2 h / \partial k_y^2$ and evaluate it at $k_{y*}$, giving

\begin{equation}
	\begin{alignedat}{2}
	& \frac{\partial ^2 h} {\partial k_y^2} \bigg|_{k_x, k_y*} = \frac{L}{K_0} \frac{\Im[\Psi_{yy0}']}{ \big| \Psi_{yy0}' \big|^2} \frac{ {K}_{x\mu*}^4}{ {K}_{\mu*}^4} \bigg( 1 + \frac{K_0^2 \sin^2\alpha_0}{ {K}_{x\mu*}^2 } \bigg)^2 \approx \frac{L}{K_0} \frac{\Im[\Psi_{yy0}']}{ \big| \Psi_{yy0}' \big|^2} = \frac{\Im[\Psi_{yy0}]}{ \big| \Psi_{yy0} \big|^2} . \\
	\end{alignedat}
	\label{d2hdky2}
\end{equation}
Putting together equations (\ref{hxy_exp}) and (\ref{d2hdky2}), we find that the exponential term in equation (\ref{prkxky_2d}), $\exp[-h(k_x, k_y)]$, is given approximately by 

\begin{equation}
	\begin{alignedat}{2}
	& \exp[-h(k_y)] \approx \exp \Bigg[ - \frac{\Im[\Psi_{yy0}]}{2 \big| \Psi_{yy0} \big|^2} (k_y + 2 K_0 \sin\alpha_0)^2 \Bigg] = \exp\left[ \frac{- 2 (k_y + 2K_0\sin\alpha_0)^2}{\Delta k_y^2 } \right] . 
	\end{alignedat}
	\label{exphxy_approx}
\end{equation}

Equation (\ref{exphxy_approx}) is the Gaussian exponential term in $k_y$ in equations (\ref{prkxky2_cart})-(\ref{ky_resolution}). In this expression, we recover the wavenumber resolution of the DBS diagnostic $\Delta k_y$ in Cartesian coordinates. Note the difference between $\Delta k_y$ in the lab frame, which is constant along the path and only depends on initial conditions, and $\Delta k_{\mu2}$, which is the wavenumber resolution in the beam frame given in \citep{valerian_ppcf_2022}, which depends strongly along the path (see equation \ref{dk2_expression}). For the resolution in $k_y$, we find $\Delta k_y^2 = 4 | \Psi_{yy0}|^2 / \Im[\Psi_{yy0}] $ (equation (\ref{ky_resolution})), which simplifies to the particular case of $\Delta k_y = 2/\rho$ for the initial conditions chosen by \cite{gusakov_ppcf_2014,gusakov_pop_2017}. Note that the value $2 / \rho$ is simply a lower limit to the diagnostic wavenumber resolution. Putting equation (\ref{exphxy_approx}) together with equation (\ref{fxymu_nonexp}), we recover the full expression for $|F_{xy,\mu}|^2$ in equation (\ref{prkxky2_cart}), including the $1/K_\mu W_{Y\mu}$ dependence outside of the exponential. Performing the integral over $k_y$, we recover equation (\ref{fkxky_2_cartesian_}) for $|F_{x\mu}|^2(k_x) = \int{\text{d}k_y} |F_{xy,\mu}|^2(k_x, k_y)$.

\section{Mapping the density between beam-aligned and Cartesian coordinates}
\label{mapping_density}  

In this appendix, we establish the relation between the beam-aligned representation of the density, as used in \citep{valerian_ppcf_2022}, and the Cartesian representation of the density that is employed by \cite{gusakov_ppcf_2014, gusakov_pop_2017}. We note that the two representations are independent of beam-tracing or scattering physics, and are simply a manifestation of two different frames of reference in which to express the density fluctuations. The relation between the density expressed in both frames is used in appendix \ref{app_amplitude_2d} to state the equivalence between the Doppler backscattered amplitude in \citep{valerian_ppcf_2022} and in \citep{gusakov_ppcf_2014, gusakov_pop_2017}. 

We start from the definition of the beam-aligned representation of the density, $\delta \hat{n}_{b}$, as given by equation (102) in \citep{valerian_ppcf_2022}. In the 2D linear layer, it takes the following form 

\begin{equation}
	\begin{alignedat}{2}
	& \delta n(\mathbf{r}) && = \int{ \text{d}k_1 \text{d}k_2 \ \delta \hat{n}_{b}(k_1, k_2) \exp\big[{i(k_1 l + k_2 Y)} \big] } . \\
	\end{alignedat}
	\label{dn_beam_app}
\end{equation}
Here $l$ and $Y$ are the two spatial dimensions that describe the small-scale fluctuations perpendicular to the background magnetic field and are aligned with the beam, $k_1$ and $k_2$ are the corresponding Fourier-conjugate wavenumber components, and we assume no variation in the direction parallel to the magnetic field ($u_{||}$ in \citep{valerian_ppcf_2022}). The coordinate $l$ is the arc length of the central ray, given by equation \eqref{raytracing_sols}. In order to improve clarity and to provide physical intuition, in this appendix we use $l$ as the parameter that the different functions depend on along the path, such as $x_c(l), \alpha(l)$, etc. instead of $\tau$ as we have used in the main body of this manuscript. 

We now turn to the Cartesian density fluctuation field $\delta n$ given by equation \eqref{density_cartesian_2d}, with $\mathbf{k}_\perp = k_x \hat{\mathbf{x}} + k_y \hat{\mathbf{y}}$. The coordinates $(x,y)$ are related to $(l,Y)$ in equation \eqref{dn_beam_app} by $ x = x_c(l) + Y \sin\alpha(l)$, $ y = y_c(l) - Y \cos\alpha(l)$ (equations \eqref{xcyc_2d}). This gives the succinct form $(x,y) = \mathbf{q} + Y \hat{\mathbf{Y}}$, where $\mathbf{q} = (x_c, y_c)$ and $k_xx + k_yy = \mathbf{k}_\perp \cdot \left( \mathbf{q} + Y \hat{\mathbf{Y}} \right)$. 

In order to relate the density fluctuation spectra between beam-aligned and Cartesian coordinates in equations \eqref{dn_beam_app} and \eqref{density_cartesian_2d}, we project one onto the other. We have

\begin{equation}
	\begin{alignedat}{2}
	& \delta \hat{n}_{b} (k_1,k_2) &&= \int \frac{ \text{d}Y \text{d}l }{(2\pi)^2} \exp{(- i k_1 l - i k_2 Y)} \ \delta {n}( \mathbf{r}) \\
	& &&= \int \frac{ \text{d}k_x \text{d}k_y }{ (2\pi)^2 } \delta \hat{n}(k_x, k_y) \int \text{d}Y \text{d}l \exp \left[ i f_{\mathbf{k}_\perp}(l,Y) \right],
	\end{alignedat}
	\label{dnb_map_dncart}
\end{equation}
where the function $f_{\mathbf{k}_\perp}(l,Y)$ is given by 
\begin{equation}
	\begin{alignedat}{2}
	f_{\mathbf{k}_\perp}(l,Y) =  \mathbf{k}_\perp \cdot \left( \mathbf{q} + Y \hat{\mathbf{Y}} \right) - k_1 l - k_2 Y .
	\end{alignedat}
	\label{fly_amp_map}
\end{equation}
In order to find a useful relation between $\delta \hat{n}_b$ and $\delta \hat{n}$, we proceed to compute the integrals in $l$ and $Y$ in equation \eqref{dnb_map_dncart}. We do so by calculating the point of stationary phase in the 2D $(l,Y)$ plane. We note that the unit vectors $\hat{\mathbf{g}}$ and $\hat{\mathbf{Y}}$ both depend explicitly on $l$, which needs to be taken into account when computing derivatives with respect to $l$. We will find useful the following relations between the unit vectors:
\begin{equation}
	\begin{alignedat}{2}
	& \frac{\text{d}\hat{\mathbf{g}}}{\text{d}l} = - \hat{\mathbf{Y}} \frac{\text{d}\alpha}{\text{d}l} , \\
	& \frac{\text{d}\hat{\mathbf{Y}}}{\text{d}l} = \hat{\mathbf{g}} \frac{\text{d}\alpha}{\text{d}l} . \\
	\end{alignedat}
	\label{unitvec_deriv_l}
\end{equation}
The point of stationary phase $(l_b, Y_b)$ is given by the joint conditions $\partial f_{\mathbf{k}_\perp} /\partial l |_{l_b,Y_b} = \partial f_{\mathbf{k}_\perp} /\partial Y |_{l_b,Y_b} = 0 $. Using equations \eqref{unitvec_deriv_l}, the derivatives of $f_{\mathbf{k}_\perp}$ with respect to $l$ and $Y$ are given by

\begin{equation}
	\begin{alignedat}{2}
	& \frac{\partial f_{\mathbf{k}_\perp} }{\partial l} = \mathbf{k}_\perp \cdot \hat{\mathbf{g}} \left( 1 + Y \frac{\text{d}\alpha}{\text{d}l} \right) - k_1, \\	
	& \frac{\partial f_{\mathbf{k}_\perp} }{\partial Y} = \mathbf{k}_\perp \cdot \hat{\mathbf{Y}} - k_2 ,
 	\end{alignedat}
	\label{dfldfy}
\end{equation}
where we used the definition of $\mathbf{g} = \text{d}\mathbf{q}/\text{d}\tau$ and $\text{d}l = g \text{d}\tau$. The point of stationary phase $(l_b, Y_b)$ is determined by

\begin{equation}
	\begin{alignedat}{2}
	& k_1 = \mathbf{k}_\perp \cdot \hat{\mathbf{g}}_b \left( 1 + Y_b \frac{\text{d}\alpha}{\text{d}l}\Big|_b \right), \\	
	& k_2 = \mathbf{k}_\perp \cdot \hat{\mathbf{Y}}_b ,
 	\end{alignedat}
	\label{kperp_k1k2}
\end{equation}
where the subscript $(.)_b$ means that functions are evaluated at $(l_b, Y_b)$. It is worth commenting on the implications from equation \eqref{kperp_k1k2}. Given the explicit dependence of $\hat{\mathbf{g}}$ and $\hat{\mathbf{Y}}$ on $l$, for a given $\mathbf{k}_\perp$, the second equation in \eqref{kperp_k1k2} states that the value of $k_2$ determines the location $l_b$ where equation \eqref{kperp_k1k2} is satisfied. Equivalently, given $\mathbf{k}_\perp$, $k_2$ is determined uniquely by the location along the path $l_b$, independently of the coordinate $Y_b$. Having determined $l_b$, the value of $k_1$ then determines the location $Y_b$ where $(k_1, k_2) $ and $\mathbf{k}_\perp$ are resonant. Or yet equivalently, given $\mathbf{k}_\perp$ and $l_b$, $k_1$ is determined by the location perpendicular to the central ray $Y_b$. 



Having calculated the location $(l_b,Y_b)$ and the relation between ${\mathbf{k}}_\perp$ and $(k_1, k_2)$, next we want to approximate $f_{\mathbf{k}_\perp}$ quadratically in $(l-l_b)$ and $(Y-Y_b)$. For that, we turn to compute the second derivatives of $f_{\mathbf{k}_\perp}$ with respect to $l$ and $Y$, which are given by

\begin{equation}
	\begin{alignedat}{2}
	& \frac{\partial^2 f_{\mathbf{k}_\perp} }{\partial l^2} = - \mathbf{k}_\perp \cdot \hat{\mathbf{Y}} \frac{\text{d}\alpha}{\text{d}l} + Y \left( - \mathbf{k}_\perp \cdot \hat{\mathbf{Y}} \left( \frac{\text{d}\alpha}{\text{d}l} \right)^2 + \mathbf{k}_\perp \cdot \hat{\mathbf{g}} \frac{\text{d}^2\alpha}{\text{d}l^2} \right), \\	
	& \frac{\partial^2 f_{\mathbf{k}_\perp} }{\partial l \partial Y} = \mathbf{k}_\perp \cdot \hat{\mathbf{g}} \frac{\text{d}\alpha}{\text{d}l} , \\
	& \frac{\partial^2 f_{\mathbf{k}_\perp} }{\partial Y^2} = 0 . \\
 	\end{alignedat}
	\label{df2ldfy2}
\end{equation}
Using equations \eqref{df2ldfy2}, we can write  
\begin{equation}
	\begin{alignedat}{2}
	f_{\mathbf{k}_\perp}(l,Y) \approx f_{\mathbf{k}_\perp}({l_b,Y_b}) + \frac{(l-l_b)^2}{2} \frac{\partial^2 f_{\mathbf{k}_\perp} }{\partial l^2}\bigg|_{l_b,Y_b} + {(l-l_b)(Y-Y_b)} \frac{\partial^2 f_{\mathbf{k}_\perp} }{\partial l \partial Y}\bigg|_{l_b,Y_b} .
	\end{alignedat}
	\label{fly_quad}
\end{equation}
Computing the Gaussian integrals in $l$ and $Y$ using equation \eqref{fly_quad}, we find the desired relation between $\delta \hat{n}(\mathbf{k}_\perp)$ and $\delta \hat{n}_b(k_1,k_2)$

\begin{equation}
	\begin{alignedat}{2}
	& \delta \hat{n}_{b} (k_1,k_2) &&= \int \frac{\text{d}k_x \text{d}k_y}{2\pi} \frac{\delta \hat{n}({k}_x, k_y)}{| \mathbf{k}_\perp \cdot \hat{\mathbf{g}}_b \ \text{d}\alpha/\text{d}l_b |} \exp\left[ i \mathbf{k}_\perp \cdot \mathbf{q}_b - i k_1 l_b \right] . \\
	\end{alignedat}
	\label{dnb_map_dncart}
\end{equation}

This is the general relation between $\delta \hat{n}_b$ and $\delta \hat{n}$ in two dimensions. In the next section, we use equation \eqref{dnb_map_dncart} to prove the equivalence between the backscattering amplitude in two dimensions using a beam-aligned $\delta \hat{n}_b$, as in \citep{valerian_ppcf_2022}, and using a Cartesian $\delta \hat{n}$, as in \citep{gusakov_ppcf_2014, gusakov_pop_2017}.

\section{Relation of the scattered amplitude $A_r$ between beam-aligned and Cartesian coordinates}
\label{app_amplitude_2d}

In this appendix, we show the equivalence between the scattered amplitude $A_r$ in beam-aligned coordinates, as in \citep{valerian_ppcf_2022}, and Cartesian coordinates, as in equation \eqref{as_kxky_dnxy_cart} and in \citep{gusakov_ppcf_2014, gusakov_pop_2017}. This shows that both the beam-tracing model of \cite{valerian_ppcf_2022} and the 2D DBS model of \cite{gusakov_ppcf_2014, gusakov_pop_2017} are equivalent to each other: they are simply expressed in different frames of reference. A similar exercise for directly mapping the DBS power between both representations requires further work, which will be the object of a future publication. 

We start from the expression of the backscattered amplitude in \citep{valerian_ppcf_2022}, equation (169), which in the 2D linear layer takes the following form

\begin{equation}
	\begin{alignedat}{2}
	& {A}_r(t) && = A_{ant} \sum_\nu \int{\text{d}k_1 \text{d}k_2 \ F_\nu(k_1, k_2) \ \delta \hat{n}_b(k_1, k_2, {u}_{||\nu}, t) \ \exp\big[i (2 {s}_\nu + k_1 {l}_\nu) \big] }, \\
	\end{alignedat}
	\label{valerian_aomega}
\end{equation}
where $F_{\nu}$ is the effective filter function in beam-aligned coordinates, and is given by 

\begin{equation}
	\begin{alignedat}{2}
	& F_\nu(k_1, k_2) = &&- \left[ \frac{ i\pi }{ \frac{\text{d}K}{ \text{d} \tau} g } \frac{ \Im[{\Psi}_{YY}] }{ \Psi_{YY} } \right]^{\frac{1}{2}}_\nu \frac{ \Big[ \bold{\hat{e}}^* \cdot ( \boldsymbol{\epsilon}_{eq} - \bold{1} ) \cdot \bold{\hat{e}} \Big]_\nu }{n_\nu} \exp \bigg( {i2\phi_{G\nu} } - \frac{i}{4} \frac{k_2^2}{\Psi_{YY\nu}} \bigg) .
	\end{alignedat}
	\label{valerian_f12mu}
\end{equation}
The notation $[{.}]_\nu$ means that functions of $\tau$ have been evaluated at ${\tau}_\nu$, the location along the central-ray path where the Bragg condition for backscattering is satisfied. In beam-aligned coordinates, the Bragg condition is 
\begin{equation}
	k_1 \approx - 2 K({\tau}_\nu).
	\label{bragg_k1_0}
\end{equation}
Note that only the lowest order contribution $\sim 1/\lambda$ to the Bragg condition has been kept here. For a given $k_1$, there can exist several ${\tau}_\nu$ associated to it. This explains the sum $\sum_\nu$ in equation (\ref{valerian_aomega}). The scattering amplitude in equation \eqref{valerian_aomega} only has contributions from backscattering events. Any forward scattering component is absent because of the Bragg condition. 

We note that ${\tau}_\nu$ differs from the $\tau_\mu$ calculated using Cartesian coordinates throughout the manuscript by a correction of order $\sim W/\lambda$, which we calculate later in this appendix. This difference between $\tau_\mu$ and $\tau_\nu$ is of limited importance in the main text of this manuscript, but it is important to consider in order to show the equivalence of the scattered amplitude in both representations. 

In order to show that the formulas for the scattered amplitude in Cartesian and beam-aligned coordinates are equivalent, we make use of equation \eqref{dnb_map_dncart} and insert it into equation \eqref{valerian_aomega} for the amplitude. By exchanging the order of integration, we have 

\begin{equation}
	\begin{alignedat}{2}
	& {A}_r(t) =&& -A_{ant} \sum_\nu \int \frac{ \text{d}k_x \text{d}k_y }{2\pi} {\delta \hat{n}({k}_x, k_y)} \int \frac{\text{d}k_1 \text{d}k_2}{| \mathbf{k}_\perp \cdot \hat{\mathbf{g}}_b \ \text{d}\alpha/\text{d}l_b |} \left[ \frac{ i\pi }{ \frac{\text{d}K}{ \text{d} \tau} g } \frac{ \Im[{\Psi}_{YY}] }{ \Psi_{YY} } \right]^{\frac{1}{2}}_\nu \\
	& && \times \frac{ \Big[ \bold{\hat{e}}^* \cdot ( \boldsymbol{\epsilon}_{eq} - \bold{1} ) \cdot \bold{\hat{e}} \Big]_\nu }{n_\nu} \exp\left( i 2\phi_{G\nu} \right) \exp\big[i f \big], \\
	\end{alignedat}
	\label{valerian_aomega_dnxy}
\end{equation}
where the phase $f$ is given by 
\begin{equation}
	f = 2 s_\nu + \mathbf{k}_\perp \cdot \mathbf{q}_b + k_1 (l_\nu - l_b) - \frac{k_2^2}{4 \Psi_{YY\nu}}.
	\label{phase_amp_map}
\end{equation}
We recall from appendix \ref{mapping_density} that the subscript $(.)_b$ corresponds to the location $(l_b, Y_b)$ where the resonance condition in equation \eqref{kperp_k1k2} takes place. We proceed to calculate the integrals in $k_1$ and $k_2$ by the method of stationary phase \citep{bender_orzag_78}. 

The phase $f$ in equation \eqref{phase_amp_map} is made of a large piece, $2 s_\nu + \mathbf{k}_\perp \cdot \mathbf{q}_b + k_1 (l_\nu - l_b) \sim L/\lambda$, and a small piece $- {k_2^2}/{4 \Psi_{YY\nu}} \sim 1$. The large piece forces $l_\nu - l_b \ll L$. One can see this by using the stationary-phase method. Indeed, $\partial f/\partial k_1 =0$ and  $\partial f/\partial k_2 =0$ give $l_b = l_\nu$ to lowest order.
 
We proceed by assuming $l_\nu - l_b \sim W$. We can then find the distance between a nearby point $l_\nu$ to $l_b$. We can expand $\hat{\mathbf{Y}}_\nu - \hat{\mathbf{Y}}_b $ as $\hat{\mathbf{Y}}_\nu - \hat{\mathbf{Y}}_b \approx (l_\nu-l_b) {\text{d} \hat{\mathbf{Y}}}/{\text{d}l}|_b $. Multiplying by $\mathbf{k}_\perp$ and using equations \eqref{unitvec_deriv_l}, \eqref{kperp_k1k2}, and ${\text{d} \hat{\mathbf{Y}}}/{\text{d}l}|_b \approx {\text{d} \hat{\mathbf{Y}}}/{\text{d}l}|_\nu$, we find 

\begin{equation}
	\begin{alignedat}{2}
	& \mathbf{k}_\perp \cdot \hat{\mathbf{Y}}_\nu - k_2 = \mathbf{k}_\perp \cdot \hat{\mathbf{g}}_\nu \frac{\text{d}\alpha}{\text{d}l_\nu} (l_\nu-l_b) + O\left( \frac{1}{L} \right),
 	\end{alignedat}
	\label{eyeyb_exp}
\end{equation}
which results in $l_\nu - l_b \sim W$ as we predicted above. Equation \eqref{eyeyb_exp} also shows that, for a given $\mathbf{k}_\perp$, we can change $k_2$ by moving in $l_b$. A similar exercise can be done for $\hat{\mathbf{g}}_\nu$, which gives $\mathbf{k}_\perp \cdot \hat{\mathbf{g}}_\nu - \mathbf{k}_\perp \cdot \hat{\mathbf{g}}_b \approx - \mathbf{k}_\perp \cdot \hat{\mathbf{Y}}_b ({\text{d}\alpha}/{\text{d}l})|_b (l_\nu - l_b) - \mathbf{k}_\perp \cdot \hat{\mathbf{g}}_b ({\text{d}\alpha}/{\text{d}l})^2|_b (l_\nu - l_b)^2/2 \sim 1/L$. In this case, we can use $\mathbf{k}_\perp \cdot \hat{\mathbf{g}}_\nu \approx \mathbf{k}_\perp \cdot \hat{\mathbf{g}}_b$. 

We now Taylor-expand $\mathbf{q}_b$ in equation \eqref{phase_amp_map} in powers of $l_b - l_\nu$. Using $\text{d}\mathbf{q}/\text{d}l = \hat{\mathbf{g}}$ and equations \eqref{unitvec_deriv_l}, we have 

\begin{equation}
	\begin{alignedat}{2}
	& \mathbf{q}_b \approx \mathbf{q}_\nu + \hat{\mathbf{g}}_\nu (l_b - l_\nu) - \hat{\mathbf{Y}}_\nu \frac{\text{d}\alpha}{\text{d}l} \Big|_\nu \frac{(l_b - l_\nu)^2}{2}.
 	\end{alignedat}
	\label{qb_qnu_exp}
\end{equation}
Next, we use equations \eqref{eyeyb_exp} and \eqref{qb_qnu_exp} to find a suitable expression of the phase in equation \eqref{phase_amp_map} in terms of powers of $l_b - l_\nu$, keeping terms up to $\sim 1$. We find

\begin{equation}
	\begin{alignedat}{2}
	& f \approx && \ 2 s_\nu + \mathbf{k}_\perp \cdot \mathbf{q}_\nu - \frac{(\mathbf{k}_\perp \cdot \hat{\mathbf{Y}}_\nu)^2}{4 \Psi_{YY\nu}} - \left( k_1 - \mathbf{k}_\perp \cdot \hat{\mathbf{g}}_\nu + \frac{ (\mathbf{k}_\perp \cdot \hat{\mathbf{Y}}_\nu) (\mathbf{k}_\perp \cdot \hat{\mathbf{g}}_\nu) \text{d}\alpha / \text{d}l_\nu}{2 \Psi_{YY\nu}} \right) (l_b - l_\nu) \\
	& && - \frac{ (\mathbf{k}_\perp \cdot \hat{\mathbf{g}}_\nu)^2 (\text{d}\alpha/\text{d}l_\nu)^2 }{4\Psi_{YY\nu} } {(l_b - l_\nu)^2}.
 	\end{alignedat}
	\label{phase_amp_nu}
\end{equation}
Note that the dependence on $k_2$ in equation \eqref{valerian_aomega_dnxy} is now hidden in $l_b$ through equation \eqref{eyeyb_exp}. 

We compute the $k_2$ integral in equation \eqref{valerian_aomega_dnxy} by changing variables of integration from $k_2$ to $l_b$ by $\text{d}k_2 = |\mathbf{k}_\perp \cdot \hat{\mathbf{g}}_b \ \text{d}\alpha / \text{d}l_b | \text{d}l_b$. The integral in $l_b$ is a complex Gaussian. We find

\begin{equation}
	\begin{alignedat}{2}
	& \int && \frac{\text{d}k_2}{|\mathbf{k}_\perp \cdot \hat{\mathbf{g}}_b \ \text{d}\alpha / \text{d}l_b |} \exp\left[ i f \right] = \left( \frac{- 4 i \pi \Psi_{YY\nu} }{ (\mathbf{k}_\perp \cdot \hat{\mathbf{g}}_\nu)^2 (\text{d}\alpha/\text{d}l_\nu)^2 } \right)^\frac{1}{2} \exp \left[ i h_\nu (k_1) \right] ,
	\end{alignedat}
	\label{dk2_integral}
\end{equation}
where the function $h_\nu (k_1) $ is given by 

\begin{equation}
	\begin{alignedat}{2}
	& h_\nu (k_1) = 2 s_\nu + \mathbf{k}_\perp \cdot \mathbf{q}_\nu + \frac{ \mathbf{k}_\perp \cdot \hat{\mathbf{Y}}_\nu (k_1 - \mathbf{k}_\perp \cdot \hat{\mathbf{g}}_\nu) }{ \mathbf{k}_\perp \cdot \hat{\mathbf{g}}_\nu \ \text{d}\alpha / \text{d}l_\nu } + \frac{ \Psi_{YY\nu} \left( k_1 - \mathbf{k}_\perp \cdot \hat{\mathbf{g}}_\nu \right)^2 }{ (\mathbf{k}_\perp \cdot \hat{\mathbf{g}}_\nu)^2 (\text{d}\alpha/\text{d}l_\nu)^2 } . \\
	\end{alignedat}
	\label{h_phase}
\end{equation}
Having completed the $k_2$ integral, equation \eqref{valerian_aomega_dnxy} now takes the following form

\begin{equation}
	\begin{alignedat}{2}
	& {A}_r(t) =&& -A_{ant} \sum_\nu \int { \text{d}k_x \text{d}k_y } \  {\delta \hat{n}({k}_x, k_y)} \int {\text{d}k_1} \left[ \frac{ \Im[{\Psi}_{YY}] }{\left(g {\text{d}K}/{ \text{d} \tau} \right)  (\mathbf{k}_\perp \cdot \hat{\mathbf{g}})^2 (\text{d}\alpha/\text{d}l)^2 }  \right]^{\frac{1}{2}}_\nu \\
	& && \times \frac{ \Big[ \bold{\hat{e}}^* \cdot ( \boldsymbol{\epsilon}_{eq} - \bold{1} ) \cdot \bold{\hat{e}} \Big]_\nu }{n_\nu} \exp\left( i 2\phi_{G\nu} \right) \exp \left[ i h_\nu (k_1) \right]. \\
	\end{alignedat}
	\label{valerian_aomega_dk1}
\end{equation}

Note how the integrand under the $k_1 $ integral sign in equation \eqref{valerian_aomega_dk1} has a non-trivial dependence on $k_1$ through the Bragg condition $k_1 = - 2 K_\nu$ in equation \eqref{bragg_k1_0}. However, the integral can be simplified by noting that $k_1 - \mathbf{k}_\perp \cdot \hat{\mathbf{g}}_\mu \sim 1/W$. This follows from the Bragg conditions \eqref{bragg_dfdt_xy} and \eqref{bragg_k1_0}. Computing the difference between equations \eqref{bragg_dfdt_xy} and \eqref{bragg_k1_0}, we find 

\begin{equation}
	\begin{alignedat}{2}
	k_1 - \mathbf{k}_\perp \cdot \hat{\mathbf{g}}_\mu \approx -2 \frac{\text{d}K}{\text{d}l_\mu} (l_\nu - l_\mu). 		\end{alignedat}
	\label{k1_kperpmu_diff}
\end{equation}
Since $l_\nu - l_\mu \sim W$ following equation \eqref{eyeyb_exp}, we Taylor-expand terms in $h_\nu$ in powers of $(k_1 - \mathbf{k}_\perp \cdot \hat{\mathbf{g}}_\mu)$. When calculating $\text{d}h_\nu/\text{d}k_1$ and $\text{d}^2h_\nu/\text{d}k_1^2$, we will make use of equation \eqref{k1_kperpmu_diff}, so that terms that are functions of $l_\nu$ can be calculated by using the chain rule 

\begin{equation}
	\begin{alignedat}{2}
	\frac{\text{d}}{\text{d}k_1} = - \frac{1}{2(\text{d}K/\text{d}l_\nu)} \frac{\text{d} }{ \text{d}l_\nu} .
	\end{alignedat}
	\label{d_dk1_l}
\end{equation}
Equation \eqref{d_dk1_l} is particularly useful for computing $\text{d} (\mathbf{k}_\perp \cdot \hat{\mathbf{g}}_\nu)/\text{d}k_1$ and $\text{d} ( \mathbf{k}_\perp \cdot \hat{\mathbf{Y}}_\nu)/\text{d}k_1$. Using equations \eqref{unitvec_deriv_l} and \eqref{d_dk1_l} , we have 

\begin{equation}
	\begin{alignedat}{2}
	& \frac{\text{d} ( \mathbf{k}_\perp \cdot  \hat{ \mathbf{g}}_\nu) }{\text{d}k_1} = \frac{ {\text{d} \alpha }/{ \text{d}l_\nu} }{2(\text{d}K/\text{d}l_\nu)} \mathbf{k}_\perp \cdot \hat{\mathbf{Y}}_\nu \sim \frac{\lambda}{W} \ll 1 , \\
	& \frac{\text{d} ( \mathbf{k}_\perp \cdot  \hat{ \mathbf{Y}}_\nu ) }{\text{d}k_1} = - \frac{ {\text{d} \alpha }/{ \text{d}l_\nu} }{2(\text{d}K/\text{d}l_\nu)} \mathbf{k}_\perp \cdot \hat{\mathbf{g}}_\nu \sim 1 ,
	\end{alignedat}
	\label{dgdk1_dydk1}
\end{equation}
where we have used $\mathbf{k}_\perp \cdot  \hat{ \mathbf{Y}}_\nu \sim 1/W$. Using equations \eqref{k1_kperpmu_diff}, \eqref{d_dk1_l}, and \eqref{dgdk1_dydk1}, the large phase term $2s_\nu + \mathbf{k}_\perp \cdot \mathbf{q}_\nu$ can be expanded as 

\begin{equation}
	\begin{alignedat}{2}
	& 2 s_\nu + \mathbf{k}_\perp \cdot \mathbf{q}_\nu = 2 s_\mu + \mathbf{k}_\perp \cdot  \mathbf{q}_\mu + \frac{1}{ 4 \text{d}K/{\text{d}l_\mu} } (k_1 - \mathbf{k}_\perp \cdot \hat{\mathbf{g}}_\mu)^2 + O \left( \frac{\lambda}{W} \right),  \\
	\end{alignedat}
	\label{2snu_knu_app}
\end{equation}
where we have used the Bragg condition $2 K_\mu + \mathbf{k}_\perp \cdot \hat{\mathbf{g}}_\mu \approx 0$. Next, we expand the third and fourth terms of $h_\nu$ in equation \eqref{h_phase}. We have 

\begin{equation}
	\begin{alignedat}{2}
	& \frac{ \mathbf{k}_\perp \cdot \hat{\mathbf{Y}}_\nu (k_1 - \mathbf{k}_\perp \cdot \hat{\mathbf{g}}_\nu) }{ \mathbf{k}_\perp \cdot \hat{\mathbf{g}}_\nu \ \text{d}\alpha / \text{d}l_\nu } = \mathbf{k}_\perp \cdot \hat{\mathbf{Y}}_\mu  \frac{k_1 - \mathbf{k}_\perp \cdot \hat{\mathbf{g}}_\mu}{ \mathbf{k}_\perp \cdot \hat{\mathbf{g}}_\mu \ \text{d}\alpha / \text{d}l_\mu }  - \frac{ (k_1 - \mathbf{k}_\perp \cdot \hat{\mathbf{g}}_\mu)^2 }{ 2 \text{d}K / \text{d}l_\mu } + O\left( \frac{\lambda}{W} \right) , \\
	& \frac{ \Psi_{YY\nu} \left( k_1 - \mathbf{k}_\perp \cdot \hat{\mathbf{g}}_\nu \right)^2 }{ (\mathbf{k}_\perp \cdot \hat{\mathbf{g}}_\nu)^2 (\text{d}\alpha/\text{d}l_\nu)^2 } =  \frac{ \Psi_{YY\mu} \left( k_1 - \mathbf{k}_\perp \cdot \hat{\mathbf{g}}_\mu \right)^2 }{ (\mathbf{k}_\perp \cdot \hat{\mathbf{g}}_\mu)^2 (\text{d}\alpha/\text{d}l_\mu)^2 } + O\left( \frac{\lambda}{W} \right) . \\
	\end{alignedat}
	\label{hnu_third_fourth}
\end{equation}
The phase $h_\nu$ now has explicit dependence on $(k_1 - \mathbf{k}_\perp \cdot \hat{\mathbf{g}}_\mu)$, and it is given by

\begin{equation}
	\begin{alignedat}{2}
	& h_\nu = 2 s_\mu + \mathbf{k}_\perp \cdot  \mathbf{q}_\mu + \frac{ \mathbf{k}_\perp \cdot \hat{\mathbf{Y}}_\mu (k_1 - \mathbf{k}_\perp \cdot \hat{\mathbf{g}}_\mu) }{ \mathbf{k}_\perp \cdot \hat{\mathbf{g}}_\mu \ \text{d}\alpha / \text{d}l_\mu } + \frac{ \text{d}^2 h_\nu }{  \text{d} k_1^2 } \bigg|_{\mu} \frac{(k_1 - \mathbf{k}_\perp \cdot \hat{\mathbf{g}}_\mu)^2}{2} + O\left( \frac{\lambda}{W} \right) , \\
	\end{alignedat}
	\label{hnu_final_k1}
\end{equation}
where $ { \text{d}^2 h_\nu }/{  \text{d} k_1^2 }\big|_{\mu} \sim L \lambda$ is given by

\begin{equation}
	\begin{alignedat}{2}
	\frac{ \text{d}^2 h_\nu }{  \text{d} k_1^2 } \bigg|_{\mu} = \frac{ 2 \Psi_{YY\mu} }{ (\mathbf{k}_\perp \cdot \hat{\mathbf{g}}_{\mu})^2 (\text{d}\alpha/\text{d}l_{\mu})^2 } - \frac{1}{2 \ \text{d}K/ \text{d}l_{\mu}} .
	\end{alignedat}
	\label{dh2dk12}
\end{equation}
With this approximation to $h_\nu$, the integral over $k_1$ in equation \eqref{valerian_aomega_dk1} becomes a Gaussian integral that can be easily evaluated. Noting that $ \mathbf{k}_\perp \cdot \hat{\mathbf{Y}}_\mu = q_{2\mu} $ from equation \eqref{rotation_q_k}, we find that the integral in $k_1$ in equation \eqref{valerian_aomega_dk1} recovers equations \eqref{as_kxky_dnxy} and \eqref{fkxky}. This is proof that the representations of the density in beam-aligned coordinates in \citep{valerian_ppcf_2022} and in Cartesian coordinates in \citep{gusakov_ppcf_2014, gusakov_pop_2017} are completely equivalent. 

We are now in a position to calculate the difference between $l_\mu$ issued from the Cartesian Bragg condition in equation \eqref{bragg_dfdt_xy} and $l_\nu$ from the beam-aligned Bragg condition in equation \eqref{bragg_k1_0}. Computing the difference between equations \eqref{bragg_k1_0} and \eqref{bragg_dfdt_xy}, we found equation \eqref{k1_kperpmu_diff}. Since according to equations \eqref{hnu_final_k1} and \eqref{dh2dk12} we have $k_1 - \mathbf{k}_\perp \cdot \hat{\mathbf{g}}_\mu \sim 1/W$, we find that $l_\nu - l_\mu \sim W$.

We note that the equivalence obtained here between the beam-aligned and Cartesian representation of the density is proven for the scattered \emph{amplitude}, and not directly for the power. It is possible to directly prove the equivalence in the scattered \emph{power} from the beam-aligned expression in $(k_1,k_2)$ in \citep{valerian_ppcf_2022} to the Cartesian expression in $(k_x, k_y)$ in equations \eqref{prkxky_cart}, \eqref{fkxky_2_cartesian_}. This requires mapping the correlation functions between beam-aligned $\hat{C}_b$ and Cartesian coordinates $\hat{C}$, which is non-trivial and will be the object of a future publication. 

It is worth commenting on the equivalence of the scattered power contributions for beam-aligned and Cartesian coordinates when regarded as a one-dimensional integral in $k_x$ or $k_1$. In this case, it is straightforward to directly map the Cartesian and beam-aligned representations of the scattered power, as follows.

The backscattered power in Cartesian coordinates from equation \eqref{prkxky2_cart} can be rewritten as an integral over the beam-aligned $k_1$ \citep{valerian_ppcf_2022}. To express the one-dimensional integral in \eqref{prkxky2_cart} in terms of $k_1$, change variables using $\text{d}k_x = \text{d}k_1 K_\mu / |K_{x\mu}| = \text{d}k_1/|\cos\alpha_\mu|$, where we used equations \eqref{sincos}. We find
   
\begin{equation}
	\begin{alignedat}{2}
	&\frac{p_r}{P_{ant}} &&\approx \pi^\frac{3}{2} K_0L \frac{e^4}{m_e^2 \epsilon_0^2 \Omega^4} \sum_{\mu} \int{\text{d}k_1 } \frac{ \langle |\delta \hat{n}_{\mu}(k_x, -2K_0\sin\alpha_0)|^2 \rangle_T }{ | {K_{x\mu}}/{K_0} | W_{Y\mu} } , \\
	\end{alignedat}
	\label{prk1_beam_}
\end{equation}   
where we have used equation \eqref{dkdtau} to write the term $|g_\mu \text{d}K/\text{d}\tau|$ as a function of $K_{x\mu}$. The dependence of $k_x$ on $k_1$ in equation \eqref{prk1_beam_} can be explained as follows. First, $k_x$ is related to the location along the path $l_\mu$ through the Bragg condition in Cartesian coordinates in equation \eqref{taumu_bar}. In turn, $l_\mu$ is related to $k_1$ through the Bragg condition when expressed in beam-aligned coordinates, $k_1 \approx - 2 K_\mu$ to lowest order, since the difference $l_\mu - l_\nu$ is of order $\sim W$ as we just saw. 

Equation \eqref{prk1_beam_} recovers the same result as in \citep{valerian_ppcf_2022} when using $\langle |\delta \hat{n}_{b,\mu}|^2 \rangle_T$. The one-dimensional change of variables between $k_x$ and $k_1$ shows that, to lowest order and at scales $\sim 1/\lambda$, the density fluctuation spectra in Cartesian coordinates is related to the beam-aligned spectrum of \cite{valerian_ppcf_2022} by a rotation of angle $\alpha_\mu$, and we have $\langle |\delta \hat{n}_{\mu}(k_x, -2K_0\sin\alpha_0)|^2 \rangle_T = \langle |\delta \hat{n}_{b,\mu}( k_x \cos\alpha_\mu - 2K_0\sin\alpha_0 \sin\alpha_\mu, 0)|^2 \rangle_T$. Noting that $k_x \cos\alpha_\mu - 2K_0\sin\alpha_0 \sin\alpha_\mu \approx -2 K_\mu \approx k_1$ by use of the Bragg conditions, we have therefore recovered the equivalence in the one-dimensional scattered power in both representations.

\bibliography{mybib11}
\bibliographystyle{jpp}

\end{document}